\documentclass[preprint]{aastex}
\usepackage{psfig}

\def\plotfiddle#1#2#3#4#5#6#7{\centering \leavevmode
    \vbox to#2{\rule{0pt}{#2}}
    \includegraphics{#1}}
\def\alwaysmath#1{\ifmmode{#1}\else{$#1$}\fi}

\received{March 9, 2003}
\accepted{}
\journalid{Astrophysical Journal}

\lefthead{Majewski et al.}
\righthead{2MASS View of the Sagittarius Dwarf Galaxy }

\begin{document}
\title{A 2MASS All-Sky View of the Sagittarius Dwarf Galaxy: I. Morphology of the
Sagittarius Core and Tidal Arms }

\author{Steven R. Majewski\altaffilmark{1}, 
M.~F. Skrutskie\altaffilmark{1}, 
Martin D. Weinberg\altaffilmark{2}, and
James C. Ostheimer\altaffilmark{1}
}

\altaffiltext{1}{Dept. of Astronomy, University of Virginia,
Charlottesville, VA, 22903-0818 (srm4n@virginia.edu, mfs4n@virginia.edu, jco9w@virginia.edu)}

\altaffiltext{2}{Dept. of Astronomy, University of Massachusetts at Amherst,
719 North Pleasant Street, Amherst, MA 01003-9305
(weinberg@astro.umass.edu)}

\begin{abstract}

We present the first all-sky view of the Sagittarius (Sgr) dwarf galaxy
mapped by M giant star tracers detected in the 
complete Two Micron All-Sky Survey (2MASS).  

Near infrared photometry of Sgr's prominent M giant population
permits an unprecedentedly clear view of the center of Sgr.
The main body is fit with a King profile of limiting 
major axis radius 30$^{\circ}$ ---
substantially larger than previously found or assumed ---
beyond which is a prominent break in the density profile 
from stars in its tidal tails; thus the Sgr radial profile
resembles that of Galactic dSph satellites.
Adopting traditional methods for analyzing dSph light profiles, 
we determine the brightness of the main body of Sgr
to be $M_V = -13.27$ (the 
brightest of the known Galactic dSph galaxies) and
the total Sgr mass-to-light ratio to be 25 in solar units.
However, we regard the latter result with suspicion and 
argue that much of the observed structure beyond the King fit
core radius (224 arcmin) may be outside the actual Sgr tidal
radius as the former dwarf spiral/irregular satellite undergoes 
catastrophic disruption over its past last orbit.  
The M giant distribution of Sgr exhibits a central density cusp 
at the same location as, but not due to,
the old stars constituting the globular cluster M54.

%
%

A striking
trailing tidal tail is found to extend from the Sgr center and 
arc across the South Galactic Hemisphere with approximately 
constant density and mean distance varying from $\sim 20-40$ kpc.
A prominent leading debris arm extends from the Sgr center
northward of the Galactic plane to an apoGalacticon $\sim45$ kpc from 
the Sun, then turns towards the North Galactic Cap 
(NGC) from where it descends back towards the Galactic plane, becomes 
foreshortened and at brighter magnitudes covers the NGC.
The leading and trailing Sgr tails lie along a well-defined orbital
plane about the Galactic Center.  The Sun lies within a kiloparsec of
that plane and near the path of leading Sgr debris; thus, 
it is possible that former Sgr stars are near or in 
the solar neighborhood.

We discuss the implications of this new view of the Sgr galaxy and its
entrails for the character of the Sgr orbit, mass, mass-loss rate, and 
contribution of stars to the Milky Way halo.  
The minimal precession displayed by the Sgr tidal debris along its 
inclinded orbit supports the notion of a nearly spherical Galactic potential.
The number of M giants in the Sgr tails is at least 15\%
that contained within the King limiting radius of the main Sgr body.
The fact that M giants, presumably formed within the past few gigayears
in the Sgr nucleus, are nevertheless so widespread along the Sgr 
tidal arms not only places limits on the dynamical age of these arms
but poses a timing problem that bears on the recent binding energy of the 
Sgr core and that is most naturally explained by recent 
and catastrophic mass loss.
Sgr appears to contribute more than 75\% of the high latitude, halo M giants,
despite substantial reservoirs of M giants in the Magellanic
Clouds.  No evidence of extended
M giant tidal debris from the Magellanic Clouds is found.

Generally good correspondence is found between the M giant, all-sky map
of the Sgr system and all previously published detections of  potential Sgr debris, 
with the exception of Sgr carbon stars --- which must be subluminous compared
with counterparts in other Galactic satellites in order to resolve the
discrepancy.

\end{abstract}


\section{Introduction}

The Sagittarius (Sgr) dwarf galaxy is a striking example of
the process of satellite disruption and assimilation long presumed
responsible for populating the Galactic halo (e.g., Searle \& Zinn
1978).  Alternatively, viewed as test particles, a sufficiently
complete spatial and kinematical sample of Sgr stars can reveal 
underlying gravitational potentials -- tracing the total mass of luminous and 
dark matter in both Sgr and the Milky Way  
(Sackett et al. 1994, Johnston et al. 1999b, Ibata et al. 2001b).  
Since the discovery of Sgr (Ibata et
al. 1994) there have followed a number of observations (reviewed below and
in Section 8) to characterize
the distribution and motion of the tidal debris and which have
aided models 
of the disruption of the satellite in the Milky Way's potential.  
Early observations were largely restricted to
small fields-of-view, but nevertheless painted the general
picture of a substantially tidally disrupted satellite
distributed across a sizeable portion of the celestial sphere.

However, many issues remain controversial and intertwined, particularly:

\begin{itemize}

\item The dark matter content in the bound Sgr system, which is integrally
tied to the long-term integrity and mass loss rate of the satellite
(e.g., Ibata \& Lewis 1998, G\`omez-Flechoso 1998, G\`omez-Flechoso, Fux \& Martinet 1999).

\item The survivability of the Sgr system in its present orbit, 
which, even if it contains substantial dark matter, should not last 
a Hubble time (Vel\'azquez \& White 1995, Johnston et al. 1995).  
Solutions to this dilemma range from a fine
tuning of the dark matter configuration within the satellite (Ibata \& Lewis
1998), to an evolving orbit for the satellite (Zhao 1998, 
G\`omez-Flechoso, Fux \& Martinet 1999), or the creation of Sgr at
later times as the daughter product of another, more major merger
(G\`omez-Flechoso, Fux \& Martinet 1999).  

\item The original mass of the Sgr satellite, which today is smaller
by an amount depending on the mass density (dark 
matter content) and orbital history of the system.    

\item The fractional contribution of Sgr stars (e.g., Ibata et al. 2001b,
 Vivas et al. 2001, Newberg et al. 2002) and clusters 
(e.g., Irwin 1999, Dinescu et al. 2000, 2001, 
Palma et al. 2002, Bellazzini et al. 2002a, 2003) to the
Galactic halo, which must also satisfy limits imposed
by the distribution of properties for Galactic halo field stars 
(cf. Unavane, Wyse \& Gilmore 1996, Majewski et al. 2002b).  

\item The shape and size of the Galactic halo.  
Tidal tails are extremely sensitive to the amount and
distribution of mass in the
Galaxy (e.g., Johnston et al.\ 1999b, Murali \& Dubinski 1999).
At least one study of the Sgr orbit (Ibata et al. 2001b) 
suggest the Milky Way dark halo to be nearly spherical, which 
places the Milky Way at an extreme of the wide range of dark halo 
flattenings (e.g., Sackett \& Pogge 1995, Olling 1997, 
Pe{\~n}arrubia et al. 2000, Sparke 2002).  
This result is at odds with (1) starcount studies, which typically 
find $c/a \sim 0.6-0.8$ (e.g., Robin, Reyl{\' e} \& Cr{\' e}z{\' e} 2000,
Siegel et al. 2002, Reid \& Majewski 1993, and references therein), (2) 
{\it dynamical} studies of halo tracers (Binney, May \& Ostriker 1987,
Amendt \& Cuddeford 1994; see also van der Marel 1991) and of HI layers
(see summary by Merrifield 2002),
(3) Galactic microlensing studies, which imply a flattened halo
(Samurovic, Cirkovic \& Milosevic-Zdjelar 1999),
and (4) expectations for very triaxial halos in models of structure
formation in the presence of Cold Dark Matter (e.g., Frenk et al. 1988, 
Dubinski \& Carlberg 1991, Warren et al. 1992)
and especially models that include gas dissipation (e.g., 
Dubinski 1994, Steinmetz \& Muller 1995).

\item The degree of visible substructure in the halo, which is
directly related to the Galactic accretion history (Tremaine 1993, Bullock,
Kravtsov \& Weinberg 2001).  For luminous halo stellar populations some mixture of Eggen,
Lynden-Bell \& Sandage (1962) and Searle \& Zinn (1978) formation pictures is typically
postulated (e.g., see review by Majewski 1993), 
but evidence is increasing that the stellar halo is only weakly phase mixed and
highly substructured (Majewski, Munn \& Hawley 1996, Vivas et al. 2001,
Gilmore, Wyse \& Norris 2002, Majewski 2003).
The contribution of Sgr to this substructure is not well established.
Recent evidence suggests that Sgr contributes of order 5\% of the halo
M giants in a correlated stream (Ibata et al. 2002a; but see Section 7.2).  

\item The degree of {\it invisible} substructure in the halo.  Cold Dark Matter (CDM)
models for the formation of galaxy halos predict the persistence of long-lived,
``sub-halos'' 
(e.g., Navarro et al.\ 1996, 1997) at a number greatly exceeding the number
of {\it luminous} Galactic satellites (Klypin et al. 1999, Moore et al. 1999).  
The degree of  coherence of tidal debris streams
provide a powerful constraint on the lumpiness of the 
Galactic halo potential (Font et al.\ 2001, Johnston et
al.\ 2002, Ibata et al.\ 2001b, Mayer et al.\ 2001).

\end{itemize}

Models for the interaction of Sgr with the Milky Way under different 
assumptions of orbit, Galactic potential, and Sgr dark matter content
make distinct and testable predictions for the appearance of the satellite 
and its debris today (e.g., Vel\'azquez \& White 1995, Johnston et al. 1995, 
Edelshohn \& Elmegreen 1997, Ibata et al. 1997, Ibata \& Lewis 1998,
G\`omez-Flechoso, Fux \& Martinet 1999,
Johnston et al. 1999a,
Jiang \& Binney 2000,
Helmi \& White 2001).  Thus, improved 
observational constraints --- e.g., on the detailed distribution 
(e.g., G\`omez-Flechoso et al. 1999, Jiang \& Binney 2000)
and degree of coherence of the Sgr debris (Ibata et al. 2002b,
Johnston, Spergel \& Haydn 2002) --- can greatly increase our understanding
of both the Milky Way and Sgr systems.

Previous studies of Sgr include
a patchwork of approaches and directions
of the sky, but, aided by the advent of large area surveys, including SDSS and
QUEST, a more global picture of the Sgr dwarf and its remains has begun to 
emerge.  For example, a survey locating 75 high latitude, halo carbon stars 
by Ibata et al. (2001b) finds more than half to lie along a great circle 
consistent with the likely Sgr orbit.  Confirmation of this great circle comes
from the location of coherent clumps of A-type (Yanny et al. 2000), RR
Lyrae (Ivezi{\' c} et al. 2000, Vivas et al. 2001), red clump (Mateo,
Olszewski \& Morrison 1998, Majewski et al. 1999a) and main sequence
turn-off stars (Mart\'inez-Delgado et al. 2001a,b, 2002, Newberg et al. 2002)
in surveys that intersect this great circle at various points.  
Other studies have suggested an even more complex {\it multiply-wrapped} 
configuration around the Galaxy (Johnston et al. 1999, Dinescu et al. 2000, 
Dohm-Palmer et al. 2001, Kundu et al. 2002).
In some cases, detections of Sgr debris far-flung from the Sgr center are 
supplemented with dynamical information (e.g., 
Majewski et al. 1999a, 
Ibata et al. 2001b, Dinescu et al. 2000, 2002, 
Kundu et al. 2002) useful for constraining the system dynamics.  

However, to date, no single, unbiased, global, empirical characterization 
of Sgr exists.  The Two Micron All Sky Survey (2MASS) remedies
this situation by offering homogeneous photometry in
bandpasses less sensitive to the effects of
reddening and with complete sky coverage.  Studies of the high density, 
inner regions 
of the Sgr galaxy from early release 2MASS data are already in hand
and delineate the bulk photometric characteristics of Sgr stars in the
2MASS bandpasses (Alard 2001, Cole 2001): Sgr apparently contains stars
with a
metal abundance of [Fe/H]=-0.5 or more (see also Bonifacio et al. 1999), 
and, as a result, contains a substantial number of M giants.  The 
combination of $J$, $H$, and $K_s$ passbands permits color-based
discrimination of M giant stars from (foreground) M dwarfs, a fact
that was exploited by Ibata et al. (2002a) to detect an excess of
halo giants defining a great circle with a pole consistent with that 
extracted from the analysis of halo carbon stars.

Here we use the complete all-sky 2MASS source extractions to
characterize the distribution of the Sgr M-giant population as
projected on the sky as well as in three dimensions.  This analysis
reveals a King profile, dSph-like appearance of the central 
region and extensive, well-defined, trailing and leading Sgr 
tidal tails in the South and North Galactic Hemispheres.  Among the 
remarkable aspects of the Sgr tidal debris stream are its coherence
and nearly constant density over 360$^{\circ}$ of orbital longitude
and that tidal debris from Sgr very likely rains down 
from the North Galactic Pole onto the solar neighborhood.

\section{2MASS Selection of Sgr M Giant Candidates }

Near-IR (NIR) colors of giant and dwarf stars are degenerate for early spectral
types ($<$K7 or $J-K_s \lesssim 0.55$) but become distinct in two-color diagrams for
the latest spectral types because of opacity effects (primarily a minimum
in the H$^{-}$ opacity at 1.6$\mu$ with modulation by H$_2^{ -}$) that have a large 
effect in the $H$ band.  The color divergence occurs for sources with 
$J-K \gtrsim 0.85$ (Lee 1970, Glass 1975, Mould 1976, Bessell \& Brett 1988, Cutri 2003). 
A preliminary selection of candidates with $J-K_s > 0.85$ was made 
from the 2MASS (Final Processing) working survey point source database covering 
$>$99.9\% of the sky. 
Sgr tidal features were evident in celestial sphere projections of
color/magnitude cuts even in this preliminary selection.  

Subsequent reddening correction eliminates intrisically
blue objects that do not satisfy our color selection criteria (below).
To account for differential reddening around the Galaxy
we have interpolated $E(B-V)$ for each of the selected stars
with algorithms and reddening data provided by Schlegel et al. (1998),
who give $E(B-V)$ values
derived from IRAS 100 micron emission all-sky maps.  Each source
was then dereddened after adopting the following selective
and total extinction laws: 

$$E(J-K_s) = 0.54 E(B-V) $$
\begin{equation}
\label{reddening}
E(J-H) = 0.34 E(B-V) 
\end{equation}
$$A(K_s) = 0.28 E(B-V). $$

\noindent For the remainder of our analysis, sources with $E(B-V)>0.555$, 
corresponding to $E(J-K_s)=0.30$, 
were excluded to avoid potential contamination from excessively reddened sources.

Noise in stellar colors smears the
distinction between dwarfs and giants.  2MASS aperture magnitudes are
more precise than the PSF-fit magnitudes for the brighter ($K_s \lesssim 12.5$,
$J < 14$) stars (Cutri et al. 2003).  
The photometry used here is exclusively aperture photometry
where the quoted photometric uncertainty was $<$0.06 mag in all three
bands.  This strong constraint on photometric accuracy imposes a
completeness limit on the 2MASS selection used here of approximately
$K_s<13.5$, substantially brighter than the 99\% survey completeness
requirement of $K_s<14.3$.

Evaluation of the ($J-H$, $J-K_s$) color-color diagram of the 
center of Sgr permits further refinement 
of the color selection criteria for Sgr M giants.  
Figure 1 shows the $(J-K_s,K_s)_o$ color-magnitude and color-color
diagrams for 2MASS point sources in $3\times3$ deg$^2$ areas centered
on the center of Sgr and in a control field centered on a point reflected 
across the Galactic $l=0$ axis at the same Galactic latitude. 
To make the Sgr red giant
branch even more clear, we include (Figure 1c) the results
of a statistical subtraction in color-magnitude space 
of the control field from the Sgr center field 
using the same method as Layden \& Sarajedini (2000) with their aspect
ratio parameter $\alpha = 5$.  Significant small-scale variations
in reddening and population densities mean that the subtraction is not perfect, 
but the position of the Sgr red giant branch (RGB) in both Figures 1c and 1f are 
obvious.\footnote{We show in 
Section 4 that the semi-major axis of Sgr is about
30$^{\circ}$ and the ellipticity ($1-b/a$) is about 0.65,  
which means that the control field here, though near the 
minor axis of Sgr, is still within a radius that contains
a measurable Sgr density.  However, that Sgr
in the field at this radius is about 1\% the center, 
so the presence of some Sgr stars in the control field does not 
effect our interpretation of the CMD here or our analysis of the 
Sgr luminosity function below in Figure 21.}
Features related to the Milky Way bulge (e.g., a ghostly RGB 
and horizontal branch red clump) are visible in the color-magnitude diagram several
magnitudes brighter than the Sgr RGB.

The initial selection for M giants is conservative, balancing a desire 
for a large statistical sample of M giants but with minimal contamination 
by other stars (improperly dereddened dwarfs and stars with large 
photometric errors; see Figure 2a), and satisfies (Figure 2):

$$ J-K_s > 0.85 $$
\begin{equation}
\label{GDsep}
J-H < 0.561(J-K_s) + 0.36
\end{equation}
$$ J-H > 0.561(J-K_s) + 0.22$$

\noindent where all magnitudes are in the intrinsic, dereddened 2MASS system. 

From a partial survey of the central regions of Sgr, Whitelock et al. 
(1999) estimate that the Sgr dwarf galaxy contains of order 100
N-type carbon stars (a slight underestimate ---  
see discussion of Figure 20 below).  Carbon stars, 
because they have extreme, easily identifiable NIR colors,
make them a potentially useful tracer of the Sgr debris stream 
(e.g., Ibata et al. 2001b). However, Sgr carbons have
a large spread in luminosity and a number are long-period 
variable (see Figure 19 below), which yields large uncertainties in 
estimated photometric parallaxes relative to the better defined
M giant color-magnitude relation.  Thus, in this paper we rely predominantly on
the much more populous M giants to trace the Sgr tidal streams, but include 
a discussion of the carbon stars in Section 8.3.

\section{Salient Features of All-Sky Maps of 2MASS M Giant Candidates }

The center of Sgr is a readily apparent feature of all-sky images of the 2MASS
point source catalogue already in the public domain (see, for example, \\ 
http://www.ipac.caltech.edu/2mass/gallery/showcase/allsky/index.html).  
Various aspects of Sgr's debris stream also become readily apparent 
in color-magnitude windows of the point source catalogue 
that highlight Sgr M giants at specific distances.  
Figure 3 shows two such windows.

Several large scale features are evident in Figure 3.  Some
are artifacts of heavy and patchy differential reddening near
the Galactic plane.  Reddening can shift early type stellar colors
into the giant star two-color locus (Figure 2).
Several prominent extensions from the disk correspond
to high latitude dust in the Galaxy also seen in IRAS 100 micron maps.

Figure 3 shows Sgr debris at varying distances in a great circle 
around the entire celestial sphere.
The most prominent M giant features, apart from the Galactic Center
and plane, are the Magellanic Clouds and the Sgr center
at ($\alpha,\delta) = (284, -30)^{\circ}$ (discussed in Section 4).
Stretching from the Sgr center itself, southward for a short span
and then northward, is a ``Southern Arc" .  
In Section 6 we explore the distance distribution of the Arc and show
that it extends physically from the main body of Sgr.
The distance modulus of the Southern Arc is more or less constant
for more than 100$^{\circ}$ from the Sgr center towards the Galactic 
anticenter.  The arc may, in fact,
cross the Galactic plane at the anticenter and cross into the Northern
Galactic Hemisphere, albeit at very low surface density in M giants.

More than a magnitude fainter than the Arc is a spike of
M giants extending prominently northward from the Galactic Center direction. 
This ``Northern Arm" has a longitudinal gradient in median $K_s$ magnitude, 
which indicates a growing distance with increasing 
Galactic latitude until reaching the North Galactic Cap.
The brighter magnitude slice in Figure 3 shows a subtle, more extended, 
``fluffy" distribution of M giant candidates situated 
near the termination point of the Northern Arm.
The fluffy concentration spans several tens of degrees, obviously wider 
than either the Northern Arm and Southern Arc, but with less surface density.
 As we show in Section 6, the bright northern fluff
represents a severely foreshortened extension of the Northern
Arm.  Indeed, in even brighter magnitude windows, this ``North Fluff" 
is still present, but is spread out over steradians.
At the brightest 2MASS magnitudes it encompasses 
nearly all of the Northern Galactic Hemisphere, as would be expected
for a large structure very near the Sun (Section 6.4).

\section{Analysis of the Sagittarius Center}

The extended, low surface brightness, central regions of Sgr lie 
nearly behind the Galactic Center, which leads to significant contamination 
by foreground Milky Way bulge and disk stars 
and obscuration by patchy foreground dust.  Early star count analyses by Ibata,
Gilmore \& Irwin (1995) and Ibata et al. (1997) explored the center and southwards
(Mateo, Olszewski and Morrison 1998).  More recently,
studies of RR Lyrae stars examined the more obscured regions north of the center
(Cseresnjes, Alard \& Guibert 2000, see also Alard 1996, Alcock et al. 1997).  
However, the integrated structure of central part of Sgr remains uncertain.  
For example, while roughly agreeing on the derived Sgr major axis scale length 
of fit exponential profiles, Mateo et al. (1998) identify a break in the 
southern profile of Sgr at a radius of 20 degrees while Cseresnjes et al. (2000)
identify one at a substantially different density in the northern 
profile at only six degrees.

Despite its rather irregular appearance in optical maps, 
Sgr is most often assumed to be a dwarf spheroidal (dSph) type galaxy,
because: (1) Sgr lacks significant HI
(Koribalski, Johnston, \& Otrupcek 1994, Burton \& Lockman 1999), 
(2) it contains both old stars and has experienced extended star 
formation epochs (cf. Mateo et al. 1995)  but 
is not presently forming strars (Bellazzini et al. 1999b), 
and (3) like other dSph galaxies, Sgr appears to have a rather 
large $M/L$ (Ibata et al. 1997).  
Alternatively, it is often postulated (Bassino \& Muzzio 1995,
Sarajedini \& Layden 1995, Layden \& Sarajedini 2000; see also discussion
by Da Costa \& Armandroff 1995) that because the center of Sgr appears 
to coincide with the globular cluster M54 (Ibata et al. 1995, 1997)
that the M54+Sgr combination may represent a {\it nucleated dwarf elliptical} 
galaxy (Zinnecker et al. 1988, Freeman 1993).  
Because Sgr exhibits
an apparent metallicity gradient with overall higher metallicity in the center
(Bellazzini et al. 1999a,b, Alard 2001 - though see Cseresnjes 2001) 
along with rather young, $\sim 0.5$ to $\sim 3$ Gyr
stellar populations there (cf. Bellazzini et al. 1999b, 
Layden \& Sarajedini 2000), Alard (2001) postulates a third scenario - that
Sgr may be more like an LMC-type
galaxy with an inner disk, or perhaps a recently disrupted disk.  
While other dSph galaxies have been shown to have metallicity gradients
(e.g., Light 1988, Da Costa et al. 1996, Hurley-Keller, Mateo \& Grebel 1999, 
Majewski et al. 1999b, Harbeck et al. 2001), Sgr 
has among the most young and metal-rich constituent populations for dSph's in the
Local group.  
Weinberg (2000) showed that the stellar disk of a dwarf spiral
without a massive dark matter halo, such as the LMC, will be heated
by a combination of resonant tidal forcing and precession to form
a spheroidal in several gigayears.   Recent simulations by Mayer et al. (2001) 
confirm this result. 
We believe that a consistent and natural interpretation of the disparate
facts that follow is that (1) because Sgr was recently undergoing significant
star formation, it must have formerly been a dwarf spiral or irregular
galaxy, and (2) a combination of mass loss from tidal encounters and star
formation activity has transformed Sgr into a dSph and brough it to 
a point of critical stability. 

The 2MASS database presents the first opportunity for a large-scale,
uniform study of the extended central parts of Sgr at NIR wavelengths where the
effects of reddening are diminished.  These data clarify a number
of the above issues regarding the nature of the Sgr galaxy.

\subsection{Radial Profile Fits to the Sgr Center}

%
%
Figure 4 illustrates the distribution of candidate
M giants around the nominal Sgr center and 
selected with $E(B-V)<0.555$, $0.95<(J-K_s)_o<1.10$ and
$10.5<(K_s)_o<12.0$.  This magnitude range excludes strong 
foreground contamination from brighter Galactic bulge/disk 
stars and a smaller number of likely disk stars at fainter magnitudes 
(see discussion of Figure 8 below).
The NIR appearance of Sgr is far smoother than that seen in
optical starcount analyses (e.g., Ibata,
Gilmore \& Irwin 1995, Ibata et al. 1997):  At NIR wavelengths
the central body of the Sgr system closely resembles a dSph
galaxy.

To show quantitatively its similarity to the appearance of dSph
galaxies, we fit the least reddened M giant
data for the Sgr center --- that with $b < -10^{\circ}$ (Figure 4)
---  with two functional forms using the Bayesian methods and 
analytical forms described by Ostheimer et al. (2003): 
a King (1962) profile and a power law plus core (PLC) 
model.  Table 1 and Figure 5 summarize these results.

Proceeding with such fits to the center of the Sgr system requires 
careful consideration of the
density contribution by the long Sgr tidal arms (Figure 3).  
A simple King profile fit to starcounts at all positions angles (Figure 5a) 
does not properly account for the unbound tidal debris in these
arms manifested as a ``break" 
from a nominal King shape at a major axis radius of about 1300 arcmin 
(even more visible in Figure 5c and Figure 12 below).
A model fit primarily along the Sgr minor axis minimizes the influence 
of these tidal arm stars (situated mainly along the major axis at large radii), 
and yields a King parameterization that
fits the observed density well to an equivalent major axis radius
of 1500 arcmin (Figure 5b).  This King parameterization is given in Table 1.
%
%
%
Transferring this King profile, derived primarily from 
Sgr stars along the minor axis, back to the average radial profile 
from stars at all azimuthal angles
(i.e., all Figure 4 data with $b < -10^{\circ}$) 
reveals more clearly the transition from the central Sgr configuration
to stars in the Southern Arc, which have a more or less constant
surface density along their extent (see Figure 13 below), but which 
present a power law  
decline when included in the azimuthal profile average (Figure 5c).
Despite the major improvement in the fit of the King profile at interior radii,
the position angle (100.5$^{\circ}$ versus 104.3$^{\circ}$), the ellipticity
($\epsilon = 1 - b/a = 0.62$ versus 0.65) and the declination of the core 
($\Delta\delta = 3$ arcmin)
change only slightly with this fit to the major-axis-truncated data. 

A single power law fit to the full azimuthally averaged 
radial density profile can only accommodate 
the general character of the density decline (Figure 5d), but not the 
detailed shape of the radial profile; for example, it ``averages over" the 
kink in the density profile discussed above.  
The mean power law fit (with $\nu = 1.44$; see Ostheimer et al. 2003 for
the precise function of the model) corresponds at large radii to an
$r^{-\gamma}$ decline with $\gamma \sim 2.88$, which is overly steep 
beyond the break near 1300 arcmin, where the data appear to decline more 
like $\gamma \sim 2$.  This $\gamma \sim 2$ decline is similar
to that observed beyond the King limiting radius of the
Carina dSph (Majewski et al. 2000).

The Table 1 King and PLC model fits yield Sgr position angles in the range of
those found by previous analyses: Mateo, Olszewski \& Morrison (1998)
obtained a position angle of $104.8^{\circ} \pm 1.2^{\circ}$ while 
Cseresnjes, Alard \& Guibert (2000) obtained a PA of $108.4^{\circ}$.
Not surprisingly, both of our model fits find the center of the Sgr system
to be of high ellipticity, $\epsilon=0.62-0.65$, though not quite as high 
as the $\epsilon=0.80\pm0.15$ previously reported (Mateo 1998).  Fits
to the centers of several other Local Group dSph galaxies yield
similar ellipticities, e.g., in And III (Caldwell et al.
1992, Ostheimer et al. 2003) and Ursa Minor (Irwin \& Hatzidimitriou 1995, 
Kleyna et al. 1998, Bellazzini et al. 2002b,
Palma et al. 2003).  Indeed, the structure of Sgr bears similarity 
to Ursa Minor (Mart\'inez-Delgado et al. 2001a, 
Palma et al. 2003) and Carina (Majewski et al. 2000) for which 
significant tidal disruption has been proposed.  Extreme ellipticities 
compared to the standard for dwarf galaxies (e.g., Sung et al. 1998)
already suggests that we are observing systems in a disrupting, 
transient state.

\subsection{Departures from a King Profile}

\subsubsection{Nucleus}

Two differences of the observed versus fitted radial profiles are noteworthy: the
presence of a break population (Section 4.2.2) and the
appearance of a ``nucleus" within about 20-25 arcminutes that is
elevated above the density trend immediately exterior to this radius.  
The centers of the King and PLC fits lie
within a few arcminutes of the center of the massive
globular cluster NGC6715 (M54) at 
$(\alpha, \delta)_{2000}=($18:55:03.3,-30:28:42).
The excess 2MASS starcounts over the flat core of the King
profile fit might be attributed to the cluster itself, however, 
(1) the core, half-light and tidal radii of M54 have been derived 
(Trager, King \& Djorgovski 1995) as 0.11, 0.46 and 7.5 arcmin ---
too small to account for the extent of the excess observed; and
(2) the 
metallicity of M54 is [Fe/H]=-1.7 and its age is 14-16 Gyr (Layden
\& Sarajedini 2000), so it should not contain stars as red as the
M giants used for the fits (and shown in Figure 4) 
and is thus ``invisible" to our survey
(the tip of the corresponding RGB is at 
$[J-K_s]_{2MASS} = 0.724$; Bertelli et al. 1994).  
The 2MASS results demonstrate that there is a nuclear condensation
in the Sgr system that is independent of the presence of the metal
poor population typically identified with the M54 globular cluster,
whether or not M54 is a distinct stellar system from Sgr.

Layden \& Sarajedini (2000) have recently shown that all metallicity 
Sgr populations are clumped around M54, with the most metal-rich
([Fe/H] $\sim -0.5$) and young (1 and 2.5-3 Gyr) stellar populations 
particularly so (see also Sarajedini \& Layden 1995, Marconi et al. 
1998, Bellazzini et al. 1999a).  
Our finding
of a cuspy distribution in the center of the Sgr profile is similar
to the findings of Cseresnjes et al. (2000) who with their various fits
also faced problems with the peaked central density of
RR Lyrae stars.  That our highest Sgr density is coincident
with the position of M54 is consistent with the findings of other large area studies
of the Sgr center (Ibata et al. 1995, 1997, Bellazzini et al. 1999a).
Whether M54 represents the actual nucleus of Sgr, or just happens to reside 
at the bottom of the Sgr potential well remains uncertain,
and M54's relation to, and potential role in creating, the much more metal rich
populations condensed around it are matters unsettled.  These
and related issues are explored in more detail by 
Da Costa \& Armandroff (1995) and Layden \& Sarajedini (2000). 

Though the overall radial profile of Sgr resembles
that of other dSphs (and even including the presence of a ``break population"
at large radii - see below), the presence of 
a nucleated center is a distinguishing feature of the Sgr system.  


\subsubsection{King Profile Break}

The azimuthally averaged 2MASS Sgr radial profile shows a ``break" 
from a King model near a semi-major axis of about 1300 arcmin (Figure 5).
This ``King + break" profile looks just like those
expected for tidally disrupted dwarf galaxies
(Johnston, Sigurdsson \& Hernquist 1999).
Moreover, the Sgr radial profile resembles those of other Galactic dSphs found to have
breaks from a King profile at large radii (see Irwin \& Hatzidimitiou 1995) ---
for example Carina (Kuhn et al. 1996, Majewski et al. 2000),
Ursa Minor (Kocevski \& Kuhn 2000, Mart\'inez-Delgado et al 2001a, Palma et al. 2003)
Sculptor (Westfall et al. 2000, 2003, Walcher et al. 2003), and Leo I (Sohn et al. 2003)
--- although we note that the Sgr major axis, King limiting radius, 
$r_{l} \sim 12.6$ kpc, is more than twice that of
any of the other Milky Way dSph galaxies (see below). 
If Sgr is a member of this homologous, Galactic dSph family, it
presents at least one case where the break population is {\it bona fide} 
tidal debris, and lends support to claims that break populations in other
dwarfs may similarly imply tidal disruption (see discussion in Majewski et al. 2002a, 
for example).  

Earlier, in a starcount analysis of fields along the major axis, 
Mateo, Olszewski \& Morrison (1998) observed a ``kink" at 
$\sim 20^{\circ}$ radius\footnote{Mateo et al.'s (1998) ``one-dimensional" 
Sgr profile matches well the representation of 2MASS density with longitude
along the major axis (see discussion of Figure 13 below).  Both analyses
show this ``kink" at about the same location. }
and speculated that it might arise from a transition from a distinct Sgr 
dwarf to tidal stream debris.  
On the other hand, Johnston et al. (1999) suggested that this kink in the 
surface density actually demarcates a transition from debris released on the 
most recent periGalactic passage and older debris (a possibility also mentioned
by Mateo et al. 1998).  More recently, Helmi et al. (2001) asserted that this 
feature in the surface density demarcates the approximate projected radius 
between still bound material and stars lost by Sgr in the last periGalacticon passage.
While it is universally accepted that the stars beyond the kink represent
unbound, tidal debris, the interpretation of the stars {\it inside} this radius
--- bound or unbound --- clearly has great bearing on what one derives for 
the mass and mass-to-light ratio of the present bound center of the Sgr system.



\subsection{Sgr Mass-to-Light Ratio Revisited}

\subsubsection{Previous Work}

The dark matter content of the Sgr dwarf remains controversial 
(Ibata \& Lewis 1998, G\`omez-Flechoso, Fux \& Martinet 1999).
The mass density of the satellite determines its long-term integrity.
Early investigations (Ibata et al. 1997) postulated that ``Sgr is being
tidally distorted and is tidally limited, but is not disrupted as yet",
and derived a Sgr model with a central mass-to-light ratio of $(M/L_V) \sim
50$ (in the present discussion, all mass-to-light ratios are in solar
units, M$_{\sun}/$L$_{\sun,V}$).  This dark matter-dominated, prolate but
tidally-limited model, where mass does not follow light, was motivated by
the apparent delicacy and short-livedness of low mass King models when placed
in the most likely Sgr orbits (Vel\'azquez \& White 1995, Johnston
et al. 1995).  It was further suggested that Sgr could not have been
significantly larger than observed today, otherwise ``we would expect to
find its `missing mass' as a substantial population of Sagittarius
dwarf debris --- globular clusters and stars --- along its
dispersion orbit" (Ibata et al. 1997).  Subsequently, with the observed extent of the Sgr
system growing and the clear indication of mass loss into tidal debris tails,
the picture of Sgr changed dramatically, with the dwarf
recognized to be ``in the process of being tidally disrupted
and assimilated into the Milky Way" (Ibata 1999).  But an apparent conundrum 
remained: Even were Sgr to contain substantial dark matter --- sufficient to 
account for the observed large central velocity dispersion and 
suggesting a global $M/L_V$ at the level of the most extreme cases among the 
Galactic dSph population --- it would not be enough to solve the puzzle of 
how the dwarf could have survived as long as it has in its present orbit.  

Fine tuning of the dark matter configuration within the satellite provides one
possible solution (Ibata \& Lewis 1998).  Such models with rigid, very extended dark 
matter haloes yield concomitantly high $M/L \sim 100$, yet still cannot match
some characteristics of the observed Sgr system and the suggested form of the
dark halo is difficult to interpret with conventional forms of dark matter, as 
pointed by Helmi \& White (2001).  

More prosaic alternatives, albeit ones requiring their own dynamical fine tuning, 
have been put forward to address the dilemma of a fragile Sgr surviving a Hubble time.  
For example, Zhao (1998) proposed that Sgr has not always been in its present 
orbit, but rather it was deflected from a ``safer" orbit by an encounter 
with the Magellanic Couds several Gyr ago.  
G\`omez-Flechoso et al. (1999) suggested that as long 
as the full Galactic tidal field is experienced slowly (e.g.,
through a prolonged decay of an orbit via dynamical friction)
even a non-dark-matter-dominated satellite could survive many orbits (see
also Jiang \& Binney 2000).  Dynamical friction models require 
substantial mass loss in the satellite over the course of the orbit
transition.
Alternatively, G\`omez-Flechoso et al. (1999) propose that Sgr
may have formed in the tidal tail of a larger parent undergoing
a major merger.  

Recently, Helmi \& White (2001) claim to find two self-consistent Sgr
models, one purely stellar 
(``Model I", with intial mass $4.66 \times 10^{8}$ M$_{\sun}$
and $M/L \sim 2.25$) and the other with an extended dark halo (``Model II",
with initial mass $1.7 \times 10^{9}$ M$_{\sun}$ and $M/L \sim 15.1$)
that, when evolved in the Sgr orbit for nearly a Hubble
time, reproduce all data then available.  The tidal radius
in their Model II, 10.4 kpc, is similar to the Sgr major axis King limiting
radius found above ($r_{l} = 12.6$ kpc).  From being able to identify two viable
structural models, Helmi \& White conclude that a long-lived Sgr is 
not ``in any way anomalous".   However, it should be noted that
their models succeed by using a lighter Milky Way (asymptotic circular
velocity for the flat rotation curve of only 186 km s$^{-1}$ and mass
interior to present Sgr location of $7.87 \times 10^{10}$ M$_{\sun}$)
and more benign Sgr orbit (larger, 70 kpc apocenter and longer, $\sim 1$ 
Gyr period) than typically used by previous models.

%

\subsubsection{Standard $M/L$ Derivation } 

The derived mass and mass-to-light ratio of Sgr obviously depends on 
interpretation of the observed central structure.
In the discussion that follows, we distinguish between the true tidal radius, 
$r_{tid}$, of the system --- that distance from the center of Sgr where stars become
unbound --- and the empirically found radius, $r_{l}$, the limit 
where the best-fit King function plunges to zero.\footnote{We note that
what we call $r_l$ here is what King (1962) calls the empirical tidal radius, $r_t$, 
whereas he discusses a limiting radius, $r_{lim}$ in a manner similar
to our discussion of the true tidal radius in Section 4.3.3 below.}
For a dwarf galaxy or globular cluster in near-equilibrium in a tidal field, we expect
$r_{tid} \gtrsim r_l$.  Though we have ample reason to expect that the Sgr
dwarf is disrupting rapidly (see below),  
for simplicity we take $r_{tid} = r_l$ 
and use the 2MASS 
structural parameters to rederive Sgr's $M/L$ according to the standard
King (1966) prescription widely applied to spheroidal systems.  We do so with
the caveat that the results of such an analysis do not apply
for other ratios $f = r_{tid}/r_{l}$, as, for example, in the two
extreme cases already mentioned: (1) an extended dark matter halo 
(a constant $M/L$ is an implicit assumption of the King method) or (2) 
where the majority of the observed central Sgr structure represents unbound stars. 



Therefore, using the King profile parameters in Table 1 to represent 
the {\it bound} part of the Sgr system,
we convert the integrated light profile to a total brightness
by matching the M giant density at a specific radius to the 
equivalent surface brightness measured at the same radius.  The Table 1 King 
profile fit does not track the elevated density cusp in the center of 
Sgr so the peak surface brightness of Sgr is unrepresentative of 
the inner King brightness.  Fortunately, Mateo et al. (1995) have measured 
the surface brightness of Sgr outside the central condensation,
while Mateo, Olszewski \& Morrison (1998) have estimated the ``central" surface 
brightness as part of an extrapolation of a fit to the brightness profile that 
also ignores the cusp; both methods obtain 
$\Sigma_{o,V} = 25.2 \pm 0.3$ mag arcsec$^{-2}$.
With the latter value for the flat part of our density profile,
an integration of the King profile yields a total apparent magnitude for 
Sgr of $V_o = 3.63$, which is virtually identical to one of the results 
obtained by Mateo et al. (1998, their $V_{tot,1}$), even though
Mateo et al. integrate a two-component exponential profile fit 
to their one-dimensional cross-sectional profile of Sgr and their integration 
{\it includes} the tidal debris 
profile, whereas we fit and integrate a King profile fit to the full two-dimensional
shape of Sgr and include only the supposedly {\it bound} stars in the integration.
The two methods coincide because the Mateo et al. adoption of a 3:1 axis ratio
for the inner shape of Sgr (Ibata et al. 1997) matches well our
findings for the Sgr ellipticity, and, moreover, the tidal debris contribution 
in the area they surveyed makes a relatively minor contribution to the total luminosity.
Ignoring the central cusp likely underestimates
the total Sgr luminosity by $<5$\%.
 
A Sgr distance modulus of $(m-M)_o = 16.90 \pm 0.15$ (Ibata et al. 1997), implies 
an absolute magnitude of $M_V=-13.27$ for the bound part of the galaxy.  
Thus, the Sgr dSph is apparently the most luminous of the Milky Way family.
Ignoring the effects of stellar evolution and a variable star formation history,
we find that adding the central cusp and the M giant debris trails (Section 7.1) 
increases the minimum luminosity of the Sgr {\it progenitor} by several
tenths of a magnitude over the present Sgr brightness. 
That Sgr and the Fornax dSph
(which has $M_V=-13.2$; Mateo 1998) are of comparable luminosity is 
consistent with the currently established globular cluster specific frequency of 
the two systems: Fornax has six clusters and Sgr almost certainly has five 
(see Section 8 below), and possibly several more 
(Palma et al. 2002, Bellazzini et al. 2002a, 2003).  

To estimate the bound mass of Sgr, we use the formalism of King (1966)
as outlined by Richstone \& Tremaine (1986).  Thus, the mass
of Sgr is given by 

\begin{equation}
\label{sgrmass1}
M_{tot} = 166.5 r_c\mu/\beta
\end{equation}

\noindent where the scaling parameter $\mu$ is given by King (1966) as 
$\sim 9.38$ for on object with the observed concentration of Sgr, 
i.e. $\log(r_t/r_c) = 0.905$ (Table 1). 
The velocity parameter $\beta$ is related to the 
observed velocity dispersion, most commonly taken as the
11.4 km s$^{-1}$ value in Ibata et al.'s (1997) field ``f7" in the Sgr core.  
This dispersion yields
$\beta \sim 0.8^2/\sigma^2$ (Binney \& Tremaine 1987, see their Figure 4.11),
$M_{tot}=4.9 \times 10^8 $M$_{\sun}$, and $M_{tot}/L_{tot}=25$ in
solar units (where we adopt the total $V$ band luminosity from above as 
$2 \times 10^7 $L$_{\sun}$).  That there appears to 
be a nuclear concentration of stars (and therefore mass) 
that encompasses Ibata et al.'s field f7 raises concern that the 
velocity dispersion there may be enhanced.  Ibata et al.'s next field out
from the Sgr center, at several core radii, is ``f5", for which the 
velocity dispersion is only 9.2 km s$^{-1}$.  Adopting this
dispersion, however, leads to little difference: $\beta \sim 0.6^2/\sigma^2$,
$M_{tot}=5.8 \times 10^8 $M$_{\sun}$ and $M_{tot}/L_{tot}=29$.

These $(M/L)_{tot}$ values are two to four times
smaller than suggested by earlier studies, except
that of Mateo et al. (1998).  But it is important to point out that
the Table 1 {\it structural parameters} are also
significantly different than those that have been adopted in previous
studies and models of the Sgr system.  For example, while the stellar 
distributions in Helmi \& White's (2001) Sgr 
models have similar concentrations, $c=\log_{10}(r_t/r_c) $, to the
$c=0.90$ here, the actual scale of their {\it initial} Sgr stellar
systems are more than a factor of three smaller than found here, and
one expects the tidal radius to {\it decrease} with time. 
The models by Ibata \& Lewis (1998) and 
G\'omez-Flechoso et al. (1999) are similarly spatially compressed. 
Indeed, no model in the Ibata \& Lewis (1998) library has a tidal radius
anywhere near the $r_{l}$ derived here (their model K9, with a tidal
radius two-thirds our $r_{l}$ but a similar mass and approximately
similar [$\Psi/\sigma^2$] parameter is probably the closest match to
the observed 2MASS M giant parameters).  
Most of these models have been influenced by the original structural parameters
derived for the Sgr dwarf by Ibata et al. (1997), which yielded 
a half-light radius that is almost exactly three times smaller than 
derived here.  The large difference in the derived core radius is likely
because we have fit (Figure 5) a King profile to the entire system {\it and
that fit is generally insensitive to the localized central cusp of Sgr},
whereas Ibata et al. use the cusp to define the central surface 
brightness from which they search for a half-light decrement. 
In effect, the Table 1 profile fits to the Sgr  
dSph result in a satellite that has an overall structure that 
is much more distended 
than typically assumed; however, as we now show, this extended size begs the 
question of whether it can actually represent the limits 
of the {\it bound} Sgr core.


\subsubsection{Whither the Tidal Radius?}

Interpretations of the central parts of the Sgr system range from those where the
the bulk of Sgr within the $\sim 9$ kpc ``break" radius 
is still bound, to models (Vel\'azquez \& White 1995, 
Johnston et al. 1999, Law et al. 2003) where the bulk of the extent 
of the King profile is constituted by unbound stars.  
Indeed, recent models by Johnston, Choi \& Guhathakurta (2002) 
have shown that break-like features in the radial profiles of tidally
disrupting satellites can appear at several times the analytically 
estimated tidal radius, especially for satellites near pericenter.  
 
Based on simple dynamical arguments, we argue that a 
bound radius with $f = r_{tid}/r_{l}$ as high as unity seems unlikely.
Though only a rough guide, especially in the case of a highly elliptical
satellite on a non-circular orbit, the Roche tidal limit (e.g., 
Equation 3, King 1962)
can be used to derive a relationship between the tidal radius
and enclosed satellite mass.  In this case we 
assume a Milky Way mass interior to Sgr's present position
of $1.8 \times 10^{11}$ M$_{\sun}$ (e.g., Burkert 1997),
and normalize to the semi-major axis $r_{l} = 12.6$ kpc
to obtain the approximate mass of Sgr within the tidal radius as 

\begin{equation}
\label{sgrmass2}
m_{Sgr} = 1.6 \times 10^{11} {\rm M}_{\sun} f^3. 
\end{equation}

\noindent Clearly, under the presumption that $f=1$ we obtain an extraordinarily
heavy Sgr. 
Instead, to be conservative, we acknowledge that the Sgr core
is extremely tidally distorted and presume the {\it semi-minor} axis
to be a better representation of $r_{tid}$, so that $f=0.35$; thus we
obtain $m_{Sgr} \sim 6.9 \times 10^{9}$ M$_{\sun}$ and a total $M/L_V \sim 343$.
While a comparably high $M/L$ has recently been claimed 
for the Draco dSph based on internal velocity dispersion measures 
(Kleyna et al. 2002), it is hard to understand how Sgr 
could be so obviously losing mass into long tidal tails under these 
conditions (whereas, in contrast, several studies claim {\it no} detection of
tidal tails around Draco; Odenkirchen et al. 2001b, Aparicio et al. 2001).
Moreover, even this mass estimate is likely to be low since,
as argued by, e.g., von Hoerner (1957) and King (1962), the effective 
tidal radius should be calculated at the perigalacticon of the orbit. 
In any case, that these tidal approximation estimates are substantially at 
odds with that obtained from the King (1966) methodology 
in Section 4.3.2 indicates that $f$ cannot be near unity and casts doubt
on {\it both} analyses dependent on this assumption. 

On the other hand, analysis of the spatial and kinematical dispersion of the 
Sgr tidal tail M giants presented in this paper by Law et al. (2003) 
suggests that the present bound mass of Sgr is approximately $3 \times 10^8$ M$_{\sun}$.
If, we adopt this more modest mass for Sgr
(though a mass still at the high end among members of the Milky Way dSph 
family), then the Roche limit predicts an instantaneous Sgr $r_{tid}$ 
more like 1.5 kpc,
which is of order the measured King {\it core} radius (Table 1).
Such a physical configuration will be greatly susceptible to tides
(Pryor 1996, Burkert 1997), and, as found in the 
Johnston et al. (1999) and Law et al. (2003) Sgr models --- which have been shown 
to match the observed (e.g., Figure 13 below) Sgr surface brightness profile 
well --- the majority of the observed light profile consists of debris 
recently detached from the satellite in a major, destructive mass
loss event.  While perhaps uncomfortable to anthropic sensibilities,
``...it must clearly be possible.  
Any satellite must suffer final catastrophic disruption on some pericentric
passage, and in the case of Sagittarius we are seeing a system where this event
occurred only after a series of previous less damaging encounters had reduced 
its mass and binding energy to the point of critical stability'' 
(Vel\'azquez \& White 1995).


With the suggestion by Hayashi et al. (2003) that such tidal limit approximations
tend to overestimate the bound mass in tidally disrupting systems, it
becomes possible to contemplate that the even more centrally defined nuclear
``cusp" might represent the {\it bound} Sgr core embedded in an extensive 
cocoon of unbound stars.  
In so doing, we return the length scales of the problem to of order 
those utilized in the model studies that have queued their Sgr structural parameters
from the half-light decrements measured from the central surface brightness
of the {\it cusp}.   

\subsubsection{The M Giant Conundrum}

Reducing the actual tidal radius and binding energy of the satellite
also helps resolve a timing problem posed by the presence of stars 
as young as the Sgr M giants in extended, comparably-aged tidal tails.
As we show in Section 6.6, the bulk of the M giants
explored in this paper are likely formed relatively recently --- 
within the past several gigayears but no more than about 5 Gyr ago
(Layden \& Sarajedini 2000).  These stars are found in
tidal tails of a length that requires about several Gigayears
to form (Section 6; Law et al. 2003), thereby leaving relatively 
little time between when the M giants were created and when they 
escaped the bound Sgr system.  
The problem is exacerbated if the size scale of the bound Sgr galaxy is 
much larger than the expected, hundreds of parsecs radius for the star formation region 
--- as, for example, in the type of Sgr contemplated in Section 4.3.2 --- because
no secular diffusive mechanism can lead to the acceleration of
stars in a dSph galaxy on the required timescales.  
 


On the other hand, escape of stars formed several Gigayears ago in a 
central starburst would have been far easier if, in the course of critical
disruption, the Sgr tidal radius became of order the size of that starburst 
region; thus a present true tidal radius of order a kiloparsec in size or 
smaller is far more likely than one of order $r_{l}$, or even $0.35 r_{l}$.    
M giant escape would be enhanced were the most recent starburst spread out 
over a rotating disk rather than in a nuclear concentration.\footnote{The 
present Sgr system appears to show no minor axis rotation, however
the signal of any major axis rotation has yet to be separated
from other longitudinal velocity variations (Ibata et al. 1997).}
It is also possible that the starburst itself may have
contributed to the destruction of Sgr.  
Starburst driven galactic winds have been evoked to explain 
a number of properties of dwarf galaxies, including typically
low metallicities as a result of the loss of enriched gas.  
Clearly wholesale blowout of gas has not been a characteristic of the 
star forming processes in Sgr, given its multiple populations and 
age-metallicity relation (Layden \& Sarajedini 2000; see Table 3 below), 
but even fractional loss of 
gas in supernova winds would have contributed to drops in 
the Sgr binding energy after each starburst.  This mass loss allows the bound
part of Sgr, including the starburst region, to expand and makes  
it possible for young stars to reach the true tidal radius.  While outside
the scope of the present investigation, the formation of
{\it M giant} tidal tails (Sections 5 and 6) 
would seem to provide a rather powerful constraint
on full chemodynamical evolutionary models of the Sgr system.



Until the actual extent of the bound Sgr system is definitively
established, it will be difficult to establish its true $M/L$ and
dark matter content. Obviously, the standard methodology
of Section 4.3.2 will be an increasingly poor 
approximation as $r_{tid}$ departs from $r_{l}$,
though exactly how $M/L$ changes as $f$ decreases 
is not obvious (it depends on a proper accounting of both
the enclosed light and the actual distribution of bound mass).
One might also wonder whether the similarity of the Sgr radial profile to those
of other dSph galaxies translates to a similarity of {\it physical state} in
these other systems.
Problems with the typical $M/L$'s inferred from central velocity dispersions 
for {\it other} dSph systems 
that may not be in virial equilibrium have been discussed by, e.g., Kuhn
\& Miller (1989), Kroupa (1997), 
G\'omez-Flechoso (1998), Klessen \& Kroupa (1998), Majewski et al. (2002a) 
and G\'omez-Flechoso \& Mart\'inez-Delgado (2003).

\section{Great Circle M Giant Streams, the Sgr Orbital Plane
and a Sgr Coordinate System}

\subsection{Great Circle Cell Counts of M Giants}

To study the Sgr system over its full extent, we 
first define a coordinate system natural to the tidal debris system 
and in which projection effects are minimized.
Because Sgr and its debris lie close 
to one great circle defined by its orbit (Figure 3), 
we adopt the method of Great Circle Cell Counts (``GC3"; Johnston, 
Hernquist \& Bolte 1996) to derive an initial approximation to the
orientation of the Sgr orbital plane.
A similar approach was adopted by Ibata et al. (2002a) on the 2MASS Second 
Incremental Data Release; in their analysis of 26.4\%
of the sky Ibata et al. identified a peak in the M giant candidate 
source counts corresponding to a Sgr plane with pole 
at $(l,b)=(95,13)^{\circ}$.  Ibata et al.
(2001b) have also explored carbon star counts in great circle cells and
identified a peak at $(l,b)=(90,13)^{\circ}$ identified with Sgr.

To establish the Sgr orbital geometry we select M giant candidates with
$0.95<(J-K_s)_o<1.10$, and $E(B-V)<0.555$.  The sample is 
further limited to M giant candidates with
projected photometric parallaxes from 13 to 65 kpc 
--- the primary distance range for the
majority of material in the Southern Arc and Northern
Arm.  While limiting the volume of our GC3 assessment
of tidal streams, the above photometric parallax limit 
also reduces the contribution of ``false positive" detections 
at the faint end of the survey magnitude range (see discussion in Section 6.6).
GC3 runs with a variety of
Galactic latitude limitations were made, both including
and excluding the main body of Sgr.  By excluding the
high density central part of Sgr we give more weight to the tidal
debris in the derivation of the best fit plane, but
the results of the analysis were rather robust to
these variations as well as in variations in the
angular width of the cells and in the step size of the
poles:  For runs with various cell sizes and exclusion zones
the peak in GC3 counts yielded poles within a degree of 
$(l,b)=(273,-13)^{\circ}$. \footnote{Any great circle 
distribution on the sky produces two antipodal peaks in the 
cell counts.  Contrasting with Ibata et al. (2002a), 
we elect to identify Sgr with the peak in the South 
Galactic Hemisphere because this corresponds to the {\it angular momentum 
pole} of the satellite itself (see also Palma et al. 2002).}

Figure 6a shows the GC3 pole count analysis for a sample limited to 
$|b|>30^{\circ}$ and excluding the Large and Small Magellanic Clouds.\footnote{The 
Magellanic Clouds were removed by excluding the zone 
$260^{\circ} < l < 320^{\circ}$ and $-53^{\circ} < b < -25^{\circ}$.}
A great circle cell width of $5^{\circ}$ is used.  The pole from this
particular figure is $(l,b)=(272,-12)^{\circ}$. 
By a quirk of Nature, the Sgr debris plane is similar to 
that of the ecliptic (which has a pole [$l$,$b$] = [276,-30]).

\subsection{Best Fitting Sagittarius Plane}

\subsubsection{Fit in Galactic Cartesian Coordinates}

The GC3 methodology assumes that the debris under study is
sufficiently far away that the effects of Galactocentric parallax 
are negligible; i.e., strictly speaking, non-precessing debris streams will follow
great circles across the sky only when viewed from the Galactic Center.  However, 
parts of the Sgr debris stream come sufficiently close to the Sun
and the Galactic Center that several 
effects of perspective come into play (Figure 7).  
That (1) the Sun is not directly
in the orbital plane of Sgr, and (2) 
the Southern Arc and Northern Arm stars are at rather different
distances from us means that different GC3 poles are derived by analysis
of the two tidal tails independently: We obtain GC3 poles of
$(279,-18)^{\circ}$ and $(271.5,-11.5)^{\circ}$ 
when we divide the data set into
Northern and the Southern Galactic Hemispheres, respectively (Figures 6b and 6c).

To remove these Galactocentric parallax effects, we next search for the 
best-fitting Sgr orbital plane in the Cartesian Galactic coordinate system.
To place the survey into these coordinates, a photometric parallax is calculated
for each star using an absolute magnitude-color relation derived from the
RGB color-magnitude data shown in Figure 1c.  
The fit was restricted to the 1675 stars in the restricted 
range $0.9 \le (J-K_s)_o \le 1.10$, the primary M giant
color range explored in this paper.  With $2.5\sigma$ iterative
rejection of 158 stars, the following fit is obtained with
an RMS of 0.36 mag:

\begin{equation}
\label{colormag}
K_s = -8.650 (J-K_s)_o + 20.374
\end{equation}

The resultant distance scale (after assuming a Sgr distance
modulus of 16.90; Mateo 1998) is approximately 13\% smaller at $(J-K_s)_o = 1.0$
than one obtained from adopting the primary locus
for an [Fe/H]$\approx -0.45$ population identified by Cole (2001)
as a good match to 2MASS observations of the Sgr center:

\begin{equation}
\label{colecolormag1}
M_{K_s,CIT} = -9.43 (J-K_s)_{o,CIT} + 3.623,
\end{equation}

\noindent which translates to 

\begin{equation}
\label{colecolormag2}
K_s = -8.930 (J-K_s)_o + 20.383
\end{equation}

\noindent after transforming the Elias et al. (1982) Caltech/CTIO (CIT) system 
into the natural 2MASS filter system using the equations given by
Carpenter (2001).  Use of the color-magnitude calibration from Figure 1c 
is preferred because: (1) It 
is derived from a fit specific to the restricted color range explored 
in this paper; (2) the new relation is derived from a {\it 
background-subtracted} color-magnitude diagram of the Sgr center; (3) it
was fit in the natural 2MASS photometric system and so is free of
transformation equation uncertainties; and (4) 
it derives from a catalogue of the center of Sgr almost ten times larger than
Cole used.


Adopting this mean RGB color-magnitude relation implicitly translates the 
astrophysical scatter within the Sgr RGB into an imposed artificial scatter 
about calculated mean photometric parallaxes for Sgr features.  
The intrinsic vertical width of the Sgr RGB in Figure 1c for the 
$0.9 \le (J-K_s)_o \le 1.1$ color range determined above
($\sigma(K_s) \sim 0.36$ mag) is likely only slightly underestimated by 
excision of the weak tail to brighter $M_{K_s}$, 
more metal poor Sgr RGB stars (other possible {\it systematic} effects 
related to the relative numbers of metal weak populations are addressed in Section 6.6).
For sources with $J-K_s \sim 1$ this intrinsic
``standard candle" scatter dominates the contribution from 2MASS 
color errors until $K_s \sim 11$.  
Combining both the astrophysical scatter and the  
determined (Cutri et al. 2003) 2MASS aperture photometry uncertainties,
we estimate the imposed fractional distance spread to be
approximately $\sigma_{d}/d = 0.20$ for most stars in the survey, 
rising to $\sigma_{d}/d \sim 0.25$ for sources with $K_s \sim 13.5$. 
We also note the possibility of systematic errors in the M giant distance scale
that are linear with errors in the adopted distance of the Sgr core; 
Mateo (1998) suggests that the error on the distance to Sgr 
to be about 8\%.   


To fit the Sgr orbital plane, we first winnow the M giant sample to those within
15 kpc of the plane defined by the GC3 pole from Section 5.1.
Technically, this plane includes the Sun and not the Galactic Center, 
however, because the Sun is
{\it almost} in the Sgr orbital plane (see below), the 15 kpc
limit is more than generous enough to include all of the Sgr tidal debris.  
A restrictive color selection of $1.0 \le (J-K_s)_o < 1.1$ removes a large
amount of contamination by photometric errors in the distance range of concern for Sgr 
debris (see Section 6.6 and discussion of Figures 14 and 15), and also
lessens the effects of systematic photometric parallax errors by color.
To remove any remaining photometric error contaminants of {\it this} color-restricted
sample, and with some foreknowledge of the position
of the Sgr debris streams (Section 6), we remove stars with
$Z_{GC} \ge 50$ kpc and $Z_{GC} \le -30$ kpc.  
Finally, stars from the disk and bulge are removed with a requirement that 
$|Z_{GC}| > 11$ kpc; this also removes stars from the center of Sgr and
prevents them from biasing the fit (in the end, this
has only a minor effect on the results). \footnote{S-shape 
structures have been seen in the case of, e.g., the globular
cluster Palomar 5 (Odenkirchen et al 2001a, 2003, Rockosi et al. 2002) and 
the Ursa Minor dSph (Palma et al. 2003) -- both systems for which our 
perspective is nearly edge-on to the orbital plane.  
Part of an ``isophotal twisting" that resembles an S-shape 
is also apparent in Figure 4 and Figure 7e-f.
Because the northern
limb of the bound part of the Sgr system lies in the Galactic midplane, both
ends of the S-shape twisting of the central parts of Sgr are not evenly sampled
in our data.  Thus, including the heavy statistical weight of the 
unequally sampled inner parts of the Sgr system in our analysis would result in a 
slight biasing to the best-fitting plane.}

From this sample we determine a least-squares 
best-fitting plane by iteratively removing
$2\sigma$ outliers, redrawing the sample to those stars within
15 kpc of the new plane (and with the other limits above), 
and repeating the fit.  From a final sample of 1161 stars, from which
695 lie within $2\sigma$ (and where the RMS is 1.78 kpc), 
we find the best-fitting plane in Galactic
Coordinates (defined where the Sun is at $X_{GC}=0$ and this axis is positive 
towards the Galactic anticenter, analogous to the left-handed system
described in Mihalas \& Binney, \S6.1) to be:

\begin{equation}
\label{plane}
0.064 X_{GC} + 0.970 Y_{GC} + 0.233 Z_{GC} + 0.776 {\rm  kpc} = 0.
\end{equation}

\noindent The derived errors on the coefficients are
(0.002, 0.008, 0.002, 0.038), respectively.  This plane
corresponds to a Galactocentric orbital pole of 
$(l_{GC},b_{GC})=(273.8, -13.5)^{\circ}$,
only
slightly different from that obtained with the GC3 analysis.
The pole derived here is independent of any distance scale errors ---
such errors only affect the distance to the plane from the Sun 
(the fourth constant in the equation, 0.776 kpc) and from the Galactic Center.
Note that the best-fitting plane was not constrained to 
include the Galactic Center; if we assume
the Galactic Center to lie at $(X_{GC}, Y_{GC}, Z_{GC})=(-8.5, 0, 0)$ kpc,
this point lies 0.23 kpc from the plane.  
Had this plane been more inclined to the $X_{GC}$ axis, 
defining it would have given a new estimate of
the distance to the Galactic Center, which presumably lies
in the Sgr orbital plane; unfortunately very little leverage on this is 
offered in the present configuration.  However this technique may be applicable
to other extended halo tidal streams orbiting the Galactic Center found in the future.

\subsubsection{The Flatness and Proximity of the Sgr Plane}

Our proximity to the Sgr plane is a rare coincidence.
If we adopt the solar position as $3.8^{\circ}$ from the Sgr-Galactic
plane line of nodes, we are closer to the Sgr orbital plane during less than
4\% of our own orbit around the Galaxy.  This number drops to 2\%
when we consider that it is only on this side of the Galactic Center that 
the Sgr leading debris arm (as traced by M giants) apparently gets near the Solar 
Circle (see Sections 6.4 and 9).  Our 0.78 kpc distance from the Sgr plane
is less than half the RMS spread in Sgr debris about the plane fit above, 
so that if Sgr debris passes within a few kpc of the Solar Circle, we are very likely
to be amidst that debris.  Sections 6.4 and 9 address 
the implications of this unusual time in Solar System history.

That the Sun is not {\it precisely} in the orbital plane 
leads to perspective effects shown in Figure 7.  
Only those stars used in the derivation of the best fitting plane and
lying within $2\sigma$ are shown.\footnote{Equivalent 
side views of the Milky Way and Sgr with no restriction to $<2\sigma$ 
outliers are shown later in Figure 16, and show that the relative
thinness of the distribution is not simply contrived by the present analysis. }
Figures 7a and 7b
compare the slightly different perspectives offered by a projection on the
$X_{GC}-Z_{GC}$ and the plane obtained by rotation of $3.8^{\circ}$, which
allows a direct edge-on view of Sgr.  Because of the variation in distance of
Sgr tidal debris from the Sun, a slight ``bowing" of the apparent Sgr
orbital plane is removed when viewed more edge-on.  This bowing
(seen more obviously in Figure 16 below)
explains the differences in derived GC3 poles for Northern
and Southern Hemisphere GC3 analyses in Section 5.1.

The coherence of the Sgr debris tightly to one plane highlights
how little precession the Sgr system experiences for the 1-2 Gyr of orbit
traced by the observable debris --- no more than a few degrees
total (Law et al. 2003).
Orbital precession in tidal tails is acutely sensitive to the shape of
the halo potential (cf. Johnston, Sackett \& Bullock 2001, Mayer et al. 2002),
and the flatness of the
Sgr debris stream strongly points to a {\it spherical} mass potential
for the Milky Way to at least $\sim50$ kpc.  Our results here concur with and 
{\it strengthen} the similar arguments previously made by Ibata et al. (2001b)
because even tighter coherence of the Sgr stream is demonstrated after properly 
removing Galactocentric parallax effects.
A quantitative analysis of the halo flattening
from these results is presented in Law et al. (2003).

Figure 7c, a projection of the $|Z_{GC}| > 11$ kpc portions of the
tidal arms onto the Galactic plane, illustrates the tilt in the Sgr
orbital plane.
Figure 7d shows the variation in the width of the Sgr stream when projected 
onto the sky.  As may be seen, the debris stream is most foreshortened and 
spans the largest angle when it is near us, 
in the general direction of the North and South Galactic Poles.
Figure 7d demonstrates how much of the high latitude celestial sphere
contains lines of sight that intercept the Sgr stream
(especially accounting for the fact Figure 7d does not 
display $>2\sigma$ outliers from the Sgr midplane); particularly at the
Galactic poles, Sgr debris is hard to avoid!  Figure 12, discussed
later, shows $>2\sigma$ outliers from the Sgr plane and 
makes this point even more clearly. 
We review various proposed detections 
of Sgr debris in this context in Sections 8 and 9.

\subsubsection{Sagittarius Spherical Coordinate Systems}

Determining the Sgr orbital plane, as done in Section 5.2.1, permits us to
derive a more natural spherical coordinate system for the interpretation of
Sgr tidal debris - one with the equator defined by the Sgr debris midplane.  
Two such systems (Table 2) --- one heliocentric and one Galactocentric --- are useful.  
In the first, Sgr latitudes, $B_{\sun}$, 
are defined by the Sgr debris {\it projected on the sky
as viewed from the Sun}.  We adopt a debris midplane (equator) corresponding to 
a pole $(l,b)$ given by the $(l_{GC},b_{GC})=(273.8, -13.5)^{\circ}$ pole derived above. 
Sgr longitudes, $\Lambda_{\sun}$, are defined to 
increase in the direction of trailing Sgr debris, with the prime meridian, 
$\Lambda_{\sun} = 0^{\circ}$, defined by the longitude of the center of the King profile
fit to the center of Sgr determined in the previous section.  
This first coordinate system is entirely observationally based, but, being 
Sun-centered, preserves Galactocentric parallax effects.

The second, {\it Galactocentric} spherical coordinate system --- where the equator is 
defined by the Sgr plane in Equation 8 --- while immune from 
Galactocentric parallax effects, is, however, subject to scale and 
random errors in the determination of photometric parallaxes.  
We define a Galactocentric
($\Lambda_{GC}, B_{GC}$) system, with $\Lambda_{GC}=0$ taken as centered on
Sgr, as before.  Because the plane does not actually contain the Galactic
Center (it was not constrained to do so), 
we take as the center of the ($\Lambda_{GC}, B_{GC}$) system
the point in the plane closest to $(X_{GC}, Y_{GC}, Z_{GC}) = (-8.5, 0, 0)$ kpc, which
is $(X_{GC}, Y_{GC}, Z_{GC}) = (-8.51, -0.22, -0.05)$ kpc.  

Table 2 gives the Eulerian rotation angles (under the ``x-convention";
see, for example, Goldstein 1980),
($\phi, \theta, \psi$), and the Cartesian Galactic coordinates of the centers of rotation used 
to define the two Sgr ($\Lambda, B$) coordinate systems used here.  
Note that these systems as adopted are right handed, and therefore determined
from the left-handed, Cartesian Galactic system after the 
translation $X_{GC} \rightarrow -X_{GC}$.  For some illustration presented here
it is convenient and intuitive to leave the intersection of the Sgr and Galactic plane
horizontal, and this is achieved by setting the third Euler angle rotation to $\psi=180^{\circ}$. 
This results in the following new Cartesian Sgr coordinate systems corresponding to
each of the spherical coordinate systems derived above:

$$ X_{Sgr,\sun} =  d_{\sun}\cos(\Lambda_{\sun}+14.42^{\circ})\cos(B_{\sun}) $$
\begin{equation}
\label{solarcart}
 Y_{Sgr,\sun} =  d_{\sun}\sin(\Lambda_{\sun}+14.42^{\circ})\cos(B_{\sun}) 
\end{equation}
$$ Z_{Sgr,\sun} =  d_{\sun}\sin(B_{\sun}) $$

\noindent where $d_{\sun}=(X_{Sgr,\sun}^2 + Y_{Sgr,\sun}^2 + Z_{Sgr,\sun}^2)^{0.5}$ 
is the distance of the star from the Sun, and a second system (distinct from the 
normal Galactic $[X_{GC},Y_{GC},Z_{GC}]$ coordinates used above)

$$ X_{Sgr,GC} =  d_{Sgr,GC}\cos(\Lambda_{GC}+21.60^{\circ})\cos(B_{GC}) $$
\begin{equation}
\label{GCcart}
 Y_{Sgr,GC} =  d_{Sgr,GC}\sin(\Lambda_{GC}+21.60^{\circ})\cos(B_{GC}) 
\end{equation}
$$ Z_{Sgr,GC} =  d_{Sgr,GC}\sin(B_{GC}) $$

\noindent where $d_{Sgr,GC}=(X_{Sgr,GC}^2 + Y_{Sgr,GC}^2 + Z_{Sgr,GC}^2)^{0.5}$
is the distance from the center of the ($\Lambda_{GC}, B_{GC}$) system as
given above and in Table 2.\footnote{David R. Law has written a suite of
codes for converting between different Galactic and Sgr coordinate systems.
These computer routines are available at http://www.astro.virginia.edu/\~drl5n/Sgr/.}

Figure 7e makes use of the latter coordinate system to show a projection of Sgr
perpendicular to the best fitting Sgr plane.
The canting of the Sgr major axis with respect to the best-fitting plane and in 
the direction of the normal to the Galactic plane can be seen.  
Indeed, the angle of this tilt  
is nearly identical to the angle between the Sgr major axis and the 
normal to the Galactic plane (see Figure 7f),
or a little more than about $6^{\circ}$ in each case.  This canting is the
rationale for removing the Sgr center from our calculation of the
best-fitting plane above.  Figures 7e and 7f shows how the beginning
of the Sgr tidal stream emanates from the main body more or less
evenly to either side of the debris midplane, despite the tilt of main body of Sgr. 
Figures 7e and 7f provides a slight qualification to the usual 
assumption (e.g., Lynden-Bell 1982) 
that the ellipticity of satellite systems are aligned 
with the direction of orbital motion and should therefore
point in the direction of their tidal tails.  This observed canting 
may provide an additional constraint on dynamical models 
because torquing of the tidal elongation 
depends on the details of the non-circular satellite orbital trajectory 
with respect to the Galactic potential. 


\subsection{No Magellanic Cloud M Giant Streams}

No other strong GC3 peak appears in the M giant candidate pole counts in agreement   
with the preliminary analysis of 2MASS M giants by Ibata et al. (2002a).
This GC3 result only applies however, 
for that part of the halo within $\sim75$ kpc, for streams with substantial extent
above $|b| = 30^{\circ}$, and for tracers obeying the other specific
M giant photometric criteria employed here (for example, Equations 2).
However, this result is reconfirmed for {\it all} 
late type giants (M giants and carbons) by our 
analysis of M giant streams in Cartesian coordinates in Sections 7 and 8.3.  
In addition, no GC3 peak corresponds to tidal 
debris from the Magellanic Clouds, even though the Magellanic Clouds are copiously
populated by such stars (e.g., Nikolaev \& Weinberg 2000), and are, by far, the 
predominant reservoirs of late type giants in the Galactic halo.  Previous analysis of 
a sample of halo carbon stars by Ibata et al. (2001b) suggested the existence
of a Magallanic carbon star stream.  We note that while our GC3 analysis
specifically leaves out the region around the Magellanic Clouds (excluding
the zone $-25^{\circ} > b >-53^{\circ}$, $260^{\circ} < l < 312^{\circ}$) 
to avoid the interference 
of a large great circle band dominating Figure 6, any roughly coherent tidal streams
extending more than about 25 degrees from the Clouds should be apparent as a 
GC3 peak in that figure.  Analysis of the distribution
of 2MASS starcounts by van der Marel (2001) shows the Large Magellanic Cloud to
be elongated by Galactic tidal forces, but the lack of any GC3 peaks associated with
the Magellanic Clouds suggests that any tidal forces on them either are not sufficient 
to create extended streams of 
extratidal stars, or at least that young, metal rich populations are not
presently participating in such streams.  

The planar coherence of the Sgr debris and the implied sphericity 
of the Galactic potential suggests that other tidal streams in the outer
Galaxy should also face little precessional smearing, remain spatially coherent 
for at least several gigayears, and therefore be evident as great circles on the sky.  
Ibata et al. (2002a) have argued that the lack of any other discovered 
M giant GC3 streams means that the present accretion rate
of luminous, low-mass satellites must be very low, and that most of the 
luminous part of the Milky Way halo must have been in place more than 3 Gyr ago, 
before the accretion of Sgr.  However, this conclusion
applies only to systems sufficiently metal-rich to produce M giants.
Most halo globular clusters and Galactic dSph galaxies contain
few if any such stars because they are dominated by old, metal-poor populations
(note, as just one example, the total absence of the four Sgr globular clusters
in the M giant distribution of the central part of Sgr shown in Figure 4).  Indeed,
the Sgr center and the Magellanic Clouds are the only readily identifiable, 
{\it intact} stellar systems away from the Milky Way
disk within the full sky, 2MASS M giant distribution explored here.
Thus, the lack of other M giant streams places no limit on the present accretion
rate of older, more metal-poor systems.

\section{Analysis of Observed Tidal Features in the Sagittarius Plane} 

\subsection{Tidal Tails}

The celestial sphere projection of M giants in Figure 3 gives only a rudimentary
sense of the relative distances of Sgr tidal features by their apparent 
brightness. 
Figure 8 shows 
the {\it planar} distribution of the dereddened $K_s$ magnitudes of 
M giant candidates with $(J-K_s)_o \ge 1.0 $ and 
$-10^{\circ} \le B_{\sun} \le +10^{\circ}$ as a function of 
Sgr longitude.  
Figure 9 presents the same distribution in a polar projection.
Figures 8 and 9 show directly observed quantities, so are free
of interpretation.
Both figures show the more complex character of the Southern Arc and Northern Arm,
and give proof of their contiguous connection:
While the two features are on average at different mean $K_s$ magnitudes, 
projections of the magnitude-longitude 
trends through the Zone of Avoidance show that the two features meet at the Sgr center, and
that they represent the leading (Northern Arm) 
and counterpart trailing (Southern Arc) tidal tails.

In Figures 10 and 11 we present the M giant planar distributions
in terms of derived distances from the
Sun and distances from the Galactic
Center (tehcnically, the latter is the center of the GC system given in 
Table 2).  
The M giant photometric parallaxes derived from 
the absolute magnitude-color relation given in Equation 5.  

Figures 8-11 make clear the leading/trailing tail structure of the Sgr dwarf and 
the rosette nature of its orbit.  Figure 11, which shows the distribution
of stars projected onto the presumed Sgr orbital plane, 
gives a particularly clear impression of the rosette shape.  We
fit this distribution to a model of the Sgr dwarf in the Galactic 
potential in a subsequent contribution (Law et al. 2003), but as a
general guide to understanding the overall structure of the tidal arms
illustrated in Figure 11 we 
call attention to Ibata \& Lewis (1998) model K6-a shown in their
Figure 3 (a model highlighted more clearly in Figure 3 of Ibata et al. 2001b).
Though shown in the slightly different (canted by about $13.5^{\circ}$) 
$X_{GC}-Z_{GC}$ plane, the overall appearance of the Ibata \& Lewis
K6-a model illustration bears great resemblance to the M giant distribution 
shown in Figure 11 (see also Figure 14, particularly panel c). 
Another useful interpretive guide is Figure 8 of Helmi \& White (2001), which
shows one of their models in a coordinate system similar to that shown in
the top panel of Figure 10.

\subsection{Trailing Tidal Debris}

As may be seen in Figures 8 and 9, the center thread
of the Southern Arc M giants varies by only about a magnitude across the
Southern Galactic Hemisphere.  The actual mean photometric parallax 
of this trailing debris tail is (1) roughly 25 kpc where it attaches to the center of Sgr, 
(2) slightly less than 20 kpc when it achieves its closest distance to us near the 
SGP, and then (3) gets progressively more distant towards 
the Galactic anticenter (e.g., $\sim 40$ kpc at $\Lambda_{\sun} \sim 160^{\circ}$).
The Galactocentric distance of the Southern Arc ranges from $R_{GC}=16$ kpc
at the Sgr center and a similar distance when it passes beneath the Galactic
Center ($\Lambda_{GC} \sim 70^{\circ}$) to $\sim 50$ kpc at the Galactic 
anticenter. 

The disposition of this trailing tidal arm at longitudes even 
farther from the Sgr center is less clear.
Inspection of Figures 8 through 11 (and particularly the bottom panel
of Figure 10) suggests that the trailing arm
crosses the Galactic plane, since there appears to be a continuation of
the sweeping Southern Arc north of the Galactic plane, and an
overdensity of points
near ($\Lambda_{\sun}, K_s) \sim (185^{\circ}, 13-14$).
Unfortunately, this is where our selection of M giants becomes both incomplete 
and noisy (see discussion of Figure 15 below).  A large
number of stars appear at $(K_s)_o > 13.0$ at all longitudes, but their
reality as M giants, much less Sgr M giants, must be considered highly uncertain
and remains to be verified spectroscopically.  We address the issue of the length of
the tidal tails further in Section 6.4 below.

\subsection{Leading Tidal Debris}

The Northern Arm can be seen (Figures 8-11) to represent the leading
tidal debris tail of Sgr.  Figures 10 and 11 show that the approximate
center of the locus of the leading tidal debris
arm reaches a mean apoGalacticon distance of about 40 kpc around $\Lambda_{GC}
= 280^{\circ}$ ($l \sim 350^{\circ}, b \sim 45^{\circ}$). 

Figure 11 makes clear the relationship between the diffuse 
North Galactic Cap (NGC) ``fluff" and the Northern Arm:  
The diffuse NGC material apparently represents an extension of the
Northern Arm, which together constitute one ``Northern Loop" 
around the Galactic Center and returning back towards the Galactic plane.  
The NGC material is more spread out on the celestial sphere
simply because it is closer and foreshortened along the line of sight.  Figures
8-10 show the looping Northern Arm spreading
across the NGC, covering a large angular range when it gets to the 
smallest distances from us (see top panel of Figure 10).
The top panel of Figure 10 gives the strongest impression that
debris from the leading arm of Sgr orbits back toward the Galactic plane 
near the Solar Circle: Because no distance scaling problem can move stars 
{\it on the celestial sphere}, that material is seen to either side
of $\Lambda_{\sun} = 256^{\circ}$ (the direction closest to the North Galactic
Pole) points to the likelihood of Sgr material falling to either side 
of the Solar Circle.

\subsection{The Sagittarius Leading Arm Near the Solar Neighborhood}

In Section 5.2 it has been found that the Sun lies within a kiloparsec
of the Sgr orbital plane, a distance well within the width of the
Sgr tidal debris stream; thus the actual proximity of Sgr debris
to us depends on the length of the leading arm and where it
crosses the Galactic plane on this side of the Galactic Center if it
is long enough to do so.  For a variety of reasons, 
whether and where the northern tidal arm crosses the Galactic 
plane toward the southern hemisphere must still be considered 
uncertain, because the stars that look to be nearby parts of the
leading arm in Figures 8-11 might also be contributions
of M giants from the Galactic Intermediate Population II/thick disk, 
bulge, or inner halo, or might even be other substructure.  
The following summary points suggest the {\it plausibility} of
Sgr debris near the Sun, but further work is needed to confirm
this scenario:

1) In the fit to the Sgr plane in Section 5, we obtained an RMS
residual of nearly 2 kpc.  While there is a $\sim20$\% distance smearing 
imposed from the intrinsic spread about the adopted
color-magnitude relation, it is clear that Sgr debris girdles the Sgr
orbital midplane with a total width of 4-8 kpc or more.  This
is supported by the fact that the Southern Arc (at a distance of about
20 kpc) is 10-20 degrees or more wide on the sky (e.g., Figures 3, 7d and 
12).  Simplistically assuming a cross-section for the tidal arms 
too far from circular in shape would then yield a {\it depth} of the Sgr arms 
{\it within} the orbital plane (e.g., that projection shown in Figure 11) 
of about the same order of magnitude. 
Thus, should the leading stream
be long enough to reach the Galactic plane on this side of the Galactic Center, 
and should it do so within several kiloparsecs from the Sun, then Sgr debris will pass 
through the solar neighborhood.
  
2) Though previous models (e.g., Ibata et al. 2001b, see, e.g., their Figure 3) 
derive an orbit for Sgr similar to that traced by the
rosette of debris seen here and predict 
current passage of leading arm debris through the Galactic plane 
at a mean distance of $\sim 4$ kpc outside the Solar Circle, 
our own best fitting models to the present data set (Law et al. 2003) 
obtain a passage of the center of the leading Sgr 
within two kiloparsecs of the Sun.

3) Figures 10 and 11 show the presence of $15-30$ kpc distant M giants
stretching from $\Lambda_{\sun} = 225^{\circ}$ to 
$280^{\circ}$ or more.  An even wider angular distribution at closer distances
suggests the passage of 
leading arm material both exterior {\it and} interior to the Solar Circle at these 
distances (the NGP is near $\Lambda_{\sun} = 256^{\circ}$).  
Unfortunately, increased confusion between Sgr debris and
disk, IPII/thick disk, inner halo and bulge M giants in the inner Galaxy means that 
the exact disposition of the nearby Sgr debris requires 
spectroscopic weeding of Milky Way contaminants.

4) 
The tidal debris model shown by Ibata et al. (2001b; their Figure 3) shows 
a southern extension of the downward moving Northern debris
that passes not only through the Galactic plane, but also through the
trailing debris arm and to larger distances.  Such a feature may be the origin 
of the slight excess of more distant stars (with $11.5<K_s<13$) in the 
predicted longitude range ($\Lambda_{\sun} \sim 15-65^{\circ}$).  
The lower right quadrant of Figure 14c (presented below) which matches 
the overall appearance of the Ibata et al. model, 
shows this apparent excess of more distant stars more clearly.  

Radial velocities of both very bright and faint M giant stars in each hemisphere
would be particular useful for checking whether the above features are consistent
with a vertical flow of Sgr stars 
through the nearby Galactic plane and onward, past the trailing debris arm.  
We discuss recent spectroscopic observations bearing on these subjects
in another contribution.

\subsection{Density Variation Along the Tidal Arms}

Both the length of and density variation along the
tidal debris arms of a disrupting satellite system are a function of the duration, 
strength and overall nature of the interaction with the Milky Way (Johnston 
1998).  
Figure 12 is an attempt to unwrap the Sgr tidal material into a 
ribbon around the sky to illustrate 
surface density variations in the Sgr tidal arms on the plane of the sky.
The top panel shows the ribbon in celestial coordinates.
Only stars lying within 7 kpc of the best fit plane to the Sgr debris
are shown.  A less distorted projection is one in a Sgr coordinate system
(bottom panel of Figure 12).  In our analysis of density variation with
position we concentrate on the morphologically simpler, trailing arm.  Moreover, 
because of its nearly equal distance from us as a function of
Sgr longitude, the Southern Arc of Sgr provides a facile means
by which to measure density/mass loss variations relatively free of
the effects of foreshortening.  
Therefore the natural longitude system to use is $\Lambda_{\sun}$,
not $\Lambda_{GC}$.  
To remove the bulk of Galactic disk/bulge contamination, we only show
stars more than 10 kpc from the Galactic Center 
(assuming a Galactic Center distance of 8.5 kpc), more than 15 kpc from us, and with
$|b| > 10^{\circ}$.  To increase the S/N of the tail density, 
we include bluer M giants by opening our selection criterion to 
$0.95 \le (J-K_s)_o \le 1.10$. 

In Figure 13 we show the 
numbers of these M giants as a function of $\Lambda_{\sun}$ position.
Counts are shown for tallies within 
{\it slabs} of various thicknesses centered on the Sgr midplane.  
To isolate those stars in
each slab associated specifically with the Southern Arc, we 
fit a quadratic function to the photometric parallaxes
of all stars in the slab as a function
of $\Lambda_{\sun}$ (with an iterative rejection of $2.5\sigma$ outliers).
For $\pm3$, $\pm5$ and $\pm7$ kpc wide 
slabs, the $\sigma$ of the Southern Arc distances are 
3.6, 3.7, and 3.9 kpc, 
respectively.  This {\it depth spread} is larger than the $\sim 2$ kpc sigma  
{\it width spread} found in our fits of the best fit plane to the Sgr 
debris, but the depth spread
is of course affected by 
the artificial, ($\sigma_d/d \sim 0.2$) spreading due to ``standard candle"
scatter (Section 5.2).

To determine a background level of non-Sgr ``contaminants"
in the Southern Arc we count M giant candidate stars in tidal-stream-like 
tubular volumes of the Galactic halo, 
but in a direction that avoids Sgr, the Magellanic Clouds and regions of large 
reddening.  Through trial and error we found an acceptable orientation by rotating 
the slabs containing the Southern Arc tubular volumes 
$35^{\circ}$ about the line of nodes represented by the intersection of Sgr and 
Galactic planes.  In this orientation, the ``Southern Arc" tubes  
now sample random halo volumes associated with the great circle
pole $(l,b)=(272,+23)^{\circ}$.  Although nearly as polar as the original Sgr plane
(and therefore presumably sampling a similar background Galactic halo density law), this
``background plane" suffers from the shortcoming that wider slabs centered on it 
become ever more contaminated by Sgr contributions near the Galactic plane. 
In one direction, this includes parts of the Sgr center, but 
because we are concerned here with
assessing the density of the more diffuse parts of Sgr, it is less 
critical to obtain an accurate accounting of the background 
near the Sgr center (which has, in any case, been done more properly in 
the radial profile fits in Section 4).  Figure 13 includes the 
derived background counts 
for the $\pm 3$ kpc wide slab as representative; the background is
typically about 10\% in the tail regions away from the Sgr center. 
Tests of various sized background slabs show that the background
level away from the Sgr center is fairly constant, at $0.33 \times$(slab width) per  
5$^{\circ}$ longitude. 
This adopted background is subtracted in the density plots shown.  

Both Figures 12 and 13 demonstrate that, for the most part, the 
trailing tidal arm of Sgr shows no substantial density variation with longitude, 
especially over the range $45 \lesssim \Lambda \lesssim \sim140^{\circ}$.  
Generally, the relative density variation along the Sgr tail is 
steadier than is observed (Odenkirchen et al. 2003) in the case 
of the tidal tails of Pal 5. 
The relatively steady Sgr tidal tail density suggests a more or 
less constant mass loss rate for the timescale represented by this 
portion of the tail.  

Nevertheless, some small variations in density do appear
in Figures 12 and 13.  The density decline at large $\Lambda_{\sun}$
is in part due to reddening and Galactic latitude limitations (note 
that the Galactic anticenter is near $\Lambda_{\sun}=166^{\circ}$).  
Figures 12 and 13 hint at a slightly higher density of M giants 
some 25-50$^{\circ}$ in longitude downstream from the Sgr center.  
A smaller, less significant overabundance also appears
at $\Lambda_{\sun} \sim 133^{\circ}$.
Figure 13 shows the more significant of these apparent overdensities
as a ``hump" in the tidal tail distribution for 
$25^{\circ} < \Lambda_{\sun} <50^{\circ}$ (recall that the King 
limiting radius of Sgr along the major axis is $30^{\circ}$,
so this excess is distinct from the central King profile).
An apparent widening in the tail at these longitudes 
is also suggested by Figure 12 as well as 
by the separation of the $z=\pm3$ kpc points from the $z=\pm5$ and
$\pm7$ kpc points in Figure 13 at this longitude 
compared to other places along the trailing debris tail.
We propose two possible explanations for the existence of 
this particular density feature:

1) 
Tidal tail caustics that correspond to the strongest 
phases of gravitational shocking during the orbit of the parent satellite
have been seen in simulations of globular cluster
disruption by Combes, Leon \& Meylan (1999).
Though transient and made up of a constantly changing set of stars, these symmetric 
(leading and trailing) lumps are seen to persist in the cluster
models for almost a Gyr in the Combes et al. models.  
For a Sgr-like system, 
dispersal within a few orbital times occurs because of the mixing of 
stars with a large range in energies and drift rates.  
Published Sgr orbits that approximately match our data (e.g., Figure 3 in 
Ibata et al. 2001b) show Sgr to have passed through periGalacticon very recently 
(within $\sim0.1$ Gyr), and an increased number of released stars
might be expected from the associated gravitational shock (e.g., Johnston et al. 1999a). 
If the ``hump" in Figure 13 is related to a periGalacticon release event, then 
a symmetrically placed feature might be expected in the leading tail; unfortunately, 
this feature, if it exists, would lie close to the Galactic plane, where 
our data become more confused, though a larger density of stars at 
$-50^{\circ}<\Lambda_{\sun}<-30^{\circ}$ is not inconsistent with the data (see 
Figure 12).  That one major ``hump" is seen in the trailing tail 
is more or less consistent with the dispersal timescale mentioned above.
However, as discussed in Section 4.3, it is possible that much more than
just the hump in the tail may be associated with the last periGalacticon
disruption event.  A detailed assessment of the overall structure of the
region where the Sgr King profile transitions into the tail and the relation
of observed features to recent mass loss requires
more careful modeling, especially given the many nuances in the
structural morphology of satellites that can be induced by tidal disruption
(Johnston, Choi \& Guhathakurta 2002).


2) If, as is suggested by the Sgr disruption model of Ibata et al. (2001b) 
as well as the various pieces of evidence within the M giant
distribution discussed in Section 6.4, the leading tidal arm
penetrates back into the South Galactic Hemisphere and crosses the
trailing arm, we would expect an increased density of stars
at about the longitudes where the excess density is observed.  
The presence of overlapping leading arm debris could also lead to
the observed widening of the apparent trailing arm at this point 
either through foreshortening or precessional displacement of the
leading arm material compared to the true trailing arm stars.
Radial velocities of stars in the hump should reveal a clear signal
of this overlap.  Early evidence from our M giant radial velocity work
suggest this may be the case (see also the discussion of overlapping Sgr
tails in this part of the sky by Johnston et al. 1999a).

In Figure 13 we have shown for comparison the Sgr longitudinal profile 
over the range $10 < \Lambda_{\sun} < 34^{\circ}$ obtained by Mateo, Olszewski \& 
Morrison (1998) for main sequence turn-off stars.  The detailed shape of their profile
is remarkably consistent with the 2MASS M giant profile over the longitudinal
range of overlap.  However because the main body of Sgr is canted
somewhat with respect to the mean trend of the debris, and because 
Mateo et al. extrapolated their outer fields from the direction of 
the major axis of the Sgr center, 
their outer fields progressively fall away from the 
center of the debris stream (see Figure 12 and Figure 7f, where 
the $\Lambda_{\sun} = 34^{\circ}$ Mateo et al. point corresponds to a location
about 12 kpc below the Sgr center).

\subsection{Length of, and Possible Population Variation Along, the Tidal Arms}

The lengths of the Sgr tidal debris arms are of interest not
only because they bears on the question of the duration of the mass loss process, 
but because of the issue of whether the leading tail
is long enough to reach the solar neighborhood.
As described earlier, uncertainty over the length of the leading tail 
is complicated by  
contamination by thick disk, bulge and other 
M giants at low Galactic latitudes, and the
possibility of tail overlap below the Galactic plane.  It would be
useful, therefore, if the length of the Southern tail could serve as a guide.
Section 6.2 offered evidence that the Southern Arc may extend to 
the Northern Hemisphere at the Galactic anticenter, but confidence in this
result is limited by increased magnitude errors 
at large $K_s$, which makes selection of M giants both
incomplete and more contaminated by ``false positives".
Figure 14 demonstrates the latter problem with 
planar distributions of giant star candidates 
binned by $(J-K_s)_o$ color.  

Figure 14a shows
the spatial distribution of stars in the color range $0.90 < (J-K_s)_o \le 0.95$,
and quite evident is a ``shell" of excess, presumably contaminating stars introduced at 
the nominal photometric parallax limit ($\sim 35$ kpc) of stars of this color range.  
Since for the same apparent magnitude error limit redder giant candidates are projected
to greater distances, we see how the shell of excess contaminants is larger
in the $0.95 < (J-K_s)_o \le 1.00$ color bin shown in Figure 14b, and
expands outward with color until, for stars redder than
$(J-K_s)_o \sim 1.0$, the contaminating shell is outside the
distance range shown.  The distance progression of these ``contamination shells" with 
$(J-K_s)_o$ color is illustrated in Figure 15, where we have shown starcounts
as a function of distance for four $(J-K_s)_o$ color bins within a cone 
selected to be more or less free of Sgr stars and the Magellanic Clouds --- specifically
$Y_{GC} > 0$, $Z_{GC} > 0$, and  $Z_{GC} > Y_{GC}$, where the
latter limit is used to avoid much of the Galactic disk; Figure 16 is a useful aid for
orientation to this wedge.  The peak of the
``contamination shell" for $0.90<(J-K_s)_o \le 0.95$ is plainly visible at a distance 
of about 35 kpc, for example, and at 47 kpc for $0.95<(J-K_s)_o \le 1.00$.  Only
for $(J-K_s)_o > 1.05$ is the outer limit of the Northern Loop confidently 
free of significant contamination when viewed (as in Figure 14) 
in projections of density within a slab of finite width (note
that in contrast the counts shown in Figure 15 are for a volume element increasing 
as the cube of the distance).

Convolved and competing with the above technical problem of determining the 
true length of the Sgr tidal arms is a second complication arising
from stellar population considerations: 
M type red giant stars occur only in metal-rich populations.
While the Sgr center has ample numbers of sufficiently
metal-enriched stars to create a substantial M giant population, 
the Sgr metallicity gradient found by Alard (2001)
suggests that the tidal debris leaving the Sgr dwarf now 
is likely to be, on average, more metal poor 
than that remaining in the most central regions.  Moreover, Layden \& 
Sarajedini (2000) have demonstrated a clear age-metallicity 
relationship among the Sgr populations.  Thus, one might expect
a natural limit to the extent that the Sgr tidal tails {\it could} be traced
with M giants, with that limit corresponding to the oldest 
possible tidal debris that can contain M giants.  

In Table 3 we give the age-metallicity
characteristics of the three primary Sgr populations identified by
Layden \& Sarajedini (2000).  For each of these populations we use 
the isochrones and other data in Bertelli et al. (1994) to determine 
the $(J-K_s)$ color of the RGB tip, accounting for the conversion to 
the 2MASS photometric system using the equations in Carpenter (2001).  
As may be seen, the most metal-poor population in the Sgr system is
virtually invisible to the color-selected M giant candidate sample
shown in Figures 8-11 (recall the invisibility of the M54 
globular cluster in Figure 4).  Thus, uncovering the distribution of these older 
detached giant star populations requires use of earlier and intrinsically
fainter giant stars, 
but doing so with 2MASS, as shown in Figure 14a, is complicated
by the severe contamination at distances of particular interest.
On the other hand, younger populations will have had less time to 
separate from the Sgr core.
Therefore, the apparently decreasing length of both leading and trailing M giant
tidal tails as we map them with progressively
cooler giant star tracers (Figure 14) suggests 
{\it a mean stellar age/metallicity variation along the tidal 
tails}.\footnote{A tendency for the tidal arms to appear
more tightly wrapped for bluer colors may further hint at a
shift in the mean color-magnitude relation for M giants in the
tails compared to the color-magnitude relations adopted from
the Sgr center, in the sense that the tails contain more metal-poor
(brighter) M giants than the center of Sgr.}    

The actual tidal release age of any particular part of the 
Sgr arms as deduced from the apparently youngest stars 
is an upper limit because (1) the steepness of the luminosity function means that
the incidence of stars at the actual tip of the RGB is 
relative rare, so that any particular M giant is likely blueward of
the RGB-tip color for its age/metallicity population (Table 3), and 
(2) presumably stars created from residual, bound gas in the
core are not instantaneously released but must first somehow find themselves
outside the tidal boundary (see Section 4.3.4).
Therefore, it is reasonable to suspect that the M giant
population explored in this paper is actually tracing only the very 
most recently lost stars (perhaps only the last several gigayears or so) from what may be a 
longer tidal interaction and net tidal arm length.  Such youthful ages
for the length of tails reported here are consistent with Sgr disruption models
(e.g., Johnston et al. 1999a, Law et al. 2003).

Nevertheless, the highly tuned samples shown in Figures
14b-14d offer the strongest evidence 
that the trailing Sgr arm is at least long enough to presently
lie across the Galactic plane in the Northen Galactic Hemisphere 
towards the Galactic anticenter.  If so, then Sgr disruption
models (e.g., Law et al. 2003) suggest that the corresponding
leading arm can be long enough to reach the Galactic plane
on this side of the Galactic Center.
Moreover, Figure 14c (in particular)
offers tantalizing evidence that the {\it leading} Sgr arm may 
extend not only beyond the Galactic plane but beyond 
the trailing Sgr arm as well: The ``spray" of stars in the lower 
right quadrant of the orbital plane is similar to the wrapped, leading
arm stars shown in the K6-a model of Ibata et al. 2001b (see their
Figure 3) and the model shown in Figure 1 of Law et al. (2003).

\section{From Sgr to the Galactic Halo}

\subsection{Minimum Integrated Mass Loss Of Sgr}

The longitudinal profile in Figure 13 enables an estimate of the fractional 
mass of the Sgr system in its tidal arms under the assumption
that the M giants provide a suitable and equitable tracer over the entire 
Sgr + tail system.  Presuming mass in the the Sgr system to be symmetrically 
divided about its center  (so that we only use southern Hemisphere Sgr stars with 
$\Lambda > 0$ in our calculations) 
we find that the tails, defined as those Sgr stars lying outside the King
profile (Section 4, Table 1), contain about 15\% the number of stars within
the King profile.  This estimate is derived from the counts for 
stars within the $|Z_{Sgr,\sun}| = 5$ kpc slab in Figure 13.


Johnston, Sigurdsson \& Hernquist (1999) have given a formalism for 
calculating the mass loss rate in a dwarf satellite based on a measured 
profile such as that shown in Figures 5 and 13.  Using their formula (18), 
an orbital period of the Sgr system of 0.7 Gyr, and only the clearly
visible tail from the King profile 
to $\Lambda_{\sun} = 155^{\circ}$, we obtain a
mass loss rate for Sgr that is 17\% of the mass interior to the King 
limiting radius per Gyr.  A slightly different formulation by
Johnston, Choi \& Guhathakurta (2002) yields a mass loss rate
about 40\% smaller.  These order of magnitude estimates are
consistent with the the above empirically determined M giant mass fraction
in the Sgr tails if those tails correspond to mass lost over the last 
about 1.5-2 orbits; this is indeed the timescale suggested by matching
tails reproduced by Sgr disruption models (e.g., Law et al. 2003). 


As discussed above, because M giants only trace recently formed tidal debris 
they permit only estimates of a {\it lower limit} to the net stellar mass lost
by Sgr and the fractional contribution of Sgr debris to the Milky Way halo. 
Moreover, in this discussion we have ignored the issue 
of whether a sizeable fraction of the stars within the Sgr King profile represent
stars that have become unbound in the most recent periGalacticon passage
along the lines of the scenario envisaged in Section 4.3.3.  Thus, the
mass loss limit estimated above pertains primarily to stars detached prior to 
the recent periGalacticon.

\subsection{Sagittarius Stellar Contribution to the Milky Way Halo}

Figure 16 shows the distribution of 2MASS late type giants projected onto
the Galactic $Y-Z$ plane.  In this orientation, we see the Sgr orbital
plane almost ``edge-on" as the vertical spike spanning both
hemispheres, as in Figure 7a.  
The panels illustrate the same color ranges as Figure 14c-e, however, 
unlike the latter figures, in which the sample has been limited
to stars in a 14 kpc wide slab centered on the Sgr plane, 
Figure 16 shows the entire 2MASS sample, except for stars more reddened 
than $E(B-V)=0.555$.  In addition, to avoid substantial noise from 
projection of the faint magnitude ``contamination shells" (Section 6.6), 
it is necessary to remove all sources with estimated photometric parallaxes 
larger than 40 kpc, 50 kpc, and 60 kpc, respectively, in Figures 16a, 16b
and 16c. 
These distance limits effectively remove the bulk
of the faint end contaminants (see Figure 15), at the expense of slightly
truncating the most distant parts of the Northern Loop.
Figure 16 illustrates that, apart from
the Magellanic Clouds, which are the sources of the large ``finger of God" 
features\footnote{That the Magellanic Clouds are seen as ``finger of God" spikes
in Figure 16 is attributable to the fact that the color-absolute magnitude
relation we have adopted is specifically tuned to the most metal rich Sgr 
population, and is not necessarily a good description of 
the various Magellanic populations.}
in the lower right of each panel, unbound Sgr debris appears to be
the predominant source of late type giant stars in the Milky Way halo.  


On the basis of the 
data in Figure 16 and assessing only the Northern Galactic Hemisphere to 
avoid the complication of the Magellenic Clouds,
we estimate that Sgr debris represents more
than about 75\% of the high halo ($Z_{GC}>13$ kpc) in the color ranges shown.
The estimate rises to 80\% or more if the high halo is defined by
$Z_{GC}>20$ kpc.  
These estimates are based on assuming all stars within 5 kpc or so of the Sgr
plane are indeed Sgr debris, and so will be overestimated to the degree
that there are non-related halo stars in that volume; however, 
based on the apparent mean density of stars away from that plane, 
this should be a minor effect (except in the case of
chance coincidence of other M giant substructure near the Sgr plane).\footnote{The 
Section 6.5 analysis of mean contamination in the general volume around the
trailing tail found about a 10\% effect.}
Of course, these calculations do not (1) include Sgr debris stars 
lopped off the top of the Northern Loop by the distance limits 
(which would increase the fractional Sgr contribution), (2) account for any residual
contamination of the halo by false positives introduced by photometric errors, or (3) 
account for any possible increases in the number of M giants at larger radii than
the limits shown, or at Galactic latitudes lower that those analyzed.  
In addition, (4) our criterion for selecting M giants was 
guided specifically by the location of {\it Sgr}-type M giants in the NIR two-color diagram, 
though age-metallicity effects in the relevant parts of the two-color diagram are 
minor for these types of stars.  Despite these minor uncertainties,  
that Sgr debris is the major
contributor of the high latitude halo M giant population to 60 kpc 
seems a reasonable conclusion. 

The dominance of Sgr in creating the Galactic halo M giant population is reflected 
in the Great Circle Cell Counts analysis discussed in Section 5.  
However, our results differ somewhat from those of Ibata
et al. (2002a), whose analysis of the 2MASS early release data led them to 
conclude that Sgr debris represented only about 5\% of the halo M giant population.
Though the two analyses use different selection criteria to isolate M giant stars, 
we are uncertain exactly why they arrive at such substantially different limits on the M giant
contribution to the halo.  It may be that Galactic the exclusion of disk giants based on
a Galactic latitude limit, as done by Ibata et al., is not as restrictive as 
our $Z_{GC}$ criterion; but a more likely contributor to the difference is our
elimination of the excess background by the ``contamination shell" (Section 6.6) 
in both the Cartesian as well as the GC3 analysis presented in Section 5.  

However, we are in agreement with Ibata et al. (2002a) that,
apart from the presumably bound population of red stars in the Magellanic Cloud
represented by the finger of God spikes, it would appear that the Clouds 
have {\it not} been a major contributor to the halo M giant population.  

These results pertain only to Sgr contribution to the halo of the
latest type
giant stars, and say nothing about the net mass contributed to the halo, either
in the form of dark matter or in stars of all spectral types.  However, along these
lines, we find interesting the result of Vivas et al. (2001; discussed further below), 
in which almost every one of the RR Lyraes they find along the line of sight to the
apoGalacticon of the Northern Loop could conceivably be a part of Sgr tidal debris, 
possibly including even the nearby RR Lyraes, depending on the disposition of the
Sgr debris near the Sun.  In any case, that Vivas et al. find a ``hole" in their
RR Lyrae counts precisely at the distance of the interior of the Northern Loop is 
dramatic and
suggests that even for such old stars Sgr may be a dominant contributor to at least
the outer ($>25$ kpc) halo.

\section{Comparison to Previous Sgr Searches and Potential Identifications}

Dinescu et al. (2002) have summarized the various searches for extended Sgr
debris to date; their Figure 4 gives a representation of the placement
of various detections and non-detections on the celestial sphere, 
along with a great circle for the Ibata et al. (1997) Sgr
orbit, which reasonably approximates that which we have found here.
Given our new understanding of the three dimensional position of Sgr debris, 
it is worth reviewing the previous 
detections of Sgr debris in more detail here, and, in particular,
taking into account the {\it distances} of the stars that constitute
the various detections.  A comparison with other surveys is especially useful
(1) as a check on distance scales from the disparate tracers that have been used,
(2) because we are now able to place almost all previous detections into
a unified context, and (3) comparisons to surveys of other Sgr tracers provide
new insights into the Sgr disruption and debris trails.
Figure 17, which repeats the M giant distribution
of Figure 10, provides our summary comparison of the
detections by Sgr longitude and distance.  Figure 17 includes only
detections of extratidal Sgr material, and excludes the numerous studies
near the Sgr center.

\subsection{Connecting to the Sloan and QUEST Detections}

Perhaps the most striking visual impression of extended Sgr 
(and other potential) tidal debris in the halo has been that afforded by
the Sloan Digital Sky Survey.  
In several studies analyzing data from the first Sloan observations in a strip
along the celestial equator, the presence of Sgr's extended tidal arms
have made themselves known (e.g., Figure 18a).  
Figure 18b shows a slice through
the 2MASS M giants along the celestial equator, which mimics the
region of the sky surveyed by the Sloan survey on the equator.  While
the latter covers a roughly 2.5 degree wide strip along the
equator, we opened the declination range of 
our comparison image to $-10^{\circ} < \delta < +10^{\circ}$
to increase the density of plotted points for our lower density population
of M giants.

The first published results from Yanny et al. (2000), while only
in two limited angle wedges of the equatorial stripe,
nevertheless showed excess starcounts of A type stars 
in several regions that can now be firmly identified with parts of 
the Southern Arc at $\Lambda_{\sun} \sim 104^{\circ}$ 
and the far side of the Northern Loop at $\Lambda_{\sun} \sim 295^{\circ}$.
The heliocentric distances Yanny et al. (2000) infer for their two
structures are 28 kpc and 48 kpc, respectively.  These distances
generally agree with our results (Figure 17).
Though they do not comment on it, the Yanny et al. data also show an 
excess of stars $<20$ kpc away in the same direction of the sky (see, e.g.,
their Figures 18 and 19), consistent with our finding of {\it closer} M giant 
candidates at the same longitudes ($\Lambda_{\sun} \sim 295^{\circ}$).
Because of uncertainty over the mean distance of this nearby Yanny et al. clump, 
it is not represented in Figure 17.  

A similar detection of two density enhancements towards the Northern Loop
has been discovered in the study of RR Lyraes discovered in the Sloan
equatorial strip by Ivezi{\' c} et al. (2000), as well as in the QUEST RR Lyrae 
Survey (Vivas et al. 2001), which explores nearly the same region of sky 
($13 < \alpha < 16$ hours).
Both surveys comment primarily on an excess of RR Lyrae stars at 45 and 50 kpc,
respectively, a feature that we can now confidently associate with the same 
expanse of the Sgr Northern Loop identified by the Newberg et al. (2002)
and Mart{\' i}nez-Delgado et al. (2001b) surveys, and shown in the upper right
quadrant of Figure 11.  The distances of these more distant
RR Lyraes are also in reasonably good agreement with the M giant distribution 
presented here ($\Lambda_{\sun} \sim 270-310^{\circ}$, Figure 17).

As with Yanny et al.'s A stars,  both Ivezi{\' c} et al. and Vivas et al. also have in their
distance distribution of RR Lyrae stars a large number at distances 
that correspond to the $<20$ kpc M giant candidates seen at $\Lambda \sim 295^{\circ}$
in Figure 17.  Figure 4 of Vivas et al. and especially
the middle panel of Ivezi{\' c} et al.'s (2000) Figure
8 are very similar in appearance to the distribution of stars one would obtain along the 
same line of sight in our Figure 11.  Vivas et al. find
that from 16-23 kpc there is a {\it bona fide} excess of stars over
an $R_{gc}^{-3}$ law, but they attribute the majority of this excess
to be likely bound and unbound RR Lyrae stars from the tidally disrupted
Pal 5 system, while Ivezi{\' c} et al. apparently do not find an excess
over a -2.7 power law.  Clearly radial velocity
data are needed to determine whether any of the $<20$ kpc M giants
and RR Lyrae may be related to Sgr or other tidal debris interior to the Solar Circle
(e.g., wrapped up leading arm material as described in Johnston et al. 1999a and
Kundu et al. 1999), or whether they are all part of the Galactic bulge, IPII/thick
disk and/or inner halo. 

Interior to their $\sim 45-50$ kpc clumps, both 
RR Lyrae surveys also show a prominent ``hole" in their distribution that appears
to correspond to the interior of the Northern Loop.  This is an interesting
result, because Sgr disruption models (e.g., Ibata et al. 2001b, 
Law et al. 2003) predict that
trailing arm debris, if extended beyond the length limit revealed by M giants here,
should eventually reach and cross through the Northern Loop hole.  
Indeed, the cluster NGC 5364, which lies right in the middle of the Northern 
Loop hole (see Figure 17), is consistent with the position and velocity of such 
extended, wrapped Sgr trailing debris (Bellazzini
et al. 2002a, 2003).  If NGC 5634 is Sgr debris, one might expect 
to see a population of Sgr RR Lyraes along with it.  Further
work is needed to clarify this dilemma. 
 
Our analysis (Section 7.2) for the fractional contribution of
Sgr {\it M giants} to the halo pertains to a stellar species expected
only for relatively metal rich ([Fe/H]$ \gtrsim -1$) populations --- and
it is not altogether too remarkable that the relatively minor fraction of
halo stars that are that metal-rich could have come from a very small number of
contributors like Sgr.  
However, the overall distribution of {\it RR Lyrae stars} in this part of the sky,
including the contrast of near and far clumps and the intervening hole, 
appears to match closely the distribution of
the M giants.  The RR Lyrae data suggests that the dominance of younger, 
more metal-rich populations traced by Sgr M giants in the halo
extends to older, more metal-poor populations 
(at least for the outer halo and in this one direction of the sky).

The most extensive use of the SDSS for stream detections is that presented
for presumed main sequence turn off stars in a nearly complete
equatorial stripe by Newberg et al. (2002).  Figure 18a here is a reproduction of their
their Figure 1; we include on Figure 18b the azimuthal locations
of features pointed out and discussed by Newberg et al.  It is clear that
the strong Newberg et al. feature S167-54-21.5 is indeed the Southern
Arc tidal arm of Sgr, and their feature S341+57-22.5 is the far side of the
Northern Loop (as suggested by their own discussion of these features); 
the Sgr longitudes of these features are (see Figure 10) $\Lambda_{\sun} 
\sim 110^{\circ}$ and $\sim 286^{\circ}$, respectively.  But we can also
make the connection of the more diffuse clumping S297+63-20.0, which 
Newberg et al. attribute tentatively as ``a stream or other diffuse 
concentration of stars in the halo", as well as a lot of the similar-magnitude
fluff contiguously connected to this feature from $\alpha \sim 150^{\circ}$
to $\alpha \sim 210^{\circ}$, to M giants tens of kiloparsecs above the 
Galactic plane that are in the heart of the descending, foreshortened 
Northern Loop near 
$\Lambda_{\sun} = 265^{\circ}$ and stretching more generally from 
$\Lambda_{\sun} \sim 230^{\circ}$ to $\Lambda_{\sun} \sim 285^{\circ}$ 
(see Figure 17).  The consistency with the M giant debris here is noteworthy, 
and the wide spread of the S297+63-20.0 feature elicits further interest
into the question of precisely where the Northern Loop crosses the Galactic
plane near the Sun.

Newberg et al. draw attention to several other features located at the low $|b|$ 
edges of their survey wedges.   For example their feature S223+20-19.4 is discussed
in the context of a possible ``newly discovered dwarf galaxy in the Galactic plane" at 
a distance of about 11 kpc,
but they also admit the possibility that it is a metal weak, disklike structure with 
large scaleheight and scalelength.  Ibata et al. (2003) have suggested the possibility
that the Sloan detection may be a perturbation of the disk, possibly the
result of ancient warps.  To aid interpretation of this feature, 
Helmi et al. (2003) discuss models of both old, shell-like and younger, more
coherent tidal features.  Yanny et al. (2003) have shown that the feature is likely
to be a disrupted galaxy, based partly on the low velocity dispersion of 
stars within it.
Figure 18 gives the appearance of a distinct structure at the same
position and at a corresponding distance modulus ($m-M \sim 15$) if we assume 
these are M giant stars.  This structure appears to span both sides of our
Zone of Avoidance (though predominantly situated North of the Galactic plane 
in this slice through the Galaxy), 
with an overdensity of stars that also corresponds more or less to
the Newberg et al. S200-24-19.8 structure.\footnote{The apparent
overdensity in the 2MASS M giant sample corresponding to the
S200-24-19.8 structure shows more clearly when 
bluer M giants are included in the analysis.}  This feature shows up in Figure 10
as the $\sim 8-15$ kpc distant, oblong shaped feature spanning $\Lambda_{\sun}
\sim 150-200^{\circ}$ in the top panel and the $R_{GC} \sim 12-22$
kpc feature spanning $\Lambda_{GC} \sim 140-180^{\circ}$ in the bottom panel.
It is unlikely that the 2MASS 
feature is from improper dereddening, since the S223+20 structure extends to 
reasonably high
latitudes ($b>20^{\circ}$).  Figures 10 and 18 supports the reality of the Sloan
find, and in two contributions (Rocha-Pinto et al. 2003, Crane et al. 2003) this
2mass feature is explored further, with the conclusion that it is a new 
tidal tail system unrelated, but analogous, to that of the Sgr dwarf.

\subsection{Other Searches for Distant Sgr Debris}

The M giant results have already been compared to those of Mateo,
Olszewski \& Morrison (1998) in Figure 13 and Section 6.5. 
Near the Mateo et al. strip of fields, and slightly closer to the
center of the Sgr tidal debris stream is the possible detection
of Sgr red clump stars in the ASA184 field 
($[l,b]=[11,-40]^{\circ}$) by Majewski et al. (1999a).  The distance
modulus with reddening of these stars at $\Lambda_{\sun} \sim 26^{\circ}$
is projected to be $(m-M-A)_V=16.8$, which is
22 kpc assuming $A_V=0.15$; this distance is in agreement
with the M giant distribution in Figure 10.  Majewski et al.'s non-detections
in the other three fields they studied - SA184, SA107 and ASA107 -- can
be understood by comparison to the Sgr debris streams as delineated by M giants:
SA184 and SA107 are off the Sgr orbital plane (these fields were used
as control fields for ASA184 and ASA107 by Majewski et al.).
The fourth field they studied, ASA107, however {\it does} lie ($[l,b]=[353,+41]^{\circ}$) 
in the thick of the Northern Loop at 
$\Lambda_{\sun} = 300^{\circ}$ (which is why it was originally selected for 
study).  
Unfortunately, the mean distance of the loop
at this point, i.e. $\sim45$ kpc translates to an expected red clump magnitude of
$V \sim 19.5$, which was just beyond the limit of 
their study. \footnote{ Interestingly, there is an excess of stars
at this magnitude visible in the Figure 4 of Majewski et al. (1999a), 
but the excess was deemed as not statistically significant
by those authors.} This point has previously been made by Mart\'inez-Delgado
et al. (2001b).

Another Kapteyn Selected Area previously suspected to contain
Sgr debris is SA71 ($[l,b]=[167,-35]^{\circ}$).  
Dinescu et al. (2002) report an excess of
$B-V \le 1.1$ stars for $18 < V < 20$, and most prominently in
the range $18 < V < 19$.  These stars also appear to have distinct
proper motions consistent with the Sgr orbit.  SA71 lies near the 
main Sgr debris stream towards the Galactic anticenter 
($\Lambda_{\sun} \sim 128^{\circ}$) where the 
M giants are centered at about 28 kpc distance.
By assuming that their excess population corresponds to the Sgr
horizontal branch/red clump, Dinescu et al. derive a distance
for their potential Sgr debris of 29-32 kpc, in good 
agreement with the M giants.  
Dinescu et al. explore 
three other Selected Areas --- SA29, SA45, and SA118 ---  and found 
no similar Sgr-like detection.  As these authors point out, SA29 and
SA45 are considerably off the primary Sgr orbit.  However, 
in their Figure 3 the field SA118 is shown to be nearly similarly displaced
from the Ibata et al. (1997) Sgr orbit as SA71.  Our ability to pinpoint
more precisely the path of the Sgr debris allows us to determine
that, in fact, SA71 is much closer to the primary debris great
circle than is SA118; this could explain their Sgr debris
non-detection in SA118. 

Mart{\' i}nez-Delgado et al. (2002) have identified potential
main sequence Sgr stars in deep $BR$ imaging
near the globular cluster Pal 12, previously identified
by Dinescu et al. (2000) as a likely Sgr globular cluster.
Mart{\' i}nez-Delgado et al. estimate the distance 
of these stars as 17-24 kpc, depending
on assumptions about the expected absolute magnitudes of the
stars.  At this longitude ($\Lambda_{\sun} \sim 40^{\circ}$), we
find the mean M giant distance to be about 19 kpc, which is
also the same distance as Pal 12, and this is consistent with the
Mart{\' i}inez-Delgado results.   
In a similar, deep CMD search in a field in the SDSS equatorial strip,
Mart{\' i}nez-Delgado et al. (2001) also find a signal they 
tentatively associate with Northern Hemisphere Sgr dwarf material. 
The distance to the feature, which they
associated with ``the Sagittarius stream or traces of a new nearby dwarf galaxy"
is $51 \pm 12$ kpc, and $R_{GC} = 46 \pm 12$ kpc.  Their identified 
stellar population indeed corresponds to the distant part of the Northern 
Loop at $\Lambda_{\sun}=295^{\circ}$, and is only slightly farther than 
other detections in this part of the sky, including our own.
The lack of detection of
the near side of the Northern Loop by both Majewski et al. (1999a) and 
Mart{\' i}nez-Delgado et al. (2001) relates to the bright-end 
magnitude limits in both surveys.

Finally, two searches for giant stars have recently published
possible detections of Sgr debris in the Northern Hemisphere.
Dohm-Palmer et al. (2001) have found
four giant stars with similar velocities and distance in fields near the Sgr mid-plane
near $\Lambda \sim 295^{\circ}$.  At least some of these stars, 
at a typical distance of 50 kpc
and a moderate positive velocity, are a plausible Sgr Northern Loop detection
consistent with the M giant distribution.  Finally, Kundu et al. (2002) 
have found a position-velocity sequence of eight giant stars 
with unusually large negative
velocities as part of a large K giant survey.  These stars lie very
near the Sgr mid-plane and may correspond to the very near side of
the Northern Loop (see Figure 17).

We may summarize the comparisons discussed to this point as almost
uniform in agreement with regard to both the locations of Sgr debris in position
on the sky {\it and} with respect to distance (despite the   
disparate methods for identifying and gauging the Sgr debris).

\subsection{Carbon Stars}

Carbon stars have also been associated with the Sgr plane.  The
large-area APM Survey (Totten \& Irwin 1988) revealed dozens of carbons
stars having positions and radial velocities consistent with the Sgr
tidal tails and which Ibata et al. (2001a,b) used 
to define a Sgr orbital plane and a debris model that generally resembles 
the distributions of 2MASS M giants.  
However, the carbon star luminosities adopted in these studies
yield photometric parallaxes that are,
on average, $\sim 35\%$ larger than the M-giant distance scale (which has 
been shown to agree with numerous other studies; Figure 17), even
when very dusty N type stars are ignored.\footnote{The assertion of an
overestimated distance scale
assumes that the carbons near the Sgr plane shown in Figure 17 are
predominantly Sgr debris.  Totten \& Irwin (1998) have mentioned
that CH-type carbon stars ``are likely to be somewhat fainter intrinsically
than N-types and hence closer than estimated...".}

Calibration of the carbon star distance scale has been historically
complex, being complicated by variability, obscuring dust shells and
metallicity effects.  While Totten, Irwin \& Whitelock (2000) have
demonstrated good agreement between infrared, $JK$-based distance estimates
and an assumed $R$-band carbon star absolute magnitude of $M_R=-3.5$,
Demers, Dallaire and Battinelli (2002) have noted a metallicity
trend whereby the [Fe/H]$=-1.4$ carbon stars in Fornax are
0.25 mag fainter in $K_s$ absolute magnitude than [Fe/H]$\sim -0.5$
LMC carbon stars.\footnote{
Kunkel, Demers \& Irwin (1997) also find a median $R=15.2$ magnitude 
for more than 400 carbons in the LMC periphery, which, with an
LMC distance modulus of 18.55, yields an $M_R$ closer to -3.35 for 
these carbon stars.  Totten, Irwin \& Whitelock (2000) 
note a ``vertical scatter about the fitted curve [that]
covers a range of $\sim \pm 0.5$ mag, with occasional more extreme
outliers that in the main are probaby caused by variable stars."
Their data also reveal something of
a population gradient in that the bright, blue CH-type LMC carbons of
Hartwick \& Cowley (1988), which have been argued (Suntzeff et al. 1993)
to be a very young (0.1 Gyr) AGB population, lie well above their
fit color-magnitude relation, while the more ``normal" LMC carbons as well as a number
of other dSph and SMC carbons create much of the vertical scatter
0.5 mag or more {\it fainter} than the fit.  This trend echoes the Demers et al.
(2002) conclusion regarding likely metallicity effects on carbon star
luminosities.  }

The 2MASS database makes possible a new attempt to calibrate the Sgr 
carbon star color-magnitude relation and apply it to   
self-consistent photometry of presumed carbon stars in the Sgr tidal tails.  
Figure 19 highlights 95 extreme-colored ($[J-K_s]_o>1.3$) stars 
within 5$^{\circ}$ of the Sgr center that lie 
in a well-defined carbon star locus extending from the Sgr
Red and Asymptotic Giant Branches.
Stars with $(K_s)_o<8$ are most likely foreground carbon stars 
associated with the Galactic bulge. 
Though it contains several times more carbon stars, this Sgr carbon
star sequence is consistent with the Sgr carbon star locus of 
Whitelock et al. (1999).  However, both the present and the Whitelock
et al. sequences fall below the carbon star
loci determined for Milky Way satellites (Totten et al. 2000) and 
the LMC (Weinberg \& Nikolaev 2001) when these loci are 
adjusted for the distance to the Sgr core.
The mean $(K_s)_o$ of the highlighted points in Figure 19
is $9.59\pm0.06$ mag. 
In the color range $1.3 < (J-K_s)_o < 2.0$, the Sgr carbon locus
is $0.39\pm0.07$ mag fainter than the Weinberg \& Nikolaev LMC locus, 
while for $(J-K_s)_o \ge 2.0$ the Sgr locus is $0.64\pm0.10$ mag underluminous.  

Even were these vagaries in the mean calibration of the absolute magnitude
color relation worked out, 
the spatial distribution of 2MASS-selected carbon stars (Figure 20) provides a
poor estimate of Sgr morphology relative to M giants because:
(1) Sgr carbon stars are much less populous (thirty
times less numerous than $0.95 \le (J-K_s)_o < 1.10$ M giant candidates in
the same area of the Sgr center), 
%
%
(2) carbon stars have
a larger intrinsic scatter in their color-magnitude relation (an RMS of
0.59 mag in the Figure 19 carbon sample compared to 0.46 mag for the Figure 1c
M giants), and (3) a substantial number of carbon stars are long period
variables.  2MASS in particular provides mainly single epoch observations of 
a carbon star sample that likely contains a substantial fraction of 
$\Delta K > 0.4$ Mira, as well as lower amplitude, variable stars
(Whitelock et al. 1999). 

Figure 20 shows the orbital plane distribution
of Galactic carbon stars (selected as sources with $[J-K_s]_o \ge 1.3$), 
with absolute magnitudes derived 
from the Weinberg \& Nikolaev (2001) loci dimmed
by 0.5 mag (panel a) and by adopting a simple
$M_{K_s} = (9.59 - 16.90) = -7.31$ for all stars (panel b).\footnote{In 
contrast to previous plots presented here for M giants
(e.g., those shown Figure 14) that only included stars within a {\it linear} 
distance from the $\Lambda_{GC}$ plane, Figure 20
shows stars with {\it angular} ($|B_{\sun}| < 10^{\circ}$)
separations from the Sgr plane;
with the larger uncertainty in the carbon star photometric parallaxes,
we risk losing Sgr carbons with a linear constraint on 
distance from the Sgr plane.}
Comparison of Figures 20a and 20b shows that the Northern Loop 
is actually better defined and more similar to the M giant 
distribution when the constant, $M_{K_s}  = -7.31$
color-absolute magnitude relation is adopted than if one were
to calibrate from the color-magnitude relation derived from other 
Milky Way satellites. 
A ``finger of God'' effect for the Sgr center is a result of intrinsic spread 
in the color-magnitude relation (Figure 19) and source
variability.  About five or six 
dozen high latitude carbon stars lie near the Sgr plane but only loosely 
trace the M giant tidal arms (compare Figure 20 to Figure 11). 

To give some impression of the relative
contribution of carbon stars to the Galactic halo from the Sgr
dwarf, we show in Figure 20c the Galactic $Y_{GC}-Z_{GC}$
distribution of all stars with $(J-K_s)_o \ge 1.3$.  
This figure should be compared to the corresponding M giant
panels in Figure 16.
For clarity a constraint of $(K_s)_o < 11.75$ is imposed 
(without this criterion the distribution is significantly
noisier, likely due to a ``contamination
shell" problem as found for the M giants in Section 6.6). 
Outside the quadrant containing the Magellanic Clouds, Sgr appears  
to have been the predominant source of high latitude,
$R_{GC} \lesssim 75$ kpc halo field carbon stars.

\subsection{Globular Clusters}

It is presently known that four globular clusters with positions near the
Sgr center, NGC 6715 (M54), Terzan 7,
Terzan 8 and Arp2 are associated with the dwarf galaxy:  These globulars
have similar distances and radial velocities to the main body of Sgr (Ibata et al
1995).  A fifth cluster that lies in the heart of the Southern
Arc (Figure 17), Pal 12, has been shown to have orbital characteristics
that make it a reasonably good candidate for association with Sgr (Dinescu
et al. 2000; see also Mart\'inez-Delgado et al. 2002).  Several studies of
the Galactic globular cluster population have sought additional possible
cluster members of the Sgr debris streams, with a number of additional 
candidates proposed (Irwin 1999, Dinescu et al. 2001, Palma et al. 2002,
Bellazzini et al. 2002a, 2003).  The recent analysis by Bellazzini et al.
(2003), in particular, makes a strong case for several additional Sgr
clusters.  Because, as Bellazzini et al. (2003) have shown, precise knowledge
of the Sgr orbit is of great use to sorting out interesting candidates, 
we defer an analysis of connections of the Sgr debris streams and
globular clusters to a companion contribution containing velocity data 
as well as our best fit model to the M giant data.

\section{Density of Nearby Sgr Stellar Debris}

In Section 6.4 we discussed the proximity of the Sgr Northern Arm
to the Solar Neighborhood.  In Section 6.6 we argued that the
Southern Arm seemed to be at least $180^{\circ}$ long, sweeping into
the Northern Hemisphere; if so, then models of the Sgr disruption
(e.g., Law et al. 2003)
show that the Northern Loop is long enough to cross the Galactic plane 
on this side of the Galactic Center.  In Figures 14c and 14d we find
evidence that the leading arm may even cross the {\it trailing} arm in the
Southern Hemisphere.  And in Section 8 and Figure 17 we showed 
apparently confirmatory evidence from other surveys that Sgr debris  
is approaching the Solar neighborhood from the NGC.
How might the suggested local presence of debris from the Sgr dwarf 
spheroidal have impacted previous studies of the Galactic halo, many
which have been conducted with halo stars relatively near the Sun
in potentially ``Sgr-contaminated" regions of the Galaxy?
The question turns on the relative density of leading Sgr arm
debris passing through/near the solar neighborhood.

We can estimate the local density of Sgr stars by extrapolating the M giant
density just above the Galactic plane and converting that density
to other spectral types (colors) via an adopted luminosity function. 
This is most straightforward for evolved stars, where the luminosity function 
can be derived directly from 2MASS observations of the Sgr center --- for
example, the background-subtracted Sgr color-magnitude 
diagram shown in Figure 1c.  To eliminate residual, unsubtracted
contamination from non-Sgr stars and isolate the Sgr RGB we apply the 
following criterion:

\begin{equation}
\label{contam}
K_s > -7.22 (J-K_s) +17.64. 
\end{equation}

\noindent This selection effectively separates the Sgr RGB from the
residual contributions of the Galactic bulge population several 
magnitudes brighter (see Figure 1c).
%
%
%
The resulting luminosity function so calculated is shown in 
Figure 21a.  The Figure 1c color-magnitude diagram begins to ``run out"
beyond $(K_s)_o = 14.3$. 

The {\it color function} for evolved stars corresponding to the
luminosity function is shown in Figure 21b.  
From the slope of the RGB, the $(K_s)_o = 14.3$ magnitude limit means that 
the color function is complete only for RGB
stars redder than $(J-K_s)_o \sim 0.80$ --- roughly spectral
types later than K3.  

The ratios of stars of different spectral types can be computed by 
comparing counts by colors.  Bessell \& Brett (1988; 
see also Bessell et al. 1991) have given approximate colors for
stars by spectral type and luminosity class.  
Obviously metallicity effects are important,
but for a rough calculation the corresponding 2MASS color for the 
Bessell \& Brett types is simplistically adopted.
Accordingly, the 2MASS color of a
type M0III star is $(J-K_s)_o \sim 0.98$ (Carpenter 2001).
Table 4 presents the Figure 1c counts for evolved Sgr stars as determined by 
the color functions presented in Figure 21b.  Roughly, for every Sgr M giant
we expect three stars of type K3III through M0III, and more than
seven K giants of any type (a substantially conservative lower 
limit due to the incompleteness of the early K type giant counts due to the 
magnitude limit of the Figure 1 sample).  

We now extrapolate the M giant density in the Sgr leading tidal arm
to the solar neighborhood by counting
the number of M giant stars ($0.98 \le [J-K_s]_o < 1.30$)
in a 5 kpc radius cylinder centered on the Sun and
whose axis is roughly perpendicular to the Galactic plane.\footnote{The 
actual cylinder used is centered on
$[X,Z]_{Sgr,GC} = [-8.5,0]$ kpc and parallel to the $Y_{Sgr,GC}$
axis in Figure 11.}
To avoid disk contamination and primarily sample the nearby Sgr leading arm, 
only stars with $-30 < Y_{Sgr,GC} < -9$ kpc are tallied;  
70 are found, which results in an M giant density of about 0.042 kpc$^{-3}$. 
Under the assumption that all of these stars are leading arm Sgr stars
and using Table 4, this implies a nearby Sgr $>$K3 giant 
density of 0.17 kpc$^{-3}$.
Because the M:K giant ratio
decreases with age/metallicity, and, given the evidence for possible
age/metallicity/giant color variations depicted in Figure 14, 
the above K giant density is a lower limit.\footnote{For 
example, Majewski et al. (2002b) stress how the 
{\it observed} age distribution of bound populations in a steadily 
disintegrating stellar system is more heavily weighted toward younger 
populations, and does not not accurately reflect the     
balance of populations to be found in tidal debris from that stellar system.}
This density is comparable to
the density of a velocity clump of nine, mostly metal poor ([Fe/H] $< -1$) red giants 
having Hipparcos proper motions and radial velocities and located within 2.5 kpc of the
Sun discussed by Helmi et al. (1999).  These authors postulate that this clump, 
which has a velocity perpendicular to the plane consistent with that expected for nearby 
Sgr debris (roughly 225 km s$^{-1}$ downward), came from a progenitor system
that ``probably resembled the Fornax and Sagittarius dwarf spheroidal galaxies" and
that may have contributed 12\% of all metal-poor halo stars outside the Solar
Circle (a number that may be three times smaller according to a reanalysis by Chiba
\& Beers 2000).  These nine stars are distributed all over the sky with no obvious spatial
structure within the 2.5 kpc radius volume, as might be expected for a large stream passing
near the Sun.  Together, the aforementioned properties of this clump of giant stars
are enticingly consistent with Sgr leading arm in the solar neighborhood; however, 
the derived (by both Helmi et al. and Chiba \& Beers)
$L_Z$ angular momentum for this clump is apparently too large 
and the apoGalacticon for its progenitor system too small compared to expectations
for the nearly polar Sgr orbit that would have produced the M giant tidal arms
observed here.  Both inconsistencies depend to some extent on the adopted Galactic
rotation curve (mass profile) and Local Standard of Rest velocity and should be 
re-addressed with a Galactic model that self-consistently explains the Sgr debris
stream in all directions.  

The halo luminosity function is 
poorly constrained for giant stars.
Reid \& Majewski (1993; see their Figure 5) have compiled 
numerous estimates of the halo luminosity function and derive a
mean ``globular cluster" luminosity function to represent the halo.  
Adopting this function for the local halo produces a density of 
$>$K3 halo giants (assuming $M_V[\rm{K3III}] \sim 0.0$) of 
order 45 kpc$^{-3}$ --- a number that is about a factor of two higher than
Morrison's (1993) estimate (taking into account the fainter
absolute magnitude limit in her study) and so perhaps represents
an upper limit.  
To the extent that the true local halo giant density is thus described,
one might therefore conclude that if the Sgr leading arms is in the
solar neighborhood it
contributes only of order 0.4-0.8\% of the {\it local} evolved
halo stars, and would not likely have significantly impacted studies 
(e.g., Yoss et al. 1992, Morrison et al. 1993)
of nearby ``halo giants".  However, it
may well have affected more distant halo giant samples, e.g., that
by Ratnatunga \& Freeman (1989) of stars in
the SGP field SA141, a sample for which they noted
a particularly small velocity dispersion and which contains 
stars of the approximate velocity expected for the trailing Sgr
tail in this general direction. 
Assuming an effective vertical halo scaleheight 
of 3.5 kpc (Reid \& Majewski 1993) and the above local halo
giant density, 
Sgr dominates the halo K giant density by 5-6 scaleheights, 
or 17-20 kpc above the plane --- distances comparable to those
probed by the SA141 survey.  

Because of some uncertainty in the actual stellar densities of
the local halo population(s), it is difficult to assess accurately the
relative impact of Sgr debris on studies of Galactic structure.
We may, however, more directly calculate 
the actual number of Sgr stars contributing to a particular survey.
Here we focus on the magnitude limited survey of stars at
the NGP by Majewski (1992), for which a 
relatively complete radial velocity and proper motion
analysis of stars (mainly F-K dwarfs) to $V \sim 19$ is 
described in Majewski, Munn \& Hawley (1994, 1996).  To 
estimate the number of such stars that could have been
contributed by Sgr, we integrate the 15 Gyr old
(the precise age adopted has little affect on the analysis)
theoretical luminosity function for a cluster with
metal abundance $Z=4 \times 10^{-3}$ by Silvestri et al. (1998);
this luminosity function was found to give a good match to
observationally-derived RGB and main sequence luminosity functions
for the globular cluster 47 Tucanae by those authors.
Taking into account the volume completeness limits of the Majewski (1992) survey
as a function of ($M_V$), we integrate the Silvestri et al.
luminosity function from the main sequence turn off at
$M_V = 3.4$ to $M_V = 8.0$ and scale this number by
the ratio of the Sgr $>$KIII giant density above to the integral of
the luminosity function for $M_V \le 0$ (assumed to represent the 
luminosity of KIII giants).  The result
leads to an estimate that some 5-10 Sgr dwarfs should
be present in the subsample of Majewski (1992) dwarfs 
discussed by Majewski, Munn \& Hawley
(1996).  It is interesting, therefore, that the halo 
sample in that survey is constituted by three phase
space clumps with of order this number of stars each, 
and two of the clumps (and the net average of all 
halo stars in the survey) show a net negative radial
velocity, as expected for Sgr debris at the North
Galactic Pole.  A more detailed assessment of the
particular energy and momentum distribution of those
phase space clumps in the context of Sgr models
that accommodate nearby debris flow is warranted, but
clearly the several recent findings of excess numbers of
stars with a downward motion from the NGP - for example, from
the Majewski et al. survey of dwarf stars and the 
Kinman et al. (1994, 1996) studies of horizontal branch stars 
--- offer tantalizing possibilities of earlier detections
of the leading Sgr arm near the Sun.

\section{Epilogue}

The 2MASS database has been used to make the first all-sky map of the
M giant populations of the Sagittarius dwarf spheroidal galaxy system.  
The present discussion provides the first
relatively reddening-free description and analysis of the central 
regions of the dwarf, as well as extensive new information on 
the extended tail structure of tidally stripped stars.  The
latter is particularly useful for placing all previous studies of
the Sgr system into a well-defined context, and places the most
stringent constraints yet on models of the disruption of Sgr in
the Galactic potential.  
We have concentrated on an empirical description of the Sgr
system and resisted extensive interpretation via disruption model-fitting 
here because: 
(1) The degree
to which a simple {\it empirical} description of the 2MASS results nonetheless
advances our understanding of the Sgr system is manifest.  (2) Published
model fitting to previously extant data 
(especially the works by Helmi \& White 2001, Ibata et al. 2001 and 
Ibata \& Lewis 1998 to which we have frequently referred) provide a 
sufficiently accurate match to the spatial distributions described here that 
a general sense of the Sgr orbit and destruction are in hand, while further 
refinements will benefit from the addition of kinematics (see Law et al. 2003).  
(3) A survey to obtain the velocities of stars in the extended
Sgr tidal arms is underway, and first results for hundreds of M giants 
will be included in future contributions.
A Sgr system accurately characterized both
spatially and kinematically will become a powerful fiducial against which to 
delineate the structure and dynamics of the Milky Way and its halo.

The primary results from this paper may be summarized as follows:

\begin{enumerate}

\item 2MASS provides a facile means by which to explore the Sgr
dwarf galaxy and its tidal tail system, because that system contains
a prominent population of M giant stars (Figure 1) 
and such stars are readily identifiable using $JHK_s$ infrared photometry
(Figure 2; Section 2). 

\item When an M giant selection is applied to aperture photometry
from the magnitude error-limited 2MASS point source catalogue, the center
of the Sgr dwarf and both its leading and trailing tidal tails are among the 
most prominent, high latitude features observed in the sky (e.g., Figure 3; 
Section3).  

\item The central parts of the Sgr system, as traced by 2MASS M giants, exhibit
a smooth distribution resembling a dwarf spheroidal galaxy, though one
of high ellitpicity ($\epsilon > 0.6$) and large extent (Figure 4, Section 4).  
As with other dSph galaxies, radial profile fits to the center of Sgr (with the 
ends of the semi-major axis excised to minimize the contribution from the tidal 
tails) can be described by a King profile, albeit one of very large core (224 arcmin) 
and limiting (30 degrees) semi-major radii (Figure 5; Table 1; Section 4.1).  
A Power-Law + Core (PLC) fit to the radial profile is also provided, though like the
King function fit, this functional form cannot provide a good match to the
radial profile transition to the tidal tails.  However, by comparison to the 
extragalactic population of dwarf
ellipticals, the extreme ellipticity of these fits in the direction of
the tidal streams suggests the significant presence of the tidally
stripped population and suggests that these functional fits do not represent the
gravitationally bound dwarf.

\item Two departures of the observed radial profile (Figure 5) from the King (and PLC) 
fits are the presence of a central cusp (Section 4.2.1) and a break to the
tidal tails (Section 4.2.2).  The approximately half-degree radius
central cusp is coincident with the location
of the globular cluster M54, however, because that cluster is typically 
characterized as an old, metal-poor system, it cannot be contributing
M giants to the central excess of these stars above the flat part of the
King profile.  The connection between M54 and the concentration of stellar 
populations of a variety of ages at its location (Layden \& Sarajedini
2000) is still unclear, although one hypothesis (Section 4.3.3) is that
the nucleated center of Sgr may correspond to the residual bound core
of a dramatically disrupting Sgr system.  The outer break in the Sgr radial
profile to a $\sim r^{-2}$ declining population 
resembles breaks seen in the outer parts of the radial profiles of 
other dSph galaxies and which have been interpreted
as possible extratidal debris; in the case of the Sgr system this is
now established definitely to be the case.

\item The integrated brightness
of Sgr is found to be $V_o=3.63$, with the cusp adding a few more hundredths of
a magnitude of light.  If the distance modulus to Sgr is taken as 16.9, 
we find that the center of the Sgr system edges out Fornax as the brightest
of the dSph galaxies, with an absolute magnitude of $M_V = -13.27$.
These results appear to be consistent with the similarity of the
Fornax globular cluster specific frequency to that of the Sgr progenitor 
(Section 4.3.2).

\item When the observed King parameters of the 
radial profile and the Sgr central velocity dispersion are
combined in the usual King (1966) methodology we estimate
the mass of Sgr to be $5 \times 10^8$ M$_{\sun}$ and obtain a
total $M/L_V = 25$ M$_{\sun}$/L$_{\sun}$ (Section 4.3.2).  
However, since even this mass is substantially below that 
needed for a system to withstand the Galactic tidal force 
over scales of the observed King limiting radius, we argue
(Section 4.3.3) that the true mass, tidal radius and bound
fraction of the observed central Sgr system must be substantially
smaller than suggested by the King profile, and that 
Sgr is presently undergoing a major mass loss event ---
perhaps almost complete disruption --- induced by tidal shocking
from the last periGalacticon passage.  A much smaller 
radius for the bound Sgr core would help to resolve
the timing problem (``M giant conundrum") posed by the presence
of relatively recently formed stars (M giants) spread along
tidal tails of comparable age (Section 4.3.4).

\item The tidal tails of the Sgr system span both Galactic
hemispheres (Sections 6.1-6.3; Figures 3 and 8-11), 
but show relatively little evidence for precession
(Figures 6, 7 and 13).  Given 
the 13$^{\circ}$ tilt of the Sgr orbital plane, an almost
spherical halo potential is implied (Section 5.2.2).  In Section 
5.2 we provide fits to the debris (orbital) plane and define
Sgr coordinate systems based on that plane that are useful for 
interpreting the Sgr tidal system.  The trailing arm
is followed for at least $150^{\circ}$ from the Sgr center to
the Galactic anticenter, and perhaps farther, into the North
Galactic Hemisphere (Section 6.2; Figures 8-11).  For a 
large fraction of this extent, the density of the trailing debris and
its distance from the Sun are more or less constant.
The leading arm makes a Northern Loop with mean apoGalacticon 
$\sim 40$ kpc (Section 6.3; Figures 8-11) and a path
that takes it to the North Galactic Cap, from where it
arcs back down toward the Galactic disk.  

\item We find ourselves at an unusual time in Galactic history: 
For less than 2\% of the Sun's orbit around the Galaxy
are we as close to the path of the leading arm debris
as we are now.  If the leading arm is long enough to
reach the Galactic plane on this side of the Milky Way
we should expect to find Sgr stars in or near the solar
neighborhood (Sections 6.4).  The implications of this
possibility for studies of the Galactic structure near the Sun 
are discussed in Section 9; while the density of Sgr stars
would be swamped by those of other stellar populations locally, Sgr 
would dominate the halo tens of kiloparsecs above the disk.
Several previous surveys of halo stars might contain Sgr representation.
Analysis of all M giants in our sample indeed reveal Sgr
to be the prominent contributor of such stars to the high 
halo (Section 5.1, 7.2; Figures 6, 16).  A similar conclusion
holds for carbon stars (Section 8.3; Figure 20).  No evidence
for extended M giant tidal tails from the Magellanic
Clouds are seen (Section 5.3). 

\item The relatively constant density of the Sgr trailing
arm over a great part of its extent (Figures 12, 13) implies a relatively
constant mass loss rate over the last several Sgr orbits,
excluding the possible last major disruption event (Section 6.5). 
Some evidence for stellar population variations along the
arms is suggested by the changing color of M giants with
position (Section 6.6).  It is likely that older tracer stars will
map the arms to even greater length than is possible with
M giants.  The number of stars in the tidal arms is at least
15\% that within the King limiting radius (Section 7.1).
 
\item Good correspondence is found between the location
and distances of M giant tidal debris and nearly all previous
identifications of Sgr debris (Section 8; Figures 17, 18).
However, Sgr carbon stars are found to be subluminous
compared to carbon stars in other Galactic satellites (Section 8.3;
Figure 19), requiring adjustment of the previous Sgr 
carbon star distance scale.  Even so, the carbon stars
provide a much less clear picture of the Sgr system than
is offered by the 2MASS M giants (Figure 20).

\end{enumerate}

The results presented in this publication make use of data from 
the Two Micron All Sky Survey (2MASS), which is a joint project of the 
University of Massachusetts and the Infrared Processing and Analysis 
Center (IPAC), Funded by the National Aeronautics and Space Administration 
and the National Science Foundation.  The 2MASS database owes its existence 
to the dedicated work of 2MASS scientists and IPAC staff in producing data 
products of unparalleled photometric quality and uniformity.  
SRM acknowledges support from a 
Space Interferometry Mission Key Project grant, NASA/JPL contract 1228235, 
a David and Lucile Packard Foundation Fellowship, and a Cottrell Scholar
Award from The Research Corporation.  
MFS acknowledges support from NASA/JPL contract 1234021.  MDW was supported 
in part by NSF grant AST-9988146.  This work was also partially supported
by the Celerity Foundation.
SRM appreciates useful conversations with William Kunkel, Kathryn Johnston,
David Law, Neill Reid and Walter Dehnen.  
We thank Heidi Newberg for providing and giving permission 
to use the Sloan Digital Sky Survey equatorial distribution image, and
for helpful comments as referee of this paper.
Richard 
Patterson, Jeffrey Crane, Megan Kohring, Howard Powell and Kiri Xiluri
are thanked for assistance with various figures.

\begin{figure}
\plotfiddle{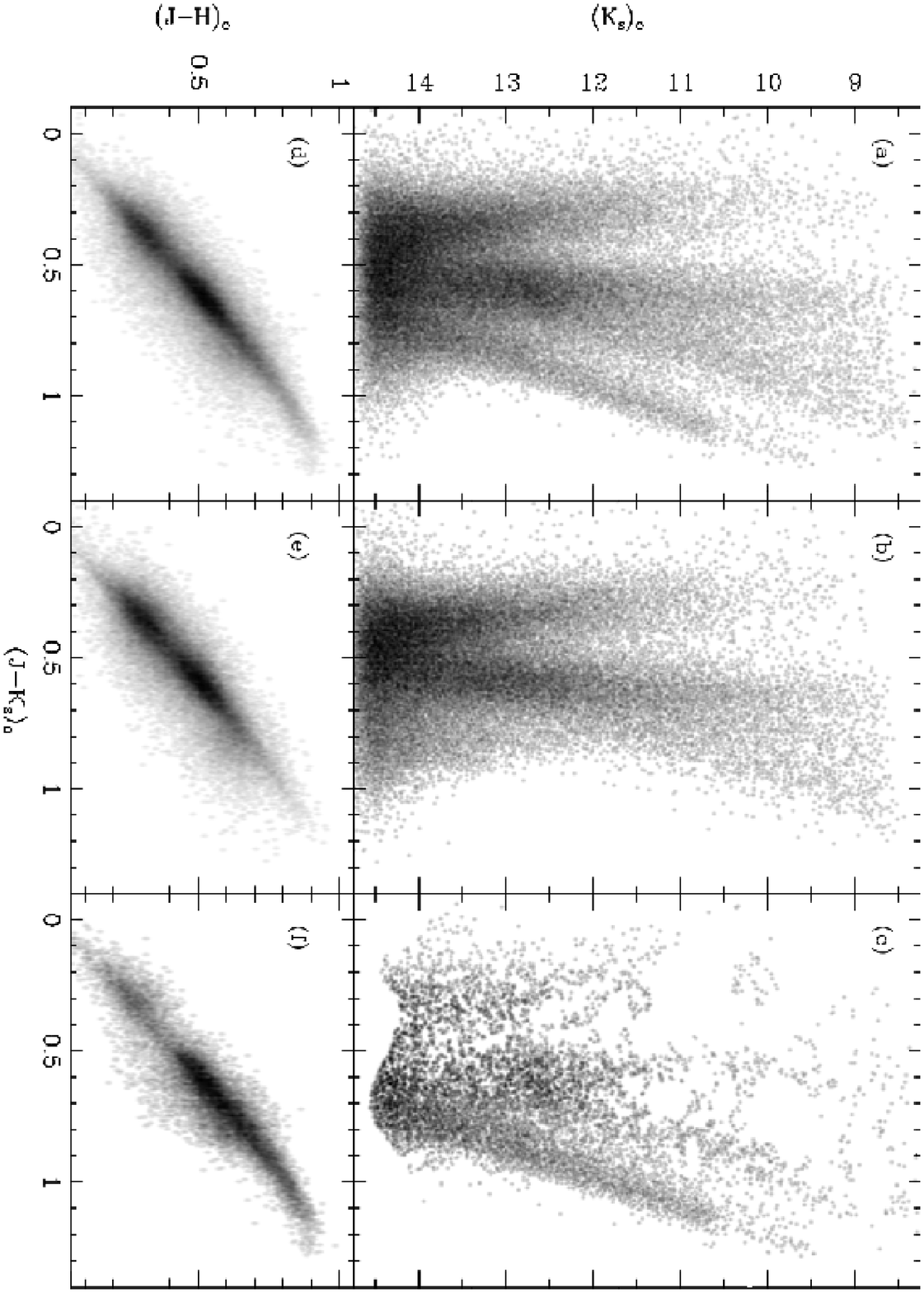}{8in}{0}{70}{70}{-220}{40}
\caption{Near-infrared $(J-K_s,K_s)$ color-magnitude diagrams of (a) the
Sgr center, (b) a control field of identical area, and Galactic coordinates
reflected about $l=0^{\circ}$, and (c) a star by star subtraction of (b) from (a).
Panels (d)-(f) show the corresponding $(J-K_s,J-H)$ two-color diagrams
for the samples shown in (a)-(c).  All sources are
dereddened using the Schlegel et al. (1998) maps.
}
\end{figure}

\begin{figure}
\plotfiddle{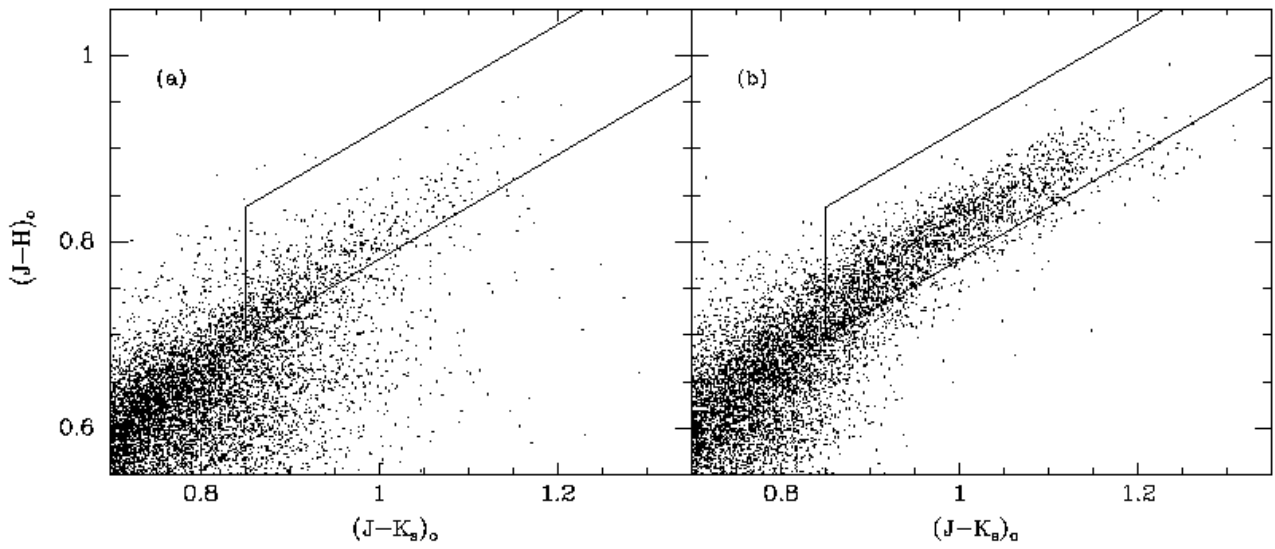}{4in}{0}{60}{60}{-200}{-80}
\caption{ The solid lines indicate the
color-color selection criteria adopted to find M giants for most
of this paper.  
Panel (a) shows the distribution
of stars in the control field, from Figure 1e, and panel (b) shows
the distribution of stars from the statistically subtracted sample
in Figure 1f.  Note that the control field, selected to be a Galactic
longitude match to the Sgr center field, 
still contains about a 1\% contribution
from the Sgr dwarf itself.
}
\end{figure}

\begin{figure}
\plotfiddle{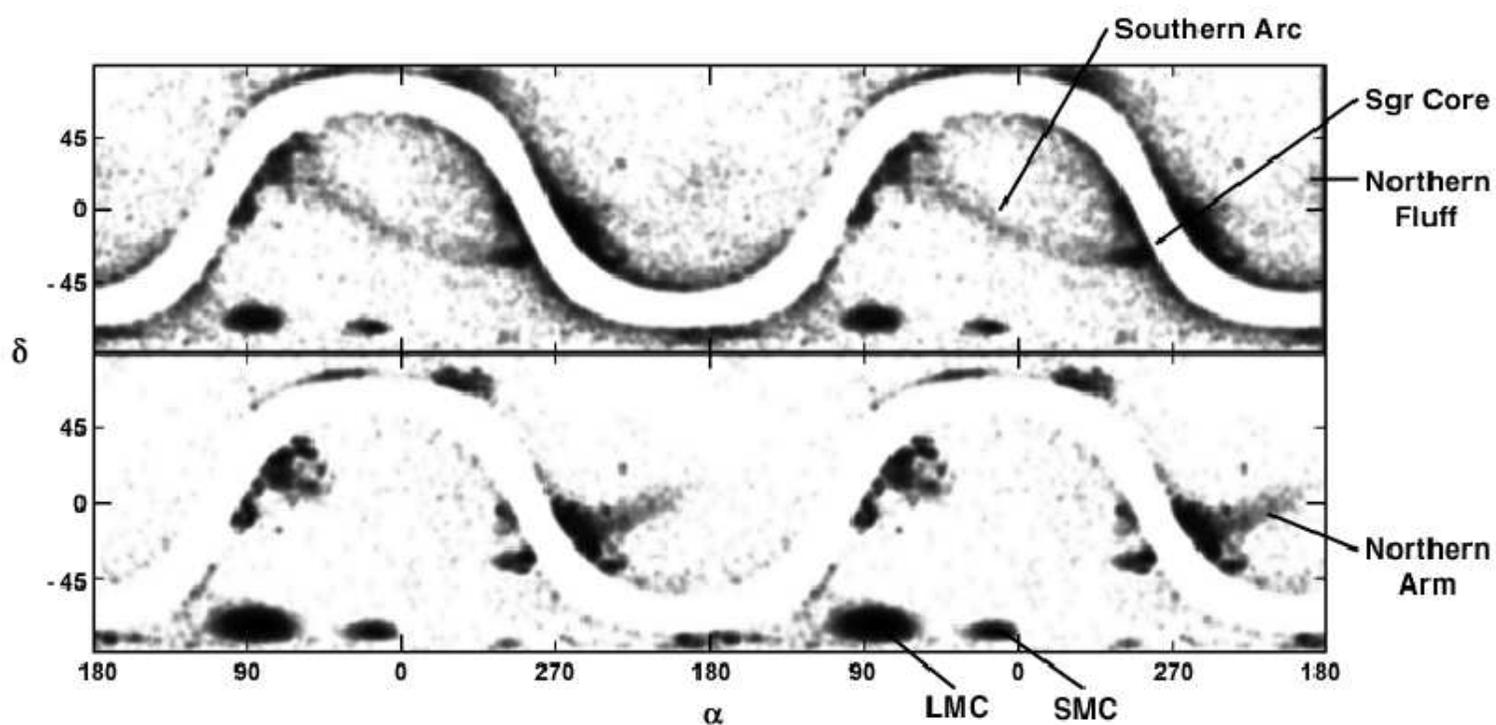}{5.5in}{270}{70}{70}{-270}{390}
\caption{Smoothed maps of the sky in equatorial coordinates for two color-magnitude
windows of the (non-dereddened) 2MASS point source catalogue filtered optimally 
to show (top) the ``Southern Arc" and (bottom) the ``Northern Arm": 
(top) $11 \le K_s \le 12$ and 
$1.00 < J-K_s < 1.05$, and (bottom) $12 \le K_s \le 13$ and 
$1.05 < J-K_s < 1.15$.  
We show two cycles 
around the sky to demonstrate the continuity of features.
}
\end{figure}

\begin{figure}
\plotfiddle{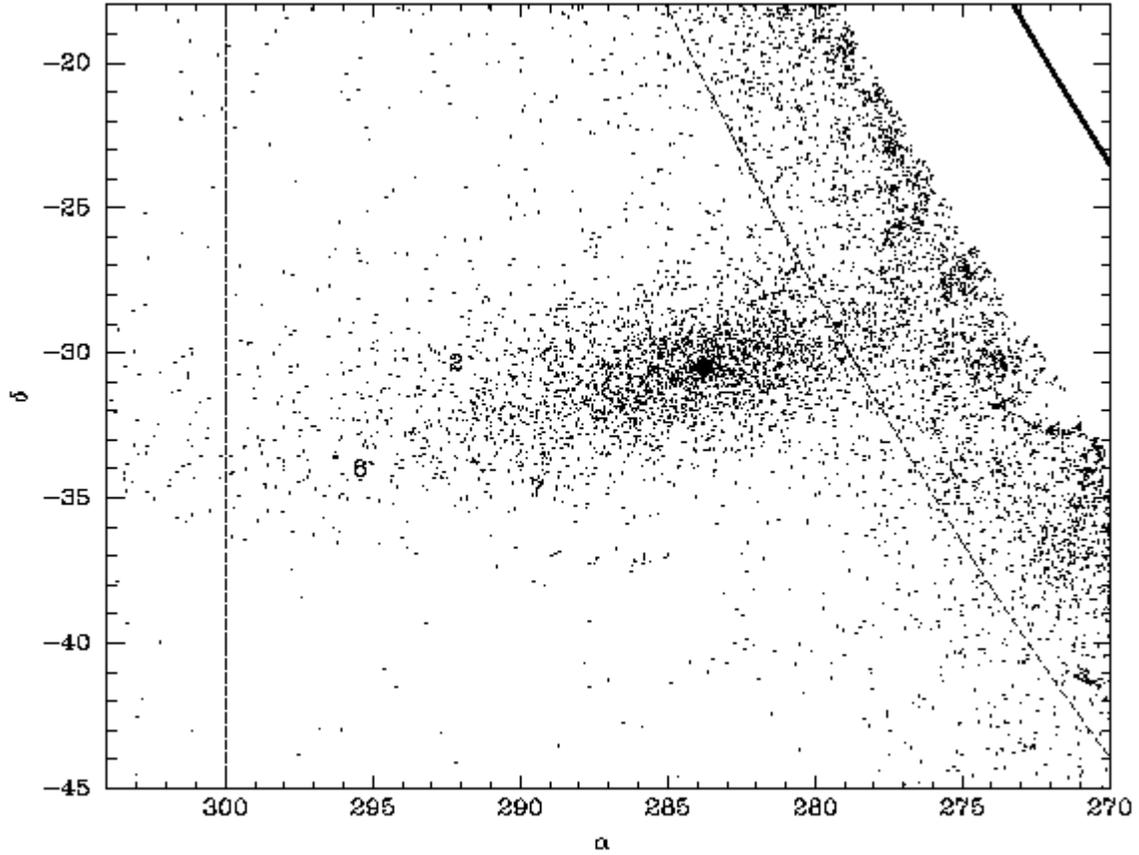}{5.5in}{90}{100}{100}{395}{-100}
\caption{View of the central parts of Sgr near where it crosses the Galactic mid-plane
(shown as the heavy line to the upper right).
Sources up to $b = -5^{\circ}$ are shown; the results of very patchy 
reddening can be seen for $b > -10^{\circ}$ (demarcated 
by the angled dashed line).  The symbols mark the locations
of globular clusters as follows: ``2" is Arp 2, ``7" is Terzan 7,
``8" is Terzan 8, and the {\it filled circle} is the location of
the cluster M 54.  The region to the left of the angled dashed line was
used in the King profile fit and the Power Law + Core fits to the
central region shown in Figures 5a and 5d, respectively.  The region delimited
by both dashed lines was used
in the King profile fits to the Sgr center shown in Figure 5b 
and given in Table 1. }
\end{figure} 
 
\begin{figure}
\plotfiddle{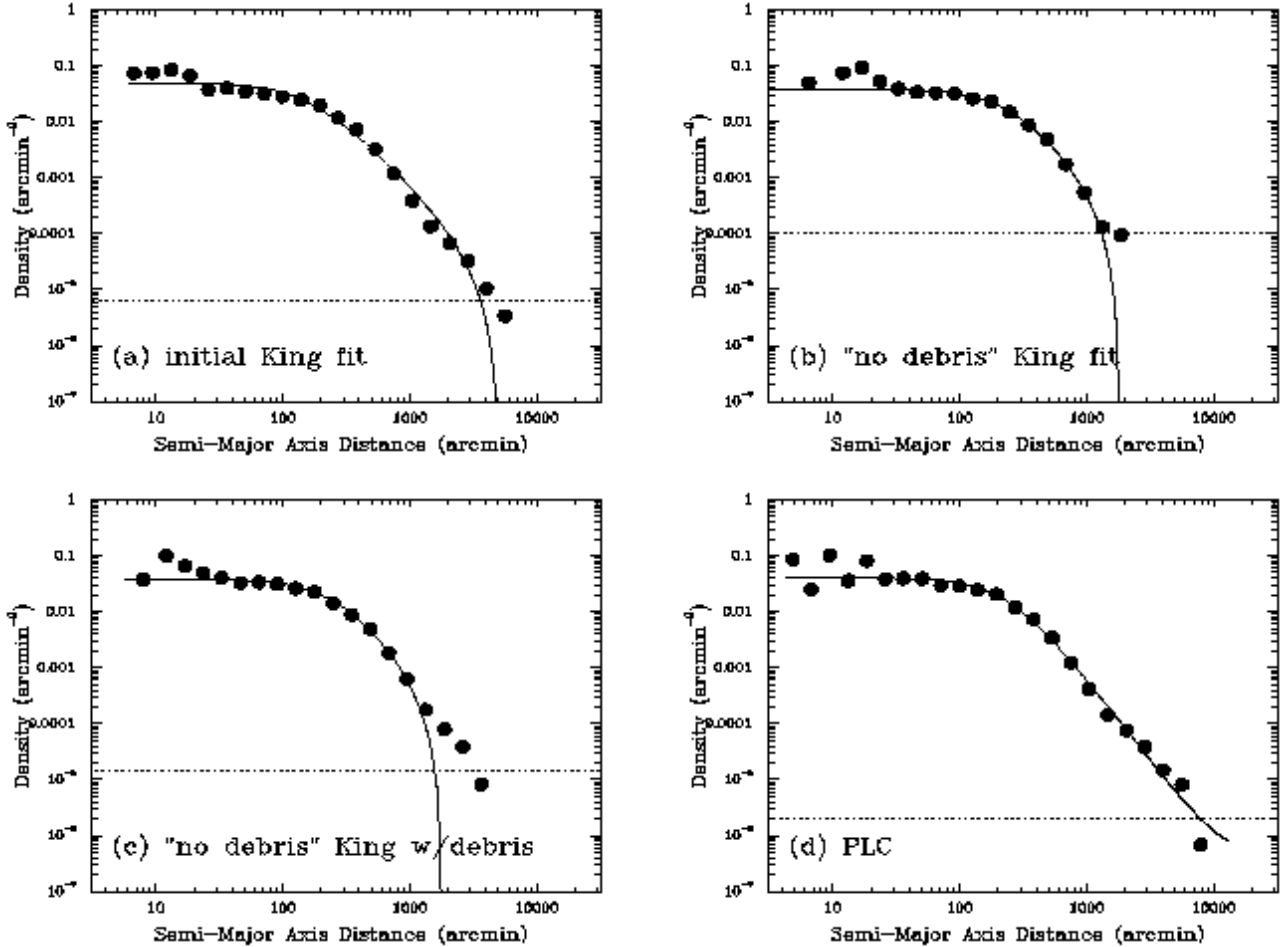}{5.5in}{90}{100}{100}{362}{-90}
\caption{Model fits to the radial profile of the Sgr main body. (a)  
Fit to the entire area shown in Figure 4, but with $b<-10^{\circ}$.  (b)  Fit
to data with an additional restriction to Figure 4 of 
$\alpha_{2000} < 300^{\circ}$ to minimize the influence of unbound stars 
forming the start of the trailing tidal arm along the major axis.  
This fit is given in Table 1.
(c) The derived fit from panel
(b), but with the full data set from panel (a).  All parameters from the panel (b)
fit are utilized, except the background level, which has been refit 
because of variations in the background
level when different Galactic latitude ranges are considered. 
(d) Power Law + Core fit to the same data as used in panel (a).  In all cases, 
the dotted lines are the derived level of the background, which has been
subtracted off the data and the fit curves.  Note that the data points
in each panel change positions due to rebinning that reflects different 
ellipticities and position angles derived from the fits.}
\end{figure} 

\clearpage

\begin{figure}
\plotfiddle{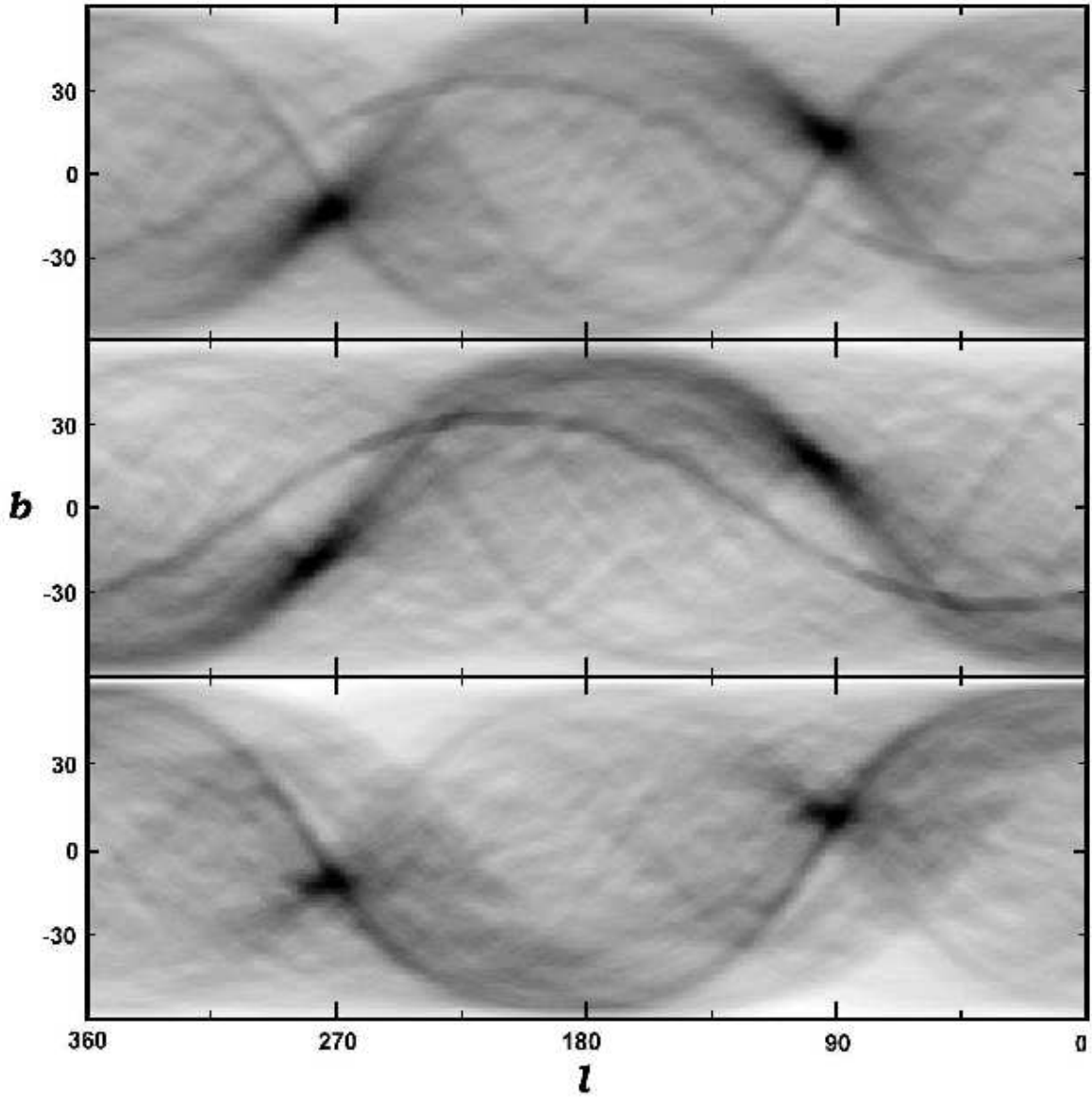}{6.1in}{0}{95}{95}{-350}{-125}
%
%
\caption{Great circle cell counts for M giant candidates in the
projected distance range 13-65 kpc and $|b|>30^{\circ}$.  The plots
are in sky-right, Galactic coordinates, from $360^{\circ}>l>0^{\circ}$ and 
$-60^{\circ} < b < 60^{\circ}$.  The top panel shows the results for both
hemispheres together, the middle panel is for inclusion 
of only Northern Hemisphere data, and the lower panel is for inclusion
of only Southern Hemisphere data.  From all panels we have removed
the Magellenic Clouds from the sample to remove the rather strong great 
circle pole families they contribute.  The darkest patches correspond
to the pole of the Sgr tidal debris stream at approximately 
$(l,b)=(272,-13)^{\circ}$ and its corresponding antipode.
No other strong peaks occur in this particular stellar sample of
M giants.  Arc-like features in the GCCC distributions result from 
various localized density peaks in the sky distribution.   }
\end{figure} 
 
\clearpage

\begin{figure}
\plotfiddle{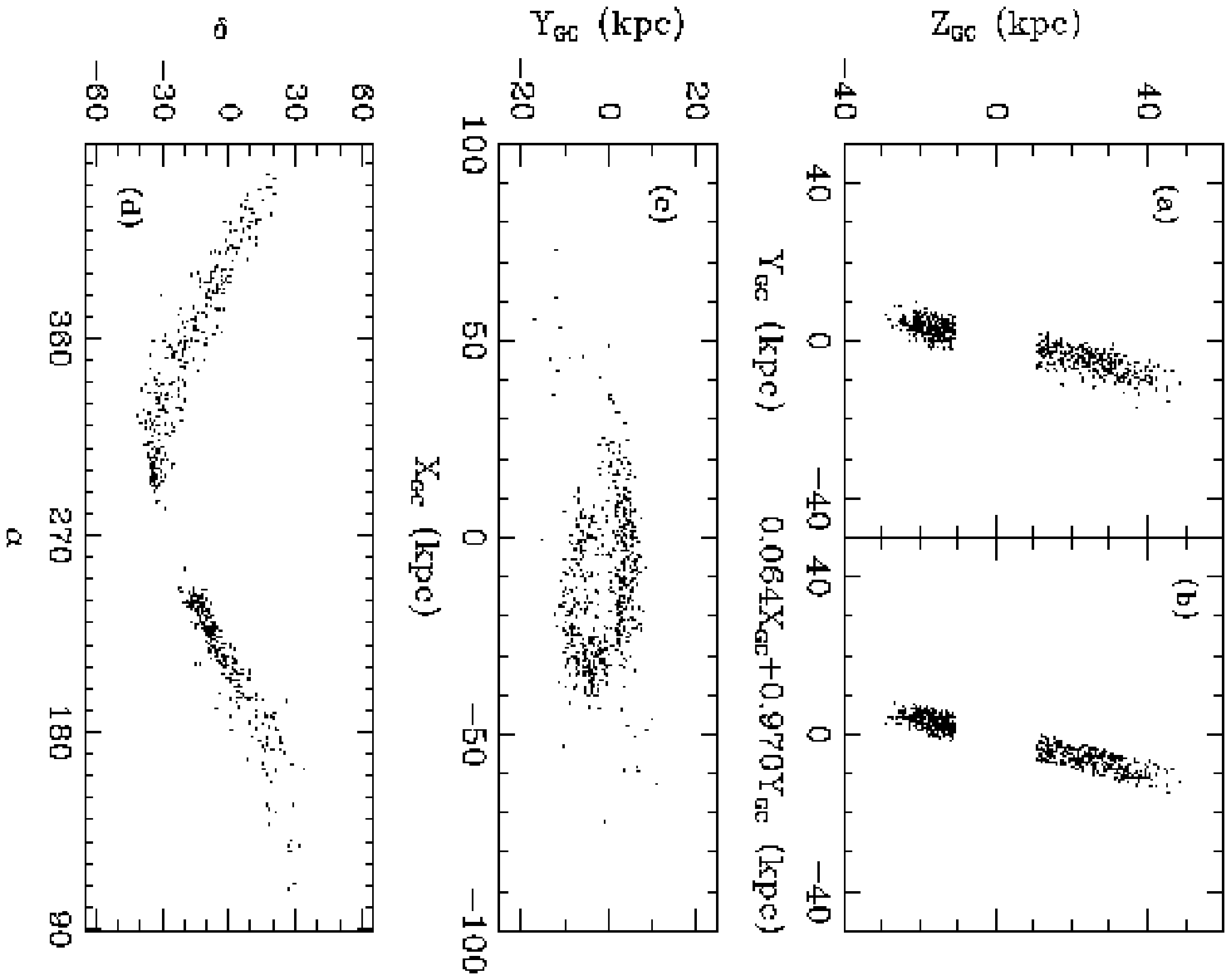}{4.7in}{90}{80}{80}{340}{-90}
\plotfiddle{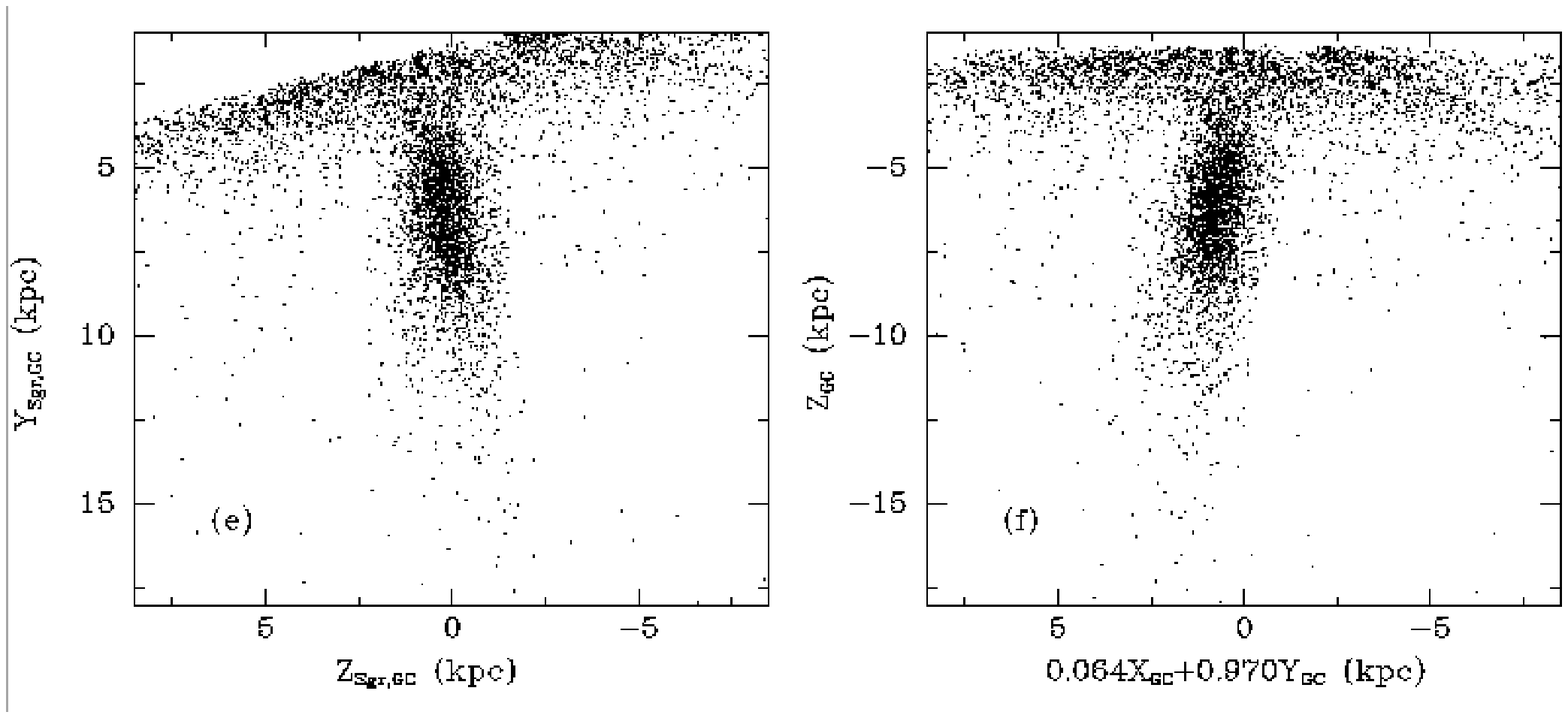}{1.7in}{0}{60}{60}{-275}{-215}
\caption{In panels (a)-(d) the points used to define the Sgr orbital 
plane are used to show various projection effects. 
(a) The Galactic $Y_{GC}-Z_{GC}$ plane.  
(b) The plane shown in panel (a) but rotated by 3.8$^{\circ}$.  This
projected plane is perpendicular to the derived best fitting plane. 
By definition, the width of the material is narrower in 
Figure 7b compared to Figure 7a.  
(c) A projection parallel to the Galactic plane.  (d) The projection
on the sky in celestial coordinates, showing the foreshortening 
effects of varying proximity to the Sun.  
In panels (e) and (f) edge-on views of the best-fitting Sgr plane
for all stars with $(J-K_s)_o>1.0$ and $E(B-V)<0.555$ are shown, restricted to stars on the
the far side of the Galactic Center ($R_{GC}\cos(\Lambda_{GC}+21.60^{\circ})>7$ kpc)
to highlight the Sgr center.  Both figures are edge-on to both the
Galactic plane and to the Sgr plane, but the coordinate system in (e) has
the best-fitting Sgr plane ($Z_{Sgr,GC} = 0$) vertical, while panel (f) is rotated 
so that the Galactic plane is ($Z_{GC}=0$) is horizontal.  Canting of the 
Sgr main body with respect to both these planes is evident. }
\end{figure}

\begin{figure}
\plotfiddle{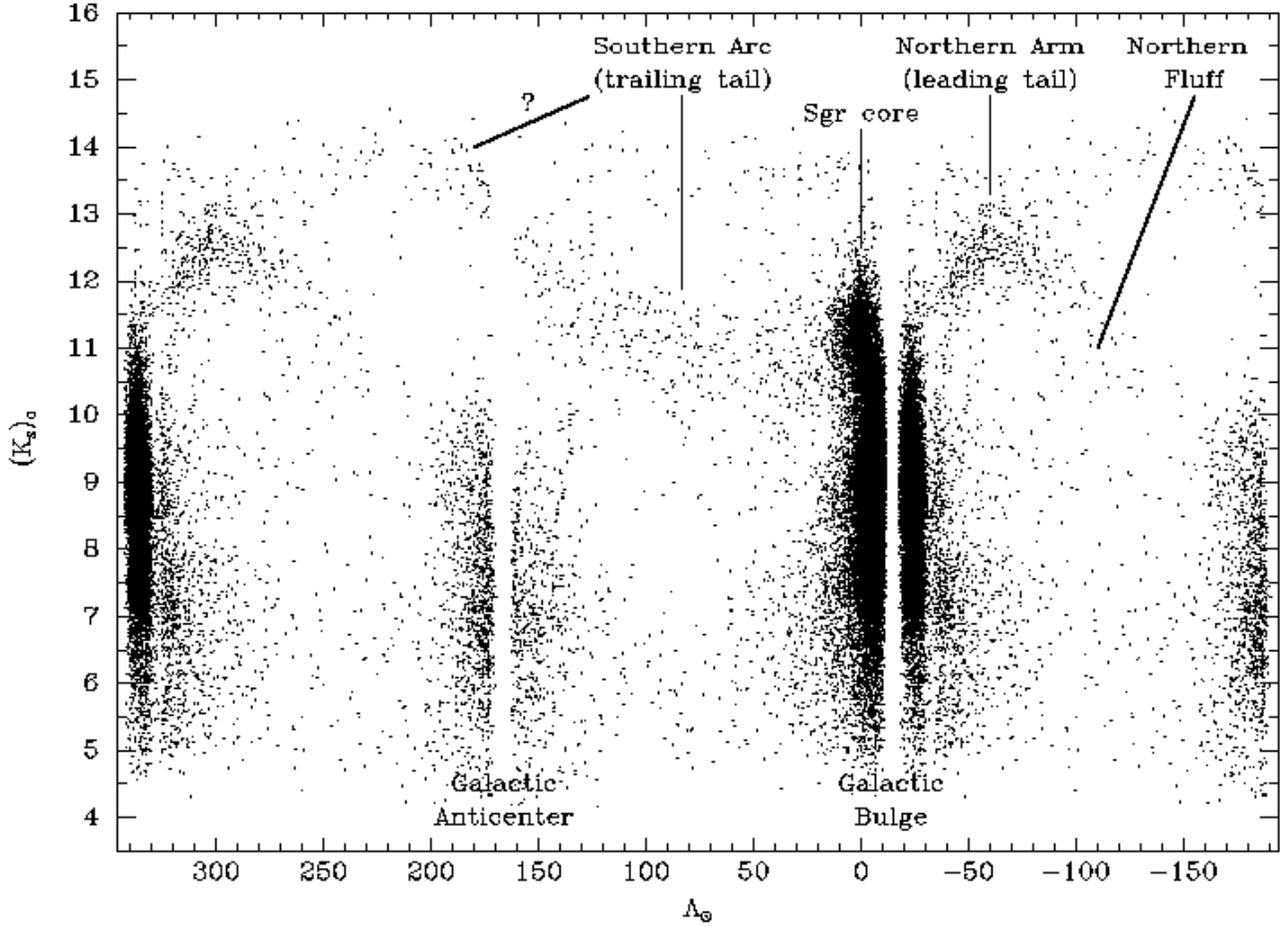}{5.5in}{0}{90}{90}{-380}{-180}
\caption{Dereddened $K_s$-band magnitudes for M giant candidates with  
$ (J-K_s)_o > 1.00$ shown as a function of Sgr longitude, $\Lambda_{\sun}$, 
along the great circle in the sky defined by the Sgr debris (Sgr orbit).  
Only candidates
within Sgr latitude range $-10^{\circ} < B_{\sun} < +10^{\circ}$ are shown. 
For clarity, we remove sources with $E(B-V)>0.555$.  The center of Sgr is 
at ($\Lambda_{\sun}, [K_s]_o) = (0^{\circ}, 11.25$ mag).  Other features 
and possible features of the Sgr debris stream are indicated. }
\end{figure}

\begin{figure}
\plotfiddle{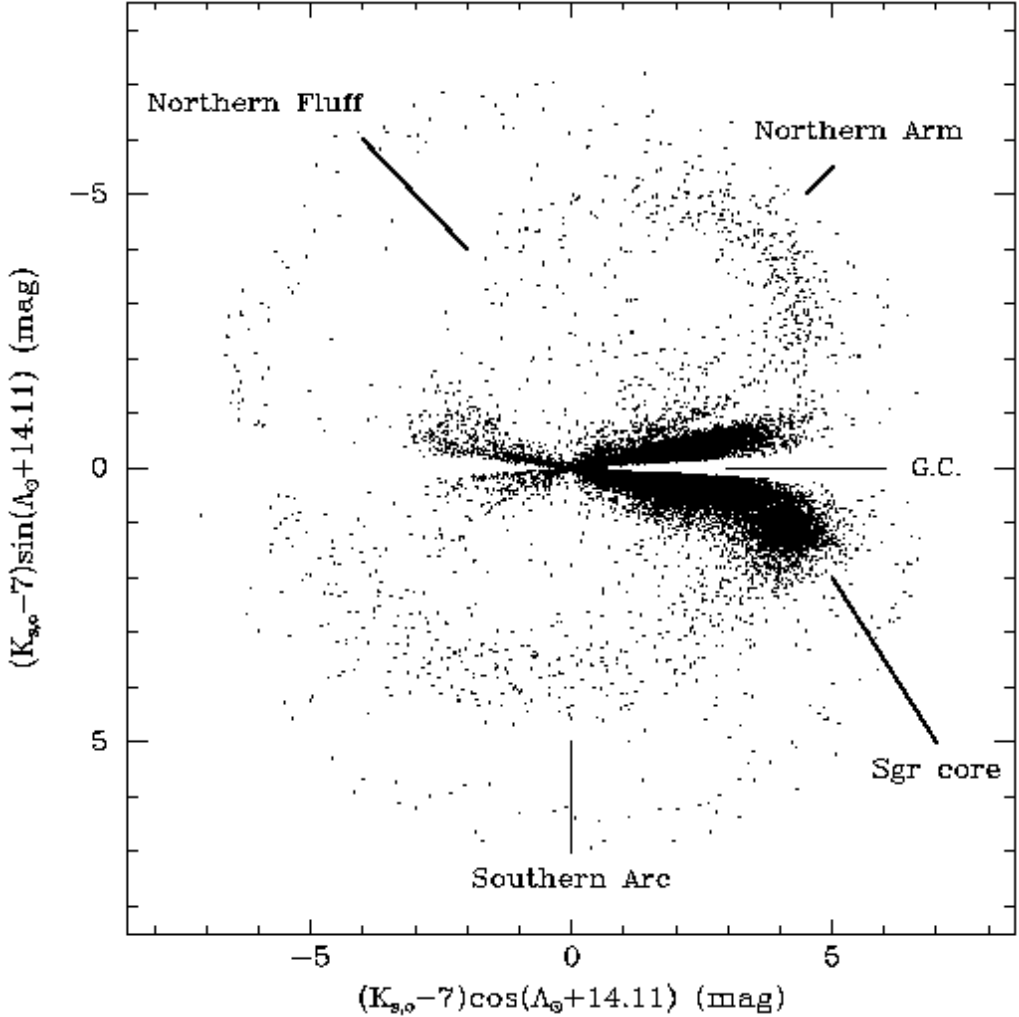}{6.5in}{0}{100}{100}{-340}{-170}
\caption{Same as Figure 8, but shown in a cross-sectional plot of the
Sgr orbital plane (that is, the approximation of that plane given by the
$[\Lambda_{\sun}, B_{\sun}]$ coordinate system), 
where $(K_s)_o$ magnitudes of M giant candidates are shown radially (after
subtraction of 7 mag).   Stars with $(K_s)_o < 7$ have been left out
of the figure.  The term $(\Lambda_{\sun} + 14.11^{\circ})$ places the 
Galactic plane horizontal across the center of the figure.  The
direction of $\Lambda_{\sun} = 0$ is towards the Sgr center (to the right
and below the Galactic plane) and $\Lambda_{\sun}$ increases counterclockwise.  
This figure shows the same sample of stars as given in Figure 8.
}
\end{figure}

\begin{figure}
\plotfiddle{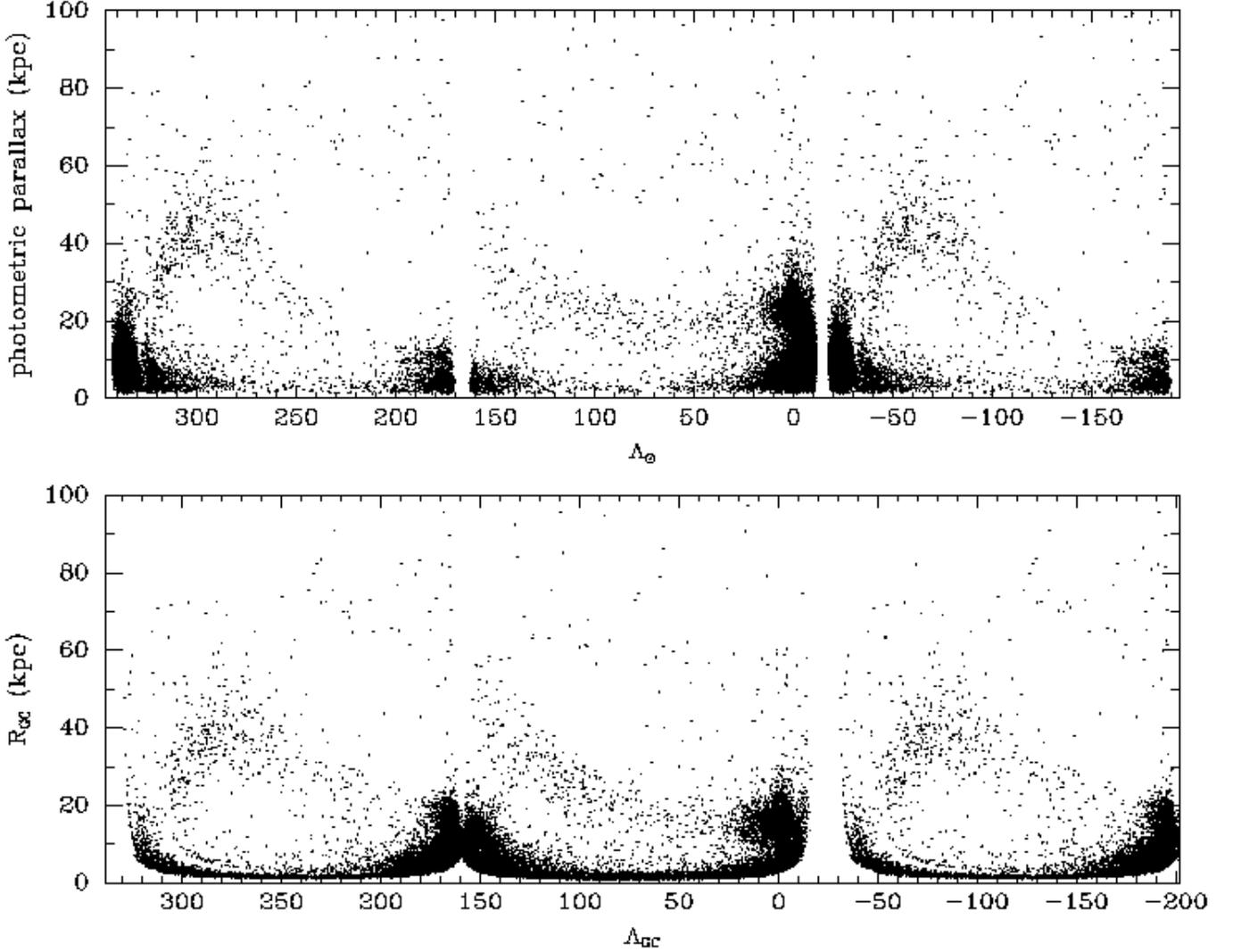}{5.5in}{270}{70}{70}{-280}{420}
\caption{({\it Top}) Same as Figure 8, but for photometric parallaxes (in kpc) 
after assigning each M giant candidate an absolute magnitude according to
its $J-K_s$ color.  
Stars within Sgr latitude range $-10^{\circ} < B_{\sun} < +10^{\circ}$ are shown. 
({\it Bottom}) The perspective from the Galactic Center point of view.  
After calculation of photometric parallaxes, distances from the 
center of the best fit plane (Equation 8) are calculated.  
For this panel, a stellar sample with $(J-K_s)_o > 1.00 $
and $E(B-V)<0.555$ is adopted, as in the top panel, but 
stars with  $-10^{\circ} < B_{GC} < +10^{\circ}$ are shown.  To remove additional
contamination at large distances (where the adopted $B_{GC}$ latitude range 
translates to a broad spatial range) we impose the additional constraint
that stars lie within 7 kpc of the best fit Sgr plane. }
\end{figure}

\begin{figure}
\plotfiddle{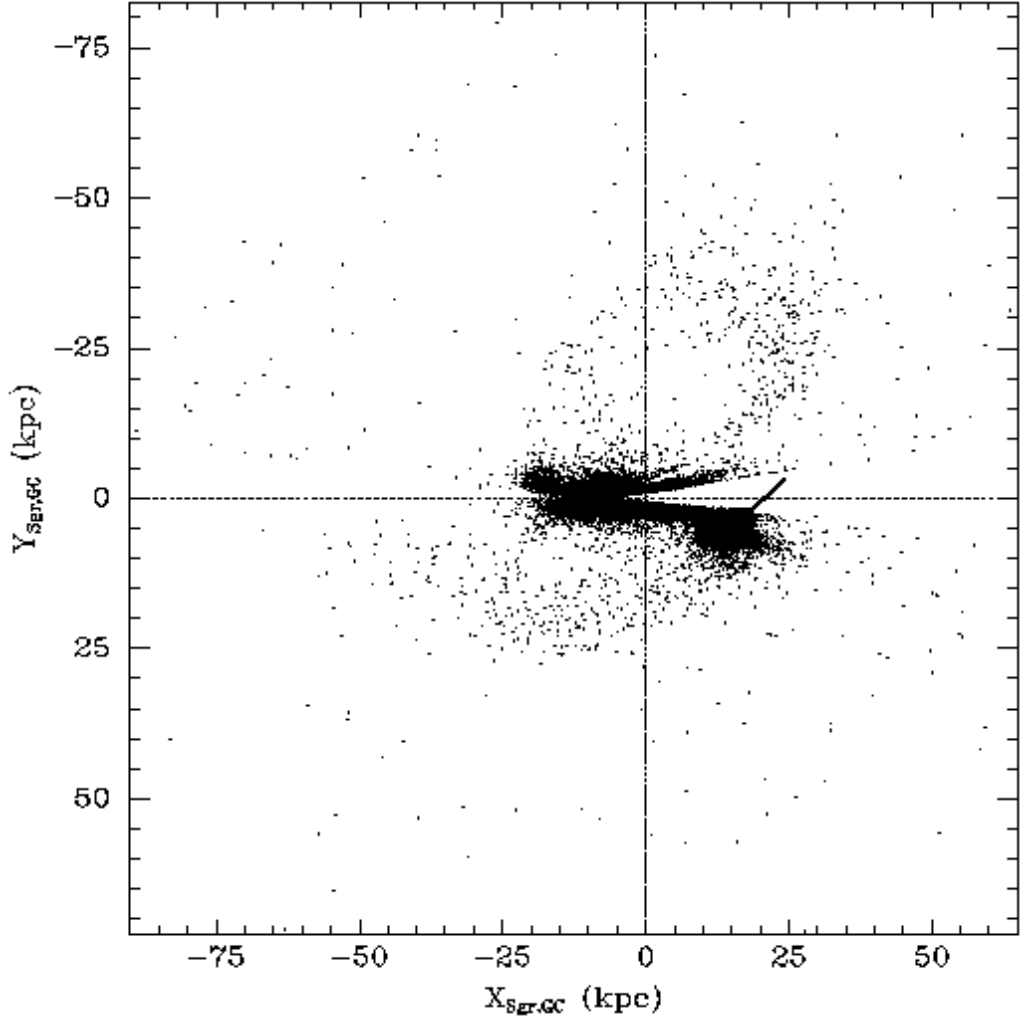}{6.0in}{270}{75}{75}{-230}{430}
\caption{Similar to Figure 9, but the radial dimension now shows distances
from the Galactic Center derived from the photometric parallaxes, 
and the plane shown is the best-fit plane from Section 5.2 (the plane shown is
slightly tilted from a traditional $[X_{GC}, Z_{GC}]$ projection -- see Table 2).  
The center of the coordinate system is actually given by
$(X_{GC}, Y_{GC}, Z_{GC}) = (-8.51, -0.21, -0.05)$ kpc, and the Sun lies
near $(X_{Sgr,GC}, Y_{Sgr,GC}) = (-8.5, 0)$ kpc (see Section 5.2).  
The stellar sample is the same as that shown in the lower panel of Figure 10. 
The nominal 
direction of motion of the main body of Sgr is shown by the angled line projecting
from the Sgr center.
The Sgr proper motion and radial velocity are from Ibata et al. (1997).
The continuity of the Northern Arm and
Southern Arc, and their association with the Sgr center, is evident in this
projection, despite obscuration by the Galactic disk.  The 
depth of Sgr features in this plot are artificially broadened 
by $\sigma_d/d \sim 0.2$ uncertainties along the line of sight from
the Sun (see Section 5.2).} 
\end{figure}

\begin{figure}
\plotfiddle{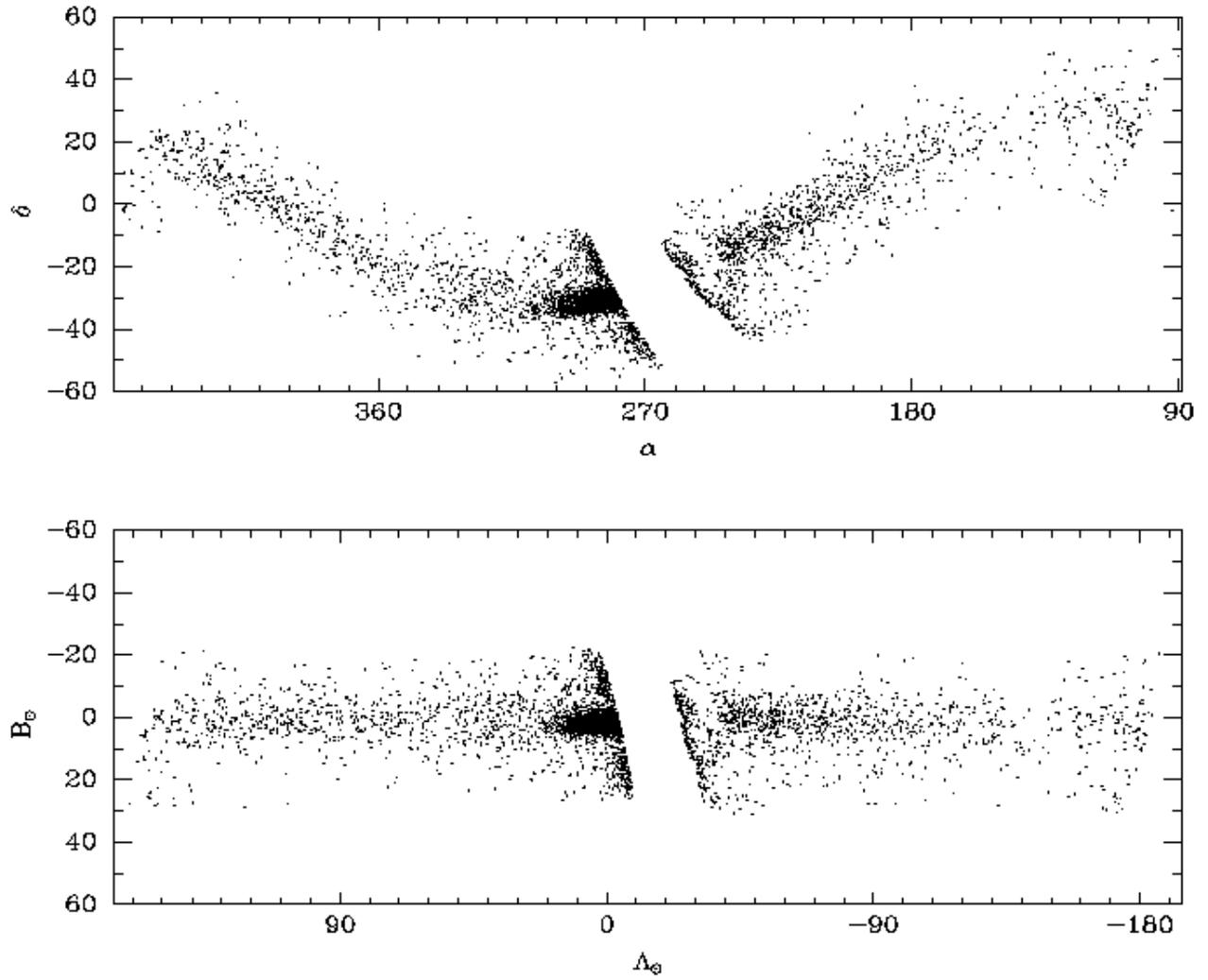}{5.5in}{270}{70}{70}{-260}{420}
\caption{The $-7 < Z_{Sgr,GC} < 7$ kpc sample explored in Figure 13 shown in
equatorial and Sgr coordinates.  
}
\end{figure} 

\begin{figure}
\plotfiddle{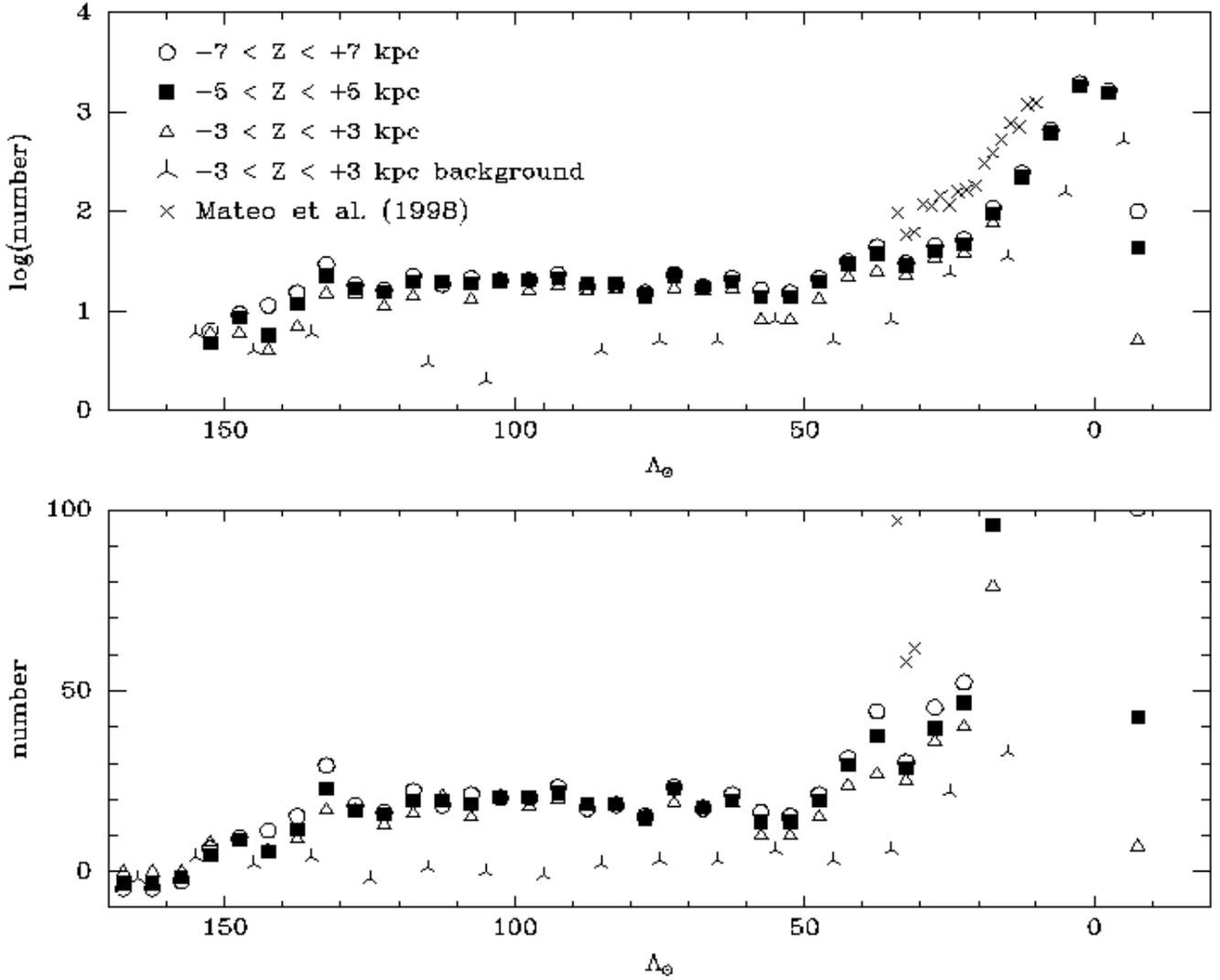}{5.5in}{270}{70}{70}{-280}{450}
\caption{Background-subtracted counts (per $5^{\circ}$ of longitude)
of $0.95 \le (J-K_s)_o \le 1.10$, trailing tail
M giants as a function of longitude $\Lambda_{\sun}$.
The {\it open circles}, {\it filled squares}, and {\it open triangles} show
counts for different allowed 
ranges of distance, $Z=Z_{Sgr,GC}$, from the best-fitting Sgr mid-plane,
whereas the {\it three pointed star} shows counts in a $Z = \pm3$ kpc 
range of distance from a ``background" plane (see Section 6.5).
To improve statistics, the background points are shown for 
$10^{\circ}$ longitude bins, rather than the $5^{\circ}$ bins
shown for the Sgr tail data.  }
\end{figure} 
 
\begin{figure}
\plotfiddle{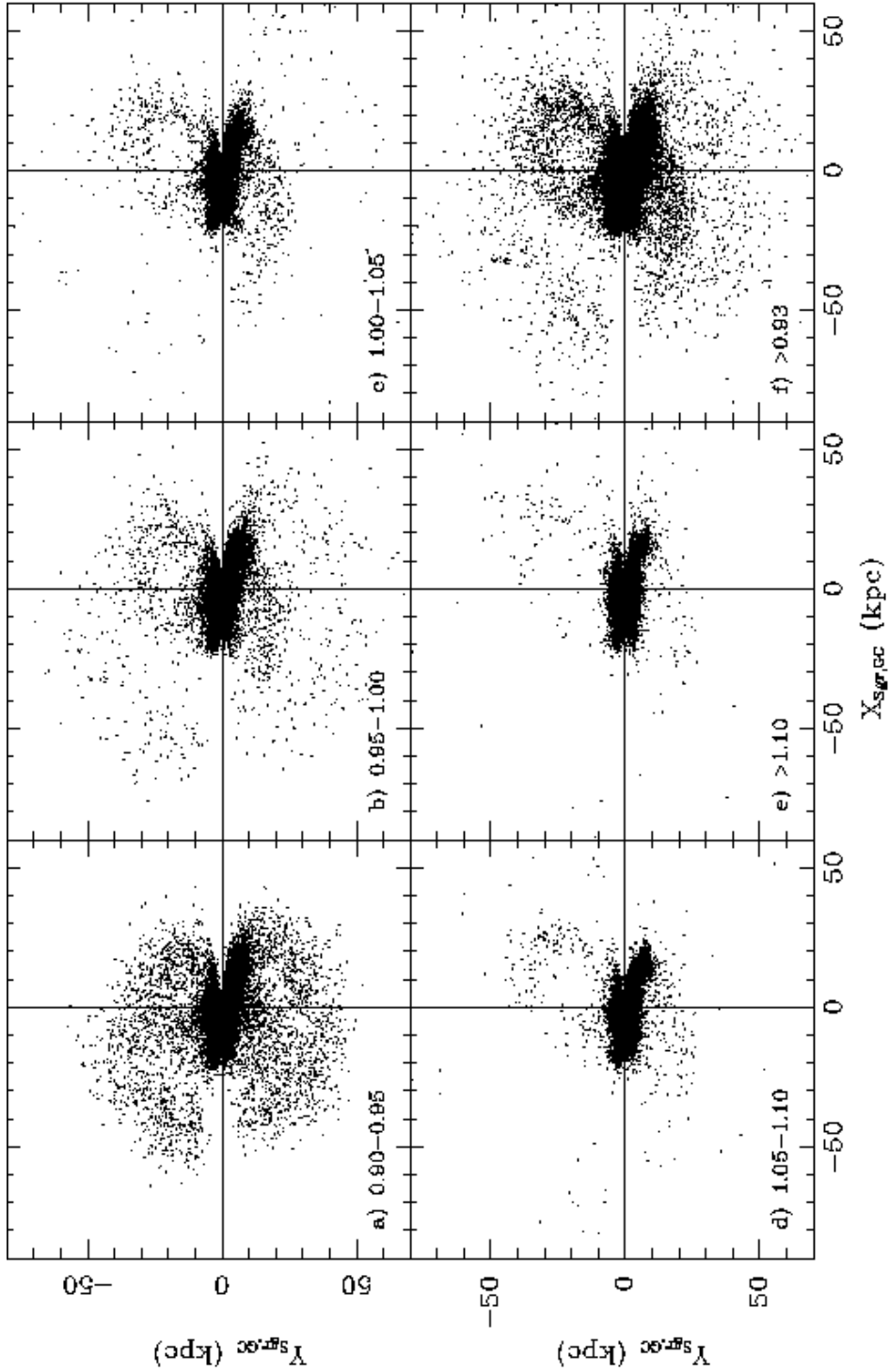}{7.5in}{0}{70}{70}{-210}{0}
\caption{ The $-7 < Z_{Sgr,GC} < 7$ kpc late type giant candidate sample shown by 
various $(J-K_s)_o$ color bins.  All stars with $E(B-V) \ge 0.555$
have been removed from the sample. The solid lines mark the approximate
location of the Galactic Center.  The apparent change in distances of
Sgr debris features with $J-K_s$ color may reflect a change in the
proportions of different age/metallicity populations among the M giants along the
tidal arms compared to those in the Sgr center that were used to define the color-absolute magnitude 
relation for photometric parallaxes.  }
\end{figure} 

\begin{figure}
\plotfiddle{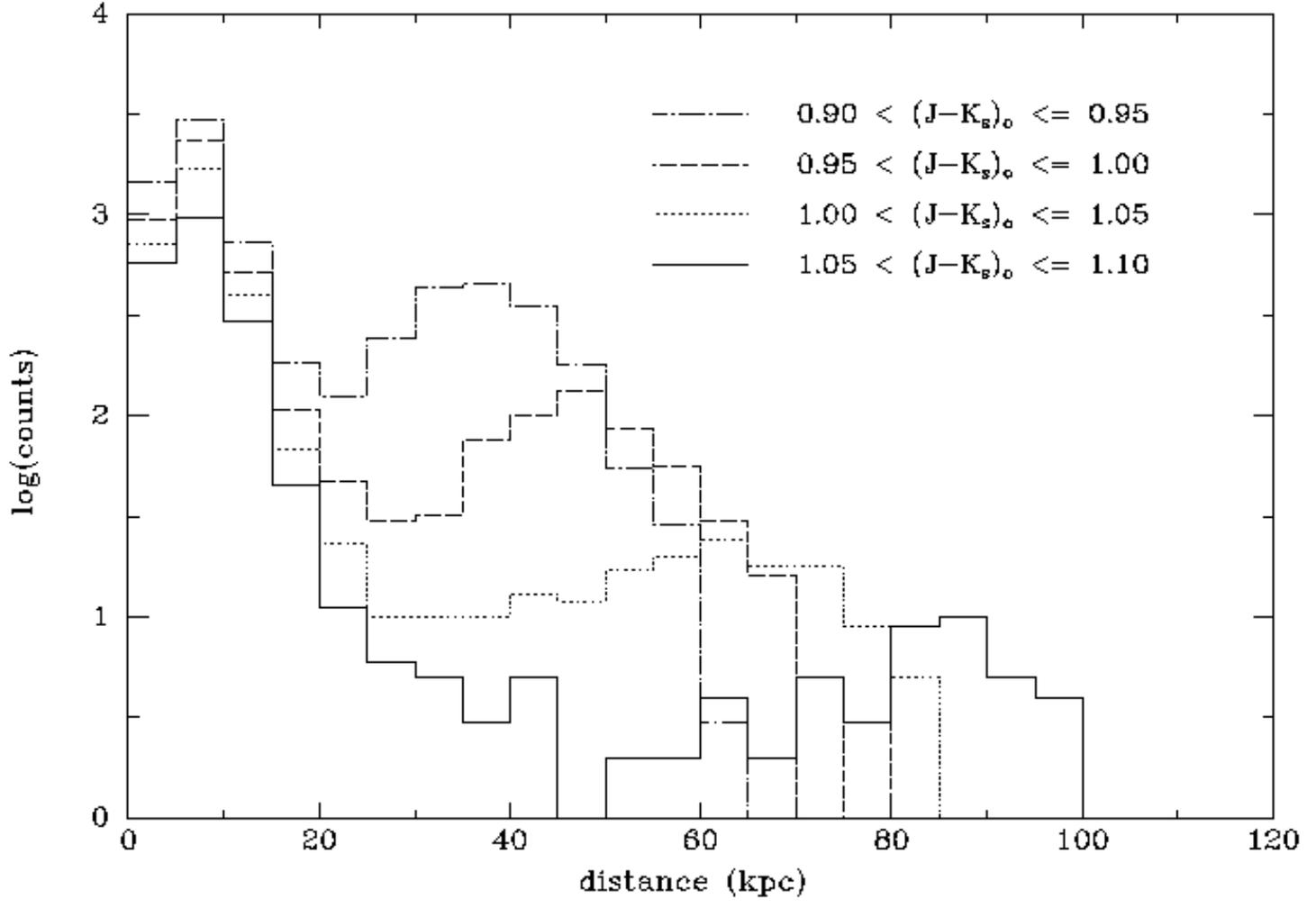}{7.5in}{270}{80}{80}{-340}{520}
\caption{ Starcounts for various $(J-K_s)$ color bins as a function of radius in a wedge
($Y_{GC}>0, Z_{GC}>0, Z_{GC}>Y_{GC}$) more or less free of stars from Sgr and the Magellanic Clouds. 
All stars with $E(B-V) \ge 0.555$ have been removed from the sample. }
\end{figure} 

\begin{figure}
\plotfiddle{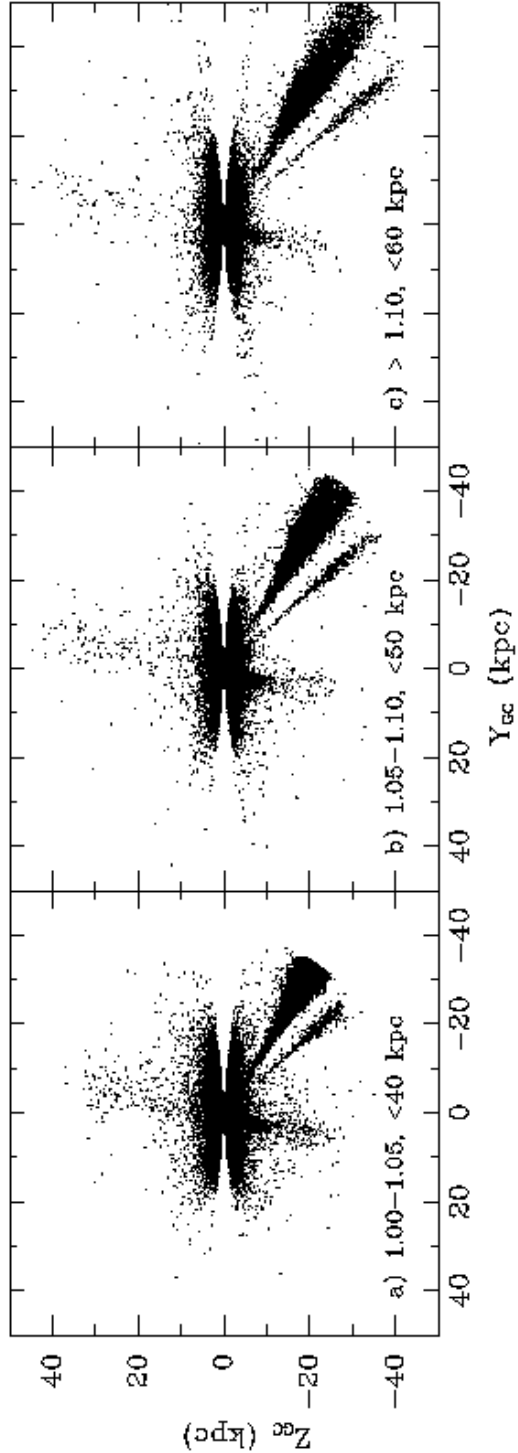}{7.5in}{0}{70}{70}{-210}{-10}
\caption{Views of the Milky Way distribution of 2MASS late type giant
candidates in projection on the Galactic $YZ$ coordinate system.
The panels show the distribution by various
$(J-K_s)_o$ color bins.  All stars with $E(B-V) \ge 0.555$
have been removed from the sample.  In order to remove the 
noise of contamination at the magnitude limits of the survey,
samples have been pruned of stars with photometric parallaxes more
than 40 kpc, 50 kpc, and 60 kpc in the $(J-K_s)_o$ samples
shown in panels a), b) and c), respectively.  Note that the
top of the Northern Loop is slightly truncated by these distance
limits. 
The ``bowing" of the 
Sgr plane is due to the Galactocentric parallax effect described
in Section 5.2.
}
\end{figure} 


\begin{figure}
\plotfiddle{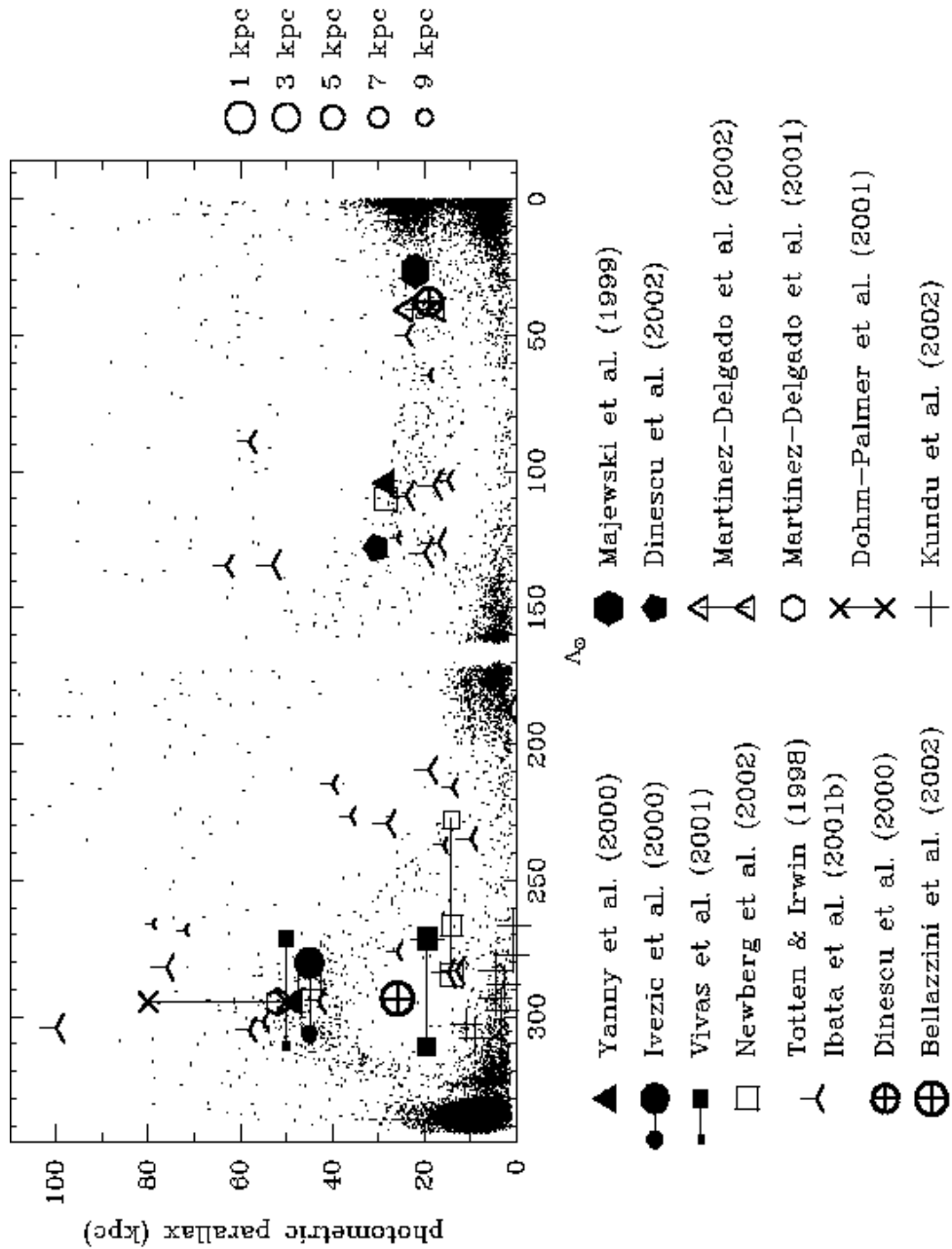}{5.75in}{0}{68}{68}{-200}{-20}
\vskip -0.25in
\caption{Summary of previous claims or suggestions of Sgr debris
detections.  Only detections near or outside the King limiting radius are shown.
Filled symbols are used for detections based on horizontal branch
stars.  Open symbols denote detections making use of main sequence
stars.  Cross-like symbols are detections based on red giant branch
or asymptotic giant branch (i.e. carbon) stars.  The clusters Pal 12
(Dinescu et al. 2000) and NGC 5634 (Bellazzini et al. 2002) are
shown by circled plus symbols.  In some cases the papers 
cited give either a range of distance, an uncertainty of
distance, or a range of longitude for their Sgr detections.  
These ranges are indicated by
solid lines connecting points.  In each case, the symbols are sized to indicate
relative proximity of the detection, at the cited distance, to
the $\Lambda_{GC}$ Sgr midplane (an approximate size scale is
shown in the legend to the right).  In the case when ranges of values
are shown, the endpoint sizes correspond to 
the relative $Z_{Sgr,GC}$ distance at that point.  
The Martinez-Delgado et al. (2002)
and Dinescu et al. (2000) symbols have both been shifted by one
degree of longitude away from each other, respectively, for clarity.  
To reproduce the Ibata et al. (2001b) carbon star sample, only
Totten \& Irwin (1998) carbon stars with $11 < R < 17$ and having
radial velocities are used, and this sample is trimmed to only
stars within 12 kpc of the Sgr plane.  Obviously dusty carbon stars for which
only an upper limit to distance has been given by Totten \& Irwin
(1998) have been left out.  }
\end{figure}

\begin{figure}
\plotfiddle{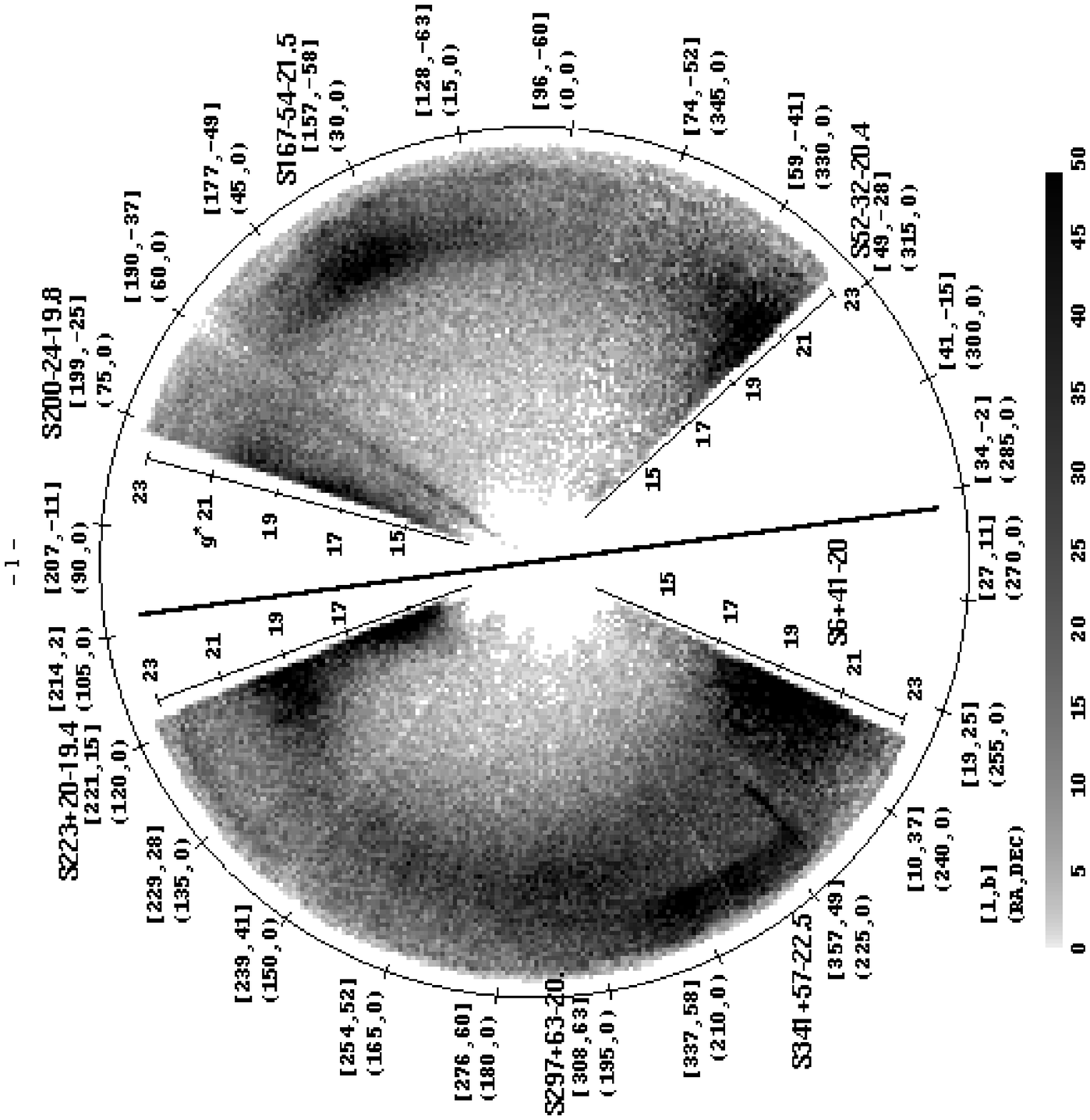}{3.65in}{270}{48}{48}{-180}{310}
\plotfiddle{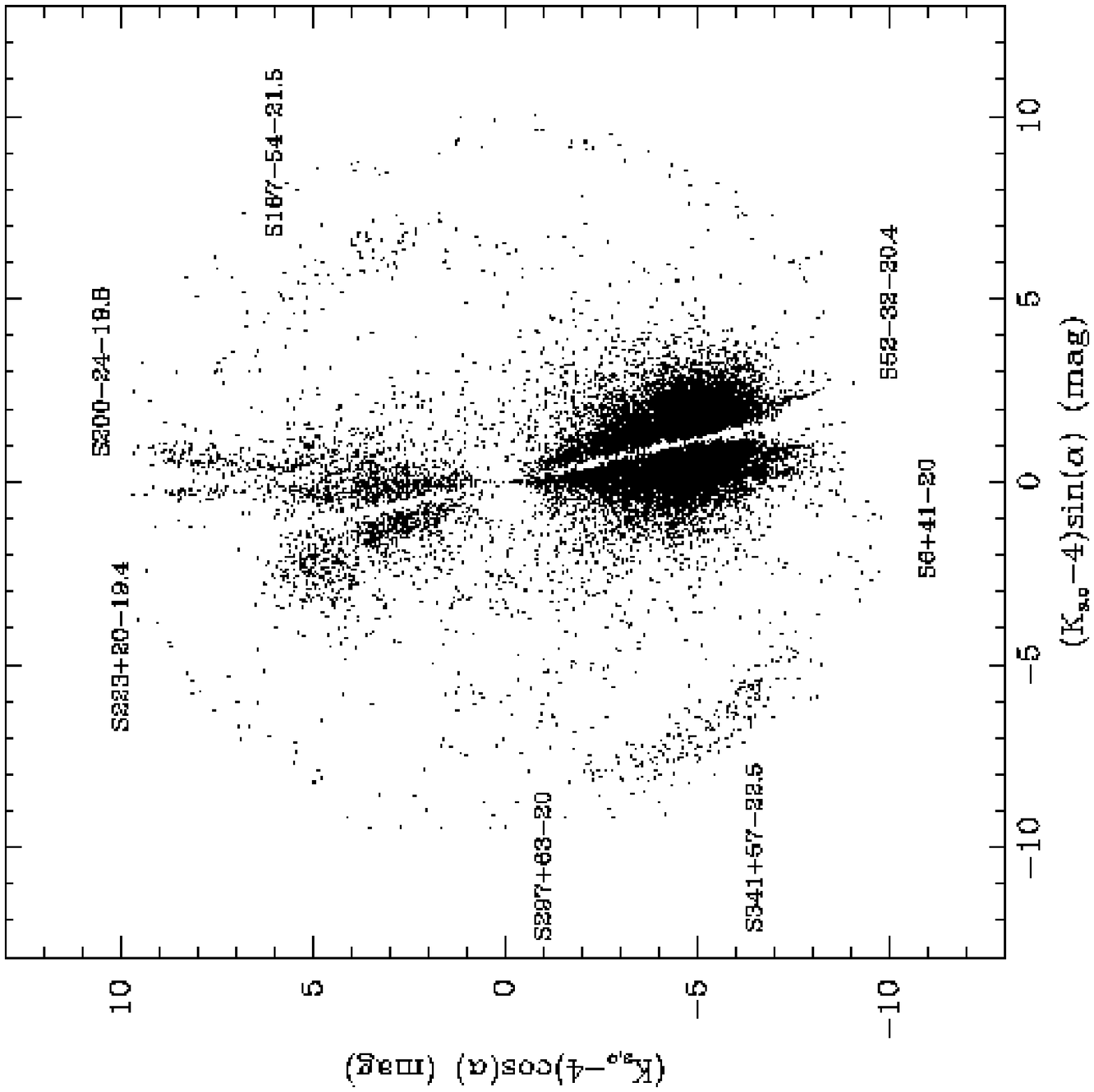}{3.5in}{270}{52}{52}{-210}{300}
\caption{{\it Upper panel:} Main sequence turnoff stars from the Sloan
Digitial Sky Survey equatorial slice by Newberg et al. (2002), reprinted
by permission of Heidi Newberg.
{\it Lower panel:} Celestial Equator slice of the 2MASS Mgiants for comparison to the 
Sloan Digital Sky Survey (Newberg et al. 2002, Figure 1).  All stars in the M giant sample 
within $10^{\circ}$ of the Celestial Equator, and having $1.00 < (J-K_s)_o < 1.10$ 
are used in this rendition.  We exclude stars with $|b|<5^{\circ}$.  The azimuthal directions
of features identified by Newberg et al. (2002) are indicated.  Two spikes
appearing at the very top of the figure are from inexact dereddening at this
low latitude.  }
\end{figure}

\begin{figure}
\plotfiddle{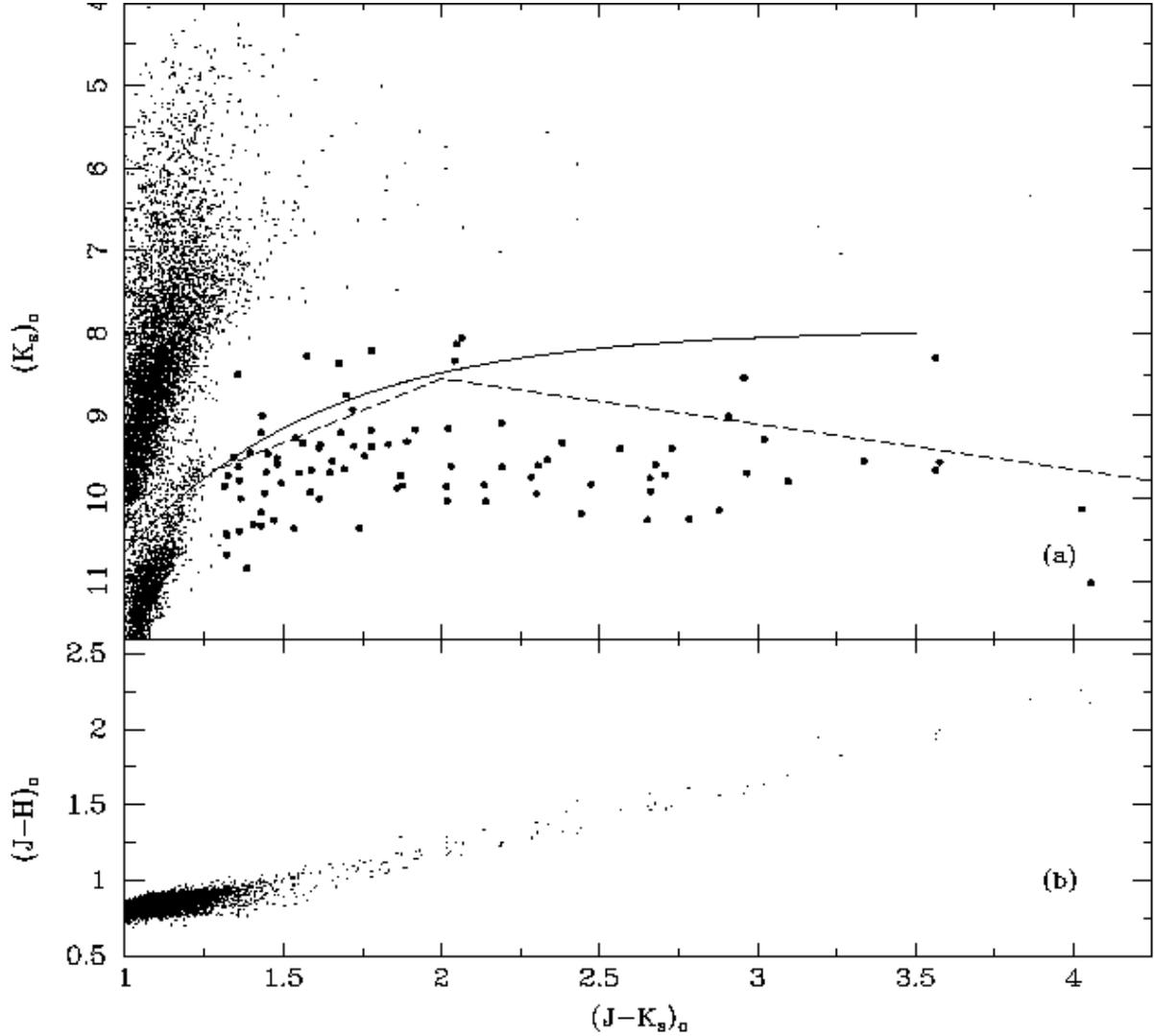}{6in}{270}{70}{70}{-280}{440}
\caption{Color magnitude diagram of stars within 5 degrees of the Sgr center (Table
1) and highlighting the carbon star population.  The {\it solid line}
is the mean carbon star color-magnitude relation from Totten, Irwin \& Whitelock
(2000) derived as a fit to the NIR photometry of a sample of carbon stars from
Milky Way satellite galaxies, converted to 2MASS colors (Carpenter 2000) 
and shifted to the Sgr distance modulus ($m-M=16.9$).
The {\it dashed line} shows the 
approximate ridge line for LMC carbon stars in Weinberg \& Nikolaev (2001),
shifted 1.65 mag brighter to account for the distance modulus difference between Sgr
and the LMC.  Points used in the various fits
discussed in the text are marked with larger points.
Stars in this plot also obey the following dereddened color
criteria: $(J-H) > 0.40(J-K_s) + 0.25$ and $(J-H) < 0.561(J-K_s) + 0.36$.  }
\end{figure}

\begin{figure}
\plotfiddle{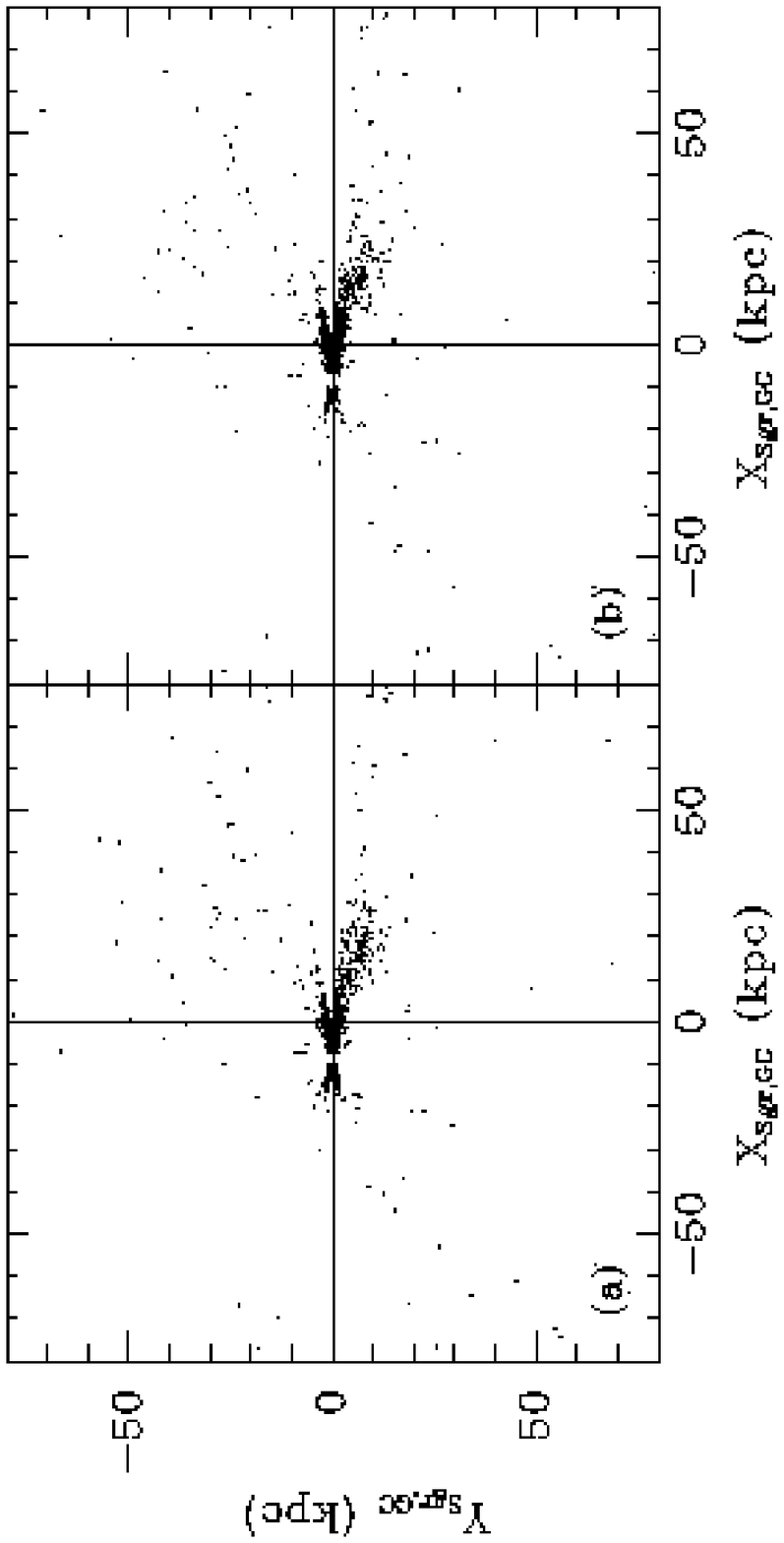}{3.5in}{270}{80}{80}{-300}{290}
\plotfiddle{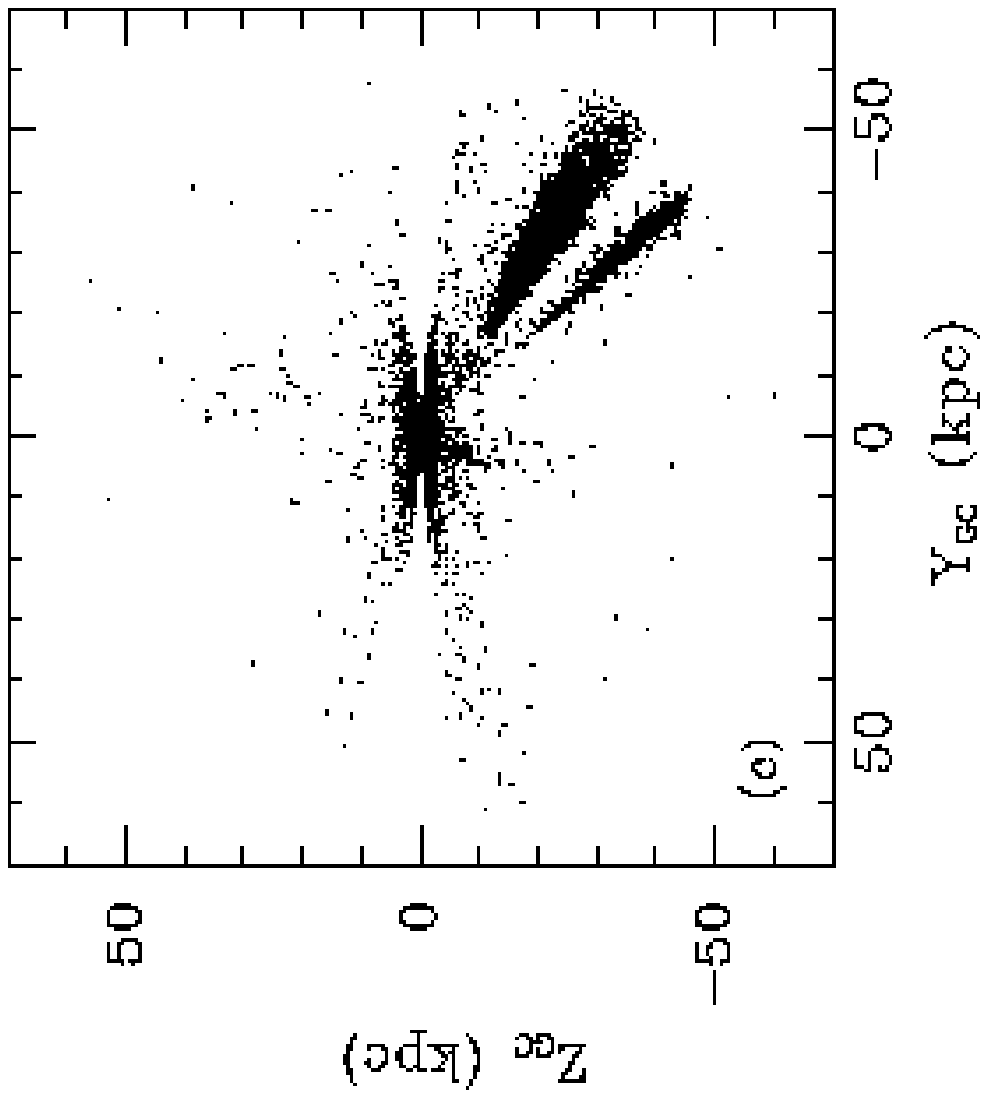}{3.0in}{270}{80}{80}{-200}{260}
\caption{Planar distribution of all carbon star candidates within $|B| = 10^{\circ}$
of the $\Lambda_{\sun}$ plane and having $(J-K_s)_o \ge 1.3$.  (a) Distribution after adopting 
the Weinberg \& Nikolaev (2001) LMC color-magnitude ridge lines, adjusted to the distance
modulus of Sgr and then dimmed an additional 0.5 mag.  (b) Distribution after
assuming all carbon stars have $M_{K_s} = -7.31$.  In both panels, a large
number of stars - likely contaminants - have projected photometric parallaxes
beyond the bounds of the region shown.  A small hole in the distribution
near the Sun is from carbon stars incompleteness at the bright end of the 
catalogue used here.   (c) Nearly edge-on view of carbon
star sample with distances as in panel (b).  In this panel all carbon star 
candiates with  $(J-K_s)_o \ge 1.3$ are shown, but, for clarity, the 
sample has been limited to $(K_s)_o < 11.75$.  
}
\end{figure}

\begin{figure}
\plotfiddle{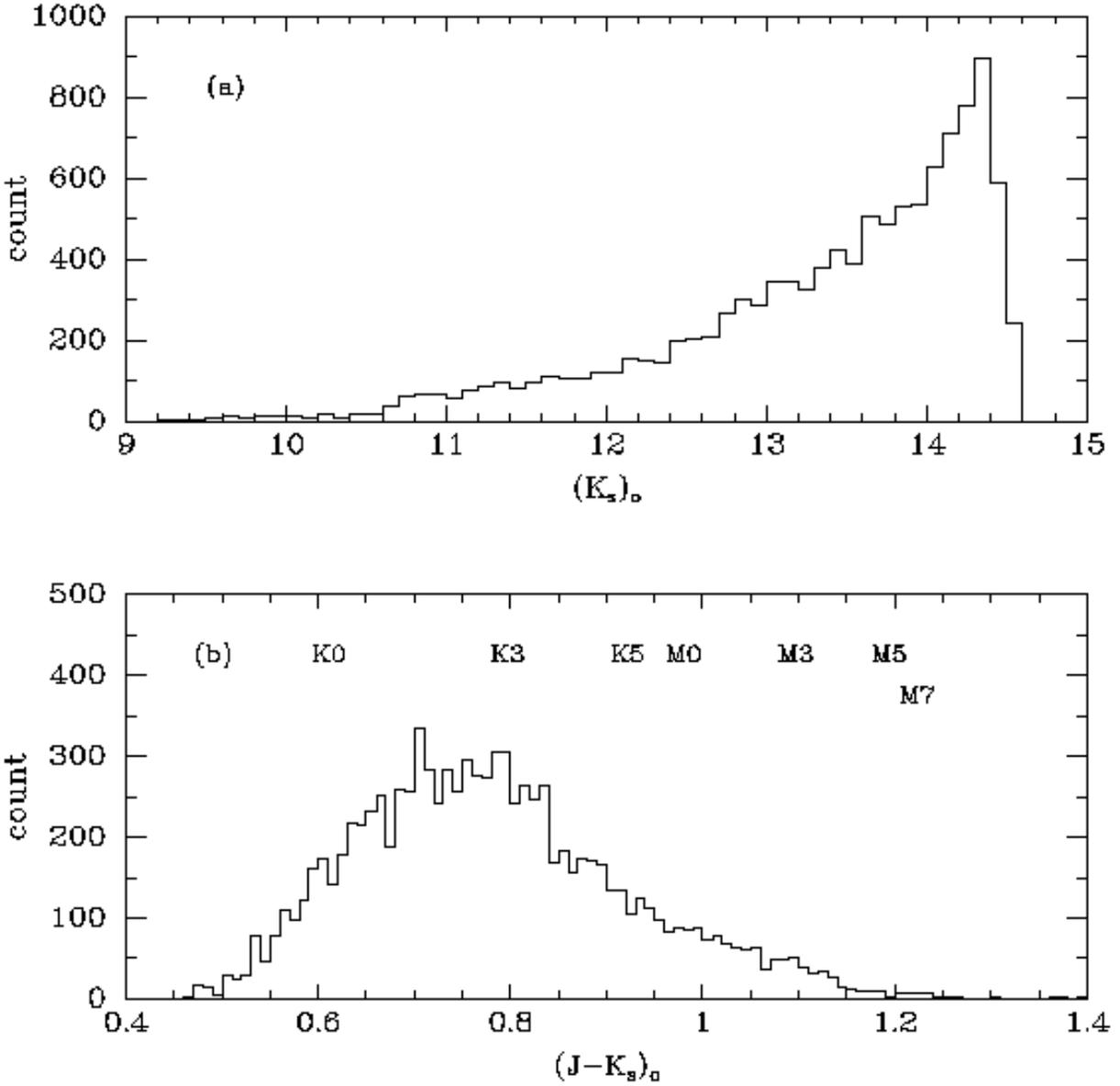}{5in}{270}{80}{80}{-290}{470}
\caption{(a) Luminosity function for Sgr RGB stars shown in
Figure 1c  
isolated by the relation $(K_s)_o > -7.22(J-K_s)_o + 17.64$.
(b) Color function for stars with $(K_s)_o < 14.3$. 
The color function becomes incomplete for RGB stars
bluer than $(J-K_s)_o \sim 0.80$.  Approximate colors 
(ignoring metallicity effects) for spectral
types for luminosity class III objects
from Bessell \& Brett (1988) are indicated (converted
to the 2MASS system using Carpenter 2001).  }
\end{figure}

\clearpage

\begin{table}
\begin{center}
\caption{Profile Fits to the Sgr Main Body}
\begin{tabular}{lcc}
\baselineskip12pt

                                &                           &                      \\
\hline
parameter                       &   King Profile Fit        &  Power Law + Core Fit \\
\hline

$\alpha_{center}$(deg)          & $283.7467\pm0.0133$       & $283.8313\pm0.0034$  \\
$\delta_{center}$(deg)          & $-30.4606\pm0.0256$       & $-30.5454\pm0.0114$    \\ 
$\alpha_{center}$(2000)         & 18:54:59.2$\pm$00:00:03.2 & 18:55:19.5$\pm$00:00:00.8 \\
$\delta_{center}$(2000)         & -30:27:38$\pm$00:01:32    & -30:32:43$\pm$00:00:41 \\     
$ l $                           &   5.6193                  &   5.5690               \\
$ b $                           & -14.0660                  & -14.1665               \\  
 background (stars arcmin$^2$)  & $1.422\times10^{-5}$      & $2.016\pm0.623 \times10^{-6}$ \\
 position angle (deg)           & $104.3 \pm 0.6$           & $100.2 \pm 0.6$      \\
 ellipticity                    & $0.65 \pm 0.01$           & $0.62 \pm 0.01$      \\
 core radius (arcmin)           & $224 \pm 12$              & $234 \pm 10$         \\
 King limiting radius (arcmin)  & $1801 \pm 112$            & ...                  \\
 Power Law index, $\nu$         &  ...                      & $1.44 \pm 0.03$      \\

\hline
\tablecomments{
The Power Law + Core fit is to the data in Figure 4 trimmed only
by $b < -10^{\circ}$.  
The King parameterization is fit to the Figure 4 data trimmed both by
$\alpha_{2000} < 300^{\circ}$ and $b < -10^{\circ}$.  This King parameterization
is shown in Figures 5b and 5c.
The fit is robust to varying the right ascension cutoff
to more conservatively exclude tidal features along the
major axis.
The errors for all parameters given have been derived using
a Metropolis Markov Chain algorithm (see Ostheimer et al. 2002).}

\end{tabular}
\end{center}
\end{table}

\begin{table}
\begin{center}
\caption{Eulerian Transformations to the Sagittarius Coordinate Systems}
\begin{tabular}{lcccccc}
\baselineskip12pt
\renewcommand\baselinestretch{1.0}

                                &                           &                      \\
System & \multicolumn{3}{c}{Euler angles}  & \multicolumn{3}{c}{rotation center}   \\ 
\tableline
        & $\phi$ & $\theta$ & $\psi^a$ & $X_{GC}$ & $Y_{GC}$ & $Z_{GC}$\\
        & (deg) & (deg) & (deg)    & (kpc)  & (kpc)            & (kpc) \\
\tableline
$(\Lambda, B)_{\sun}$ & 183.8   & 76.5    & 194.1 & 0.0              & 0.0              & 0.0 \\
$(\Lambda, B)_{GC}  $ & 183.8   & 76.5    & 201.6 & -8.51           & -0.21  & -0.05 \\
\hline
\tablecomments{a - Adopted as 180.0$^{\circ}$ for some figures in order
to keep the intersection of the Sgr and Galactic planes horizontal.}
\end{tabular}
\end{center}
\end{table}

\begin{table}
\begin{center}
\caption{Color of the RGB Tip for Sgr Populations}
\begin{tabular}{lcc}
\baselineskip12pt
\renewcommand\baselinestretch{1.0}

        &         &               \\ 
\hline
age (Gyr)   & [Fe/H]  &  $(J-K_s)_{2MASS}$ \\
\hline
10-11       & -1.3    &  0.968   \\
5           & -0.7    &  1.035        \\
0.5-3       & -0.4    &  0.665-1.114  \\

\hline
\tablecomments{RGB tip colors from Bertelli et al. (1994) with
conversions from Bessell \& Brett (1988) colors to 2MASS colors
using transformations in Carpenter (2001). }
\end{tabular}
\end{center}
\end{table}

\begin{table}
\begin{center}
\caption{Relative Counts of Stars in the Sgr Center by Spectral Type and Color }
\begin{tabular}{ccc}
\baselineskip12pt
\renewcommand\baselinestretch{1.0}

                &          &            \\ 
\hline
Spectral type & Adopted $(J-K_s)_o$ Range &  Counts  \\
\hline
$>$M0III      & $>0.980$                 &  1000      \\
K3III-M0III   & $0.797-0.980$            &  3009      \\
K0III-K3III   & $0.611-0.797$            & $>4678$    \\
\hline
\tablecomments{Color definitions from Bessell \& Brett (1988)
translated to 2MASS system using Carpenter (2001).  
}
\end{tabular}
\end{center}
\end{table}


\begin{thebibliography}{}

\bibitem[Alard(1996)]{alard1996} Alard, C.\ 1996, \aap, 458, L17 

\bibitem[Alard(2001)]{alard2001} Alard, C.\ 2001, \aap, 377, 389 

\bibitem[Alcock et al.\,(1997a)]{alcock97a}
Alcock, C. et al.\ 1997a, \apj, 474, 217


\bibitem[Amendt \& Cuddeford(1994)]{1994ApJ...435...93A} Amendt, P.~\& 
Cuddeford, P.\ 1994, \apj, 435, 93 

\bibitem[Aparicio, Carrera, \& 
Mart{\'{\i}}nez-Delgado(2001)]{2001AJ....122.2524A} Aparicio, A., Carrera, 
R., \& Mart{\'{\i}}nez-Delgado, D.\ 2001, \aj, 122, 2524 


\bibitem[Bassino \& Muzzio(1995)]{1995Obs...115..256B} Bassino, L.~P.~\& 
Muzzio, J.~C.\ 1995, The Observatory, 115, 256 


\bibitem[Bellazzini, Ferraro, \& Buonanno(1999a)]{1999MNRAS.304..633B} 
Bellazzini, M., Ferraro, F.~R., \& Buonanno, R.\ 1999, \mnras, 304, 633 

\bibitem[Bellazzini, Ferraro, \& Buonanno(1999b)]{bfb1999} 
Bellazzini, M., Ferraro, F.~R., \& Buonanno, R.\ 1999, \mnras, 307, 619 

\bibitem[Bellazzini, Ferraro, \& Ibata(2002a)]{2002AJ....124..915B} 
Bellazzini, M., Ferraro, F.~R., \& Ibata, R.\ 2002a, \aj, 124, 915 

\bibitem[Bellazzini, Ferraro, \& Ibata(2003)]{2003AJ....125..188B} 
Bellazzini, M., Ferraro, F.~R., \& Ibata, R.\ 2003, \aj, 125, 188 

\bibitem[Bellazzini et al.(2002b)]{2002AJ....124.3222B} Bellazzini, M., 
Ferraro, F.~R., Origlia, L., Pancino, E., Monaco, L., \& Oliva, E.\ 2002b, 
\aj, 124, 3222 

\bibitem[Bertelli et al.(1994)]{1994A&AS..106..275B} Bertelli, G., Bressan, 
A., Chiosi, C., Fagotto, F., \& Nasi, E.\ 1994, \aaps, 106, 275 

\bibitem[Bessell \& Brett(1988)]{1988PASP..100.1134B} Bessell, M.~S.~\& 
Brett, J.~M.\ 1988, \pasp, 100, 1134 

\bibitem[Binney, May, \& Ostriker(1987)]{1987MNRAS.226..149B} Binney, J., 
May, A., \& Ostriker, J.~P.\ 1987, \mnras, 226, 149 

\bibitem[Binney \& Tremaine(1987)]{1987gady.book.....B} Binney, J.~\& 
Tremaine, S.\ 1987, Princeton, NJ, Princeton University Press, 1987, 747 
p. 236  

\bibitem[Bonifacio, Pasquini, Molaro, \& 
Marconi(1999)]{BPMM99} Bonifacio, P., Pasquini, L., Molaro, 
P., \& Marconi, G.\ 1999, \apss, 265, 541 

\bibitem[Bullock et al.~(2001)Bullock, Kravtsov \& Weinberg]{bkw01}
Bullock, J. S., Kravtsov, A. V. \& Weinberg, D. H. 2001, \apj, 548, 33

\bibitem[Burkert(1997)]{1997ApJ...474L..99B} Burkert, A.\ 1997, \apjl, 474, 
L99 

\bibitem[Burton \& Lockman(1999)]{1999A&A...349....7B} Burton, W.~B.~\& 
Lockman, F.~J.\ 1999, \aap, 349, 7 

\bibitem[Caldwell, Armandroff, Seitzer, \& Da 
Costa(1992)]{1992AJ....103..840C} Caldwell, N., Armandroff, T.~E., Seitzer, 
P., \& Da Costa, G.~S.\ 1992, \aj, 103, 840 

\bibitem[Carpenter(2001)]{Carp2001} Carpenter, J.~M.\ 2001, \aj, 
121, 2851 

\bibitem[Chiba \& Beers(2000)]{2000AJ....119.2843C} Chiba, M.~\& Beers, 
T.~C.\ 2000, \aj, 119, 2843 

\bibitem[Cole(2001)]{cole2001} Cole, A.~A.\ 2001, \apjl, 559, L17

\bibitem[Combes, Leon, \& Meylan(1999)]{1999A&A...352..149C} Combes, F., 
Leon, S., \& Meylan, G.\ 1999, \aap, 352, 149 

\bibitem[Crane et al.(2003)]{crane2003} Crane, J.~D., Majewski, S.~R.,
Rocha-Pinto, H.~J., Frinchaboy, P.~F., Skrutskie, M.~F. \& Law, D.~R. 2003,
ApJLett, {\it in press} (astro-ph/0307505) 

\bibitem[Cseresnjes(2001)]{2001A&A...375..909C} Cseresnjes, P.\ 2001, \aap, 
375, 909 

\bibitem[Cseresnjes, Alard, \& Guibert(2000)]{2000A&A...357..871C} 
Cseresnjes, P., Alard, C., \& Guibert, J.\ 2000, \aap, 357, 871 

\bibitem[Cutri (2003)]{} Cutri, R. 2003, {\it in preparation}

\bibitem[Cutri et al.(2003)]{} Cutri, R., Skrutskie, M.F., Van Dyk, S., Beichman, C.A., 
Carpenter, J.M., Chester, T., Cambresy, L., Evans, T., et al.  2003,
Explanatory Supplement to the 2MASS All Sky Data Release, 
http://www.ipac.caltech.edu/2mass/releases/allsky/doc/ 

\bibitem[Da Costa \& Armandroff(1995)]{1995AJ....109.2533D} Da Costa, 
G.~S.~\& Armandroff, T.~E.\ 1995, \aj, 109, 2533

\bibitem[Da Costa, Armandroff, Caldwell, \& 
Seitzer(1996)]{1996AJ....112.2576D} Da Costa, G.~S., Armandroff, T.~E., 
Caldwell, N., \& Seitzer, P.\ 1996, \aj, 112, 2576 

\bibitem[Demers, Dallaire, \& Battinelli(2002)]{2002AJ....123.3428D} 
Demers, S., Dallaire, M., \& Battinelli, P.\ 2002, \aj, 123, 3428 

\bibitem[Dinescu, Majewski, Girard, \& Cudworth(2000)]{DMGC00} 
Dinescu, D.~I., Majewski, S.~R., Girard, T.~M., \& Cudworth, K.~M.\ 2000, 
\aj, 120, 1892 

\bibitem[Dinescu, Majewski, Girard, \& Cudworth(2001)]{DMGC01} 
Dinescu, D.~I., Majewski, S.~R., Girard, T.~M., \& Cudworth, K.~M.\ 2001, 
\aj, 122, 1916 

\bibitem[Dinescu et al.(2002)]{2002ApJ...575L..67D} Dinescu, D.~I.~et al.\ 
2002, \apjl, 575, L67 

\bibitem[Dohm-Palmer et al.\,(2001)]{spag5} Dohm-Palmer, R. C., Helmi,
 A., Morrison, H., Mateo, M., Olszewski, E. W., Harding, P., Freeman,
 K. C., Norris, J. \& Shectman, S. A. 2001, \apjl, 555, L37

\bibitem[Dubinski \& Carlberg(1991)]{1991ApJ...378..496D} Dubinski, J.~\& 
Carlberg, R.~G.\ 1991, \apj, 378, 496
  
\bibitem[Edelsohn \& Elmegreen(1997)]{EE97} Edelsohn, 
D.~J.~\& Elmegreen, B.~G.\ 1997, \mnras, 290, 7 

\bibitem[Eggen et al.\,(1962)Eggen, Lynden-Bell \& Sandage]{els62}
  Eggen, O. J. Lynden-Bell, D. \& Sandage, A. R. 1962, \apj, 136, 748

\bibitem[Elias, Frogel, Matthews, \& Neugebauer(1982)]{1982AJ.....87.1029E} 
Elias, J.~H., Frogel, J.~A., Matthews, K., \& Neugebauer, G.\ 1982, \aj, 
87, 1029

\bibitem[Font et al. (2001)]{font01}
 Font, A.S., Navarro, J.F., Stadel, J. \& Quinn, T. 2001,
\apjl, 563, L1

\bibitem[Freeman(1993)]{1993gcgc.work..608F} Freeman, K.~C.\ 1993, ASP 
Conf.~Ser.~48: The Globular Cluster-Galaxy Connection, 608 

\bibitem[Frenk, White, Davis, \& Efstathiou(1988)]{frenk1988} 
Frenk, C.~S., White, S.~D.~M., Davis, M., \& Efstathiou, G.\ 1988, \apj, 
327, 507

\bibitem[Gilmore, Wyse, \& Norris(2002)]{2002ApJ...574L..39G} Gilmore, G., 
Wyse, R.~F.~G., \& Norris, J.~E.\ 2002, \apjl, 574, L39

\bibitem[Glass(1975)]{1975MNRAS.171P..19G} Glass, I.~S.\ 1975, \mnras, 171, 
19P

\bibitem[Goldstein(1980)]{goldstein1980} Goldstein, H. 1980, 
Classical Mechanics, Second Edition, (Reading: Addison-Wesley), \S4.4, 
pp. 143-148

\bibitem[G{\' o}mez-Flechoso(1998)]{1998Ap&SS.263..155G} G{\' 
o}mez-Flechoso, M.~A.\ 1998, \apss, 263, 155

\bibitem[G{\' o}mez-Flechoso, Fux, \& Martinet(1999)]{GFM99} 
G{\' o}mez-Flechoso, M.~A., Fux, R., \& Martinet, L.\ 1999, \aap, 347, 77 

\bibitem[G{\' o}mez-Flechoso \& Mart{\'i}nez-Delgado (2003)]{GFMD03} 
G{\' o}mez-Flechoso, M.~A. \& Mart{\' i}nez-Delgado, D. 2003, ApJL, {\it in press}

\bibitem[Harbeck et al.(2001)]{2001AJ....122.3092H} Harbeck, D.~et al.\ 
2001, \aj, 122, 3092 

\bibitem[Helmi \& White (2001)]{HW01} Helmi, A. \& White, S.D.M. 2001, 
\mnras, 323, 529

\bibitem[Helmi, White, de Zeeuw, \& Zhao(1999)]{1999Natur.402...53H} Helmi, 
A., White, S.~D.~M., de Zeeuw, P.~T., \& Zhao, H.\ 1999, \nat, 402, 53 

\bibitem[Helmi et al.(2003)]{helmi} Helmi, A., Navarro, J. F., Meza, A.,
Steinmetz, M., \& Eke, V. R. 2003, \apjl, 592, L25

\bibitem[Hurley-Keller, Mateo, \& Grebel(1999)]{1999ApJ...523L..25H} 
Hurley-Keller, D., Mateo, M., \& Grebel, E.~K.\ 1999, \apjl, 523, L25 

\bibitem[Ibata(1999)]{Ibata99} Ibata, R.~A.\ 1999, IAU 
Symp.~186: Galaxy Interactions at Low and High Redshift, 186, 39 

\bibitem[Ibata, Gilmore, \& Irwin(1994)]{1994Natur.370..194I} Ibata, R.~A., 
Gilmore, G., \& Irwin, M.~J.\ 1994, \nat, 370, 194 

\bibitem[Ibata et al.~(1995)Ibata, Gilmore \& Irwin]{igi95}
Ibata, R. A.,  Gilmore, G. \& Irwin, M. J. 1995, \mnras, 277, 781

\bibitem[Ibata et al.~(2003)]{ibataetal2003}
Ibata, R.~A., Irwin, M.~J., Lewis, G.~F., Ferguson, A.~M.~N. \& Tanvir, N. 2003,
astro-ph/0301067

\bibitem[Ibata, Irwin, Lewis, \& Stolte(2001a)]{2001ApJ...547L.133I} Ibata, 
R., Irwin, M., Lewis, G.~F., \& Stolte, A.\ 2001a, \apjl, 547, L133 

\bibitem[Ibata \& Lewis (1998)]{IL98} Ibata, R.~A.~\& Lewis, 
G.~F.\ 1998, \apj, 500, 575 

\bibitem[Ibata, Lewis, Irwin, \& Cambr{\' e}sy (2002a)]{ILIC2002} 
Ibata, R.~A., Lewis, G.~F., Irwin, M.~J., \&  Cambr{\' e}sy, L.\ 2002, 
\mnras, 332, 921 

\bibitem[Ibata, Lewis, Irwin, \& Quinn (2002b)]{ILIQ02} Ibata, 
R.~A., Lewis, G.~F., Irwin, M.~J., \& Quinn, T.\ 2002, \mnras, 332, 915 

\bibitem[Ibata et al.(2001b)]{2001ApJ...551..294I} Ibata, R., Lewis, G.~F., 
Irwin, M., Totten, E., \& Quinn, T.\ 2001b, \apj, 551, 294 

\bibitem[Ibata et al.(1997)]{Ibata97} Ibata, R.~A., Wyse, 
R.~F.~G., Gilmore, G., Irwin, M.~J., \& Suntzeff, N.~B.\ 1997, \aj, 113, 634 

\bibitem[Irwin(1999)]{Irwin1999} Irwin, M.\ 1999, IAU Symposium, 
192, 409 

\bibitem[Irwin \& Hatzidimitriou(1995)]{1995MNRAS.277.1354I} Irwin, M.~\& 
Hatzidimitriou, D.\ 1995, \mnras, 277, 1354 

\bibitem[Ivezi{\' c} et al.(2000)]{2000AJ....120..963I} Ivezi{\' c}, 
{\v Z}., et al.\ 2000, \aj, 120, 963 

\bibitem[Jiang \& Binney(2000)]{JB00} Jiang, I.~\& Binney, 
J.\ 2000, \mnras, 314, 468 

\bibitem[Johnston(1998)]{KVJ98}
Johnston, K. V. 1998, \apj, 495, 297

\bibitem[Johnston, Choi, \& Guhathakurta(2002)]{2002AJ....124..127J} 
Johnston, K.~V., Choi, P.~I., \& Guhathakurta, P.\ 2002, \aj, 124, 127 

\bibitem[Johnston, Hernquist, \& Bolte(1996)]{JHB96} 
Johnston, K.~V., Hernquist, L., \& Bolte, M.\ 1996, \apj, 465, 278 

\bibitem[Johnston et al.(1999a)]{johnstonetal1999} Johnston, K.~V., 
Majewski, S.~R., Siegel, M.~H., Reid, I.~N., \& Kunkel, W.~E.\ 1999, \aj, 
118, 1719 

\bibitem[Johnston, Sackett, \& Bullock(2001)]{2001ApJ...557..137J} 
Johnston, K.~V., Sackett, P.~D., \& Bullock, J.~S.\ 2001, \apj, 557, 137 

\bibitem[Johnston, Sigurdsson \& Hernquist (1999)]{JSH99a}
Johnston, K. V., Sigurdsson, S. \& Hernquist, L.  1999a,
\mnras, 302

\bibitem[Johnston, Spergel, \& Haydn(2002)]{JSH02} Johnston, 
K.~V., Spergel, D.~N., \& Haydn, C.\ 2002, \apj, 570, 656 

\bibitem[Johnston, Spergel, \& Hernquist(1995)]{1995ApJ...451..598J} 
Johnston, K.~V., Spergel, D.~N., \& Hernquist, L.\ 1995, \apj, 451, 598 

\bibitem[Johnston, et al. (1999b)]{JZSH99} 
Johnston, K.~V., Zhao, H., Spergel, D.~N., \& Hernquist, L.\ 1999, \apjl, 
512, L109 

\bibitem[King(1962)]{1962AJ.....67..471K} King, I.\ 1962, \aj, 67, 471 

\bibitem[King(1966)]{1966AJ.....71...64K} King, I.~R.\ 1966, \aj, 71, 64 

\bibitem[Kinman et al.(1996)]{Kinman1996} Kinman, T.~D., Pier, 
J.~R., Suntzeff, N.~B., Harmer, D.~L., Valdes, F., Hanson, R.~B., Klemola, 
A.~R., \& Kraft, R.~P.\ 1996, \aj, 111, 1164 

\bibitem[Kinman, Suntzeff, \& Kraft(1994)]{1994AJ....108.1722K} Kinman, 
T.~D., Suntzeff, N.~B., \& Kraft, R.~P.\ 1994, \aj, 108, 1722 

\bibitem[Klessen \& Kroupa(1998)]{KK98} Klessen, R.~S.~\& Kroupa, P.\ 1998, \apj, 498, 143 

\bibitem[Kleyna et al.(1998)]{1998AJ....115.2359K} Kleyna, J.~T., Geller, 
M.~J., Kenyon, S.~J., Kurtz, M.~J., \& Thorstensen, J.~R.\ 1998, \aj, 115, 
2359

\bibitem[Kleyna et al.(2002)]{2002MNRAS.330..792K} Kleyna, J., Wilkinson, 
M.~I., Evans, N.~W., Gilmore, G., \& Frayn, C.\ 2002, \mnras, 330, 792 

\bibitem[Klypin, Kravtsov, Valenzuela, \& Prada(1999)]{KKVP99} 
Klypin, A., Kravtsov, A.~V., Valenzuela, O., \& Prada, F.\ 1999, \apj, 522, 
82 

\bibitem[Kocevski \& Kuhn(2000)]{2000AAS...197.3004K} Kocevski, D.~D.~\& 
Kuhn, J.~R.\ 2000, American Astronomical Society Meeting, 197,  

\bibitem[Koribalski, Johnston, \& Otrupcek(1994)]{1994MNRAS.270L..43K} 
Koribalski, B., Johnston, S., \& Otrupcek, R.\ 1994, \mnras, 270, L43 

\bibitem[Kroupa(1997)]{1997NewA....2..139K} Kroupa, P.\ 1997, New 
Astron.~2, 139-164, 2, 139 

\bibitem[Kuhn \& Miller(1989)]{1989ApJ...341L..41K} Kuhn, J.~R.~\& Miller, 
R.~H.\ 1989, \apjl, 341, L41 

\bibitem[Kuhn, Smith, \& Hawley(1996)]{1996ApJ...469L..93K} Kuhn, J.~R., 
Smith, H.~A., \& Hawley, S.~L.\ 1996, \apjl, 469, L93 

\bibitem[Kundu et al. (2002)]{Kundu} Kundu, A., {\it et al.} 2002, \apjl,
576, L125.

\bibitem[Kunkel, Irwin, \& Demers(1997)]{1997A&AS..122..463K} Kunkel, 
W.~E., Irwin, M.~J., \& Demers, S.\ 1997, \aaps, 122, 463 

\bibitem[Law et al.(2003)]{} Law, D.~R., Johnston, K.~V., Majewski, S.~R.
\& Skrutskie, M.~F. 2003, {\it in preparation}

\bibitem[Layden \& Sarajedini(2000)]{2000AJ....119.1760L} Layden, A.~C.~\& 
Sarajedini, A.\ 2000, \aj, 119, 1760 

\bibitem[Lee(1970)]{1970ApJ...162..217L} Lee, T.~A.\ 1970, \apj, 162, 217

\bibitem[Light (1988)] {} Light, R.M. 1988, Ph.D. Thesis, Yale Univ. 

\bibitem[Lynden-Bell(1982)]{LB1982} Lynden-Bell, D.\ 1982, The 
Observatory, 102, 202

\bibitem[Majewski\ (1992)]{Majewski1992} Majewski, S.~R.\ 1992, \apjs, 
78, 8

\bibitem[Majewski(1993)]{1993ARA&A..31..575M} Majewski, S.~R.\ 1993, \araa, 
31, 575

\bibitem[Majewski\ (2003)]{Majewski2003} Majewski, S.~R.\ 2003, 
in New Horizons in Globular Cluster Astronomy, ed. G. Piotto, 
 ASP Conf. Ser. Vol., {\it in press}. 

\bibitem[Majewski et al. (2003)]{PaperII} Majewski, S.~R., Law, D.~R.,
Johnston, K.~V., Patterson, R.~J., Kunkel, W.~E., Polak, A.~A., 
Frinchaboy, P.~M., Rhee, J. \& Hummels, C.~B. 2003, {\it in preparation.}

\bibitem[Majewski et al.\ (2002a)]{yalepaper}
  Majewski, S. R., et al. 2002a, in Yale Cosmology Workshop: The Shapes of
Galaxies and their Dark Halos, ed. P. Natarajan, (World Scientific: New Jersey), 
p. 214.

\bibitem[Majewski et al.\ (2002b)]{heidelberg}
  Majewski, S. R., et al. 2002b, in Modes of Star Formation and the Origin
of Field Populations, eds. E. K. Grebel \& W. Brandner, ASP Conf. Ser. Vol. 285, 
p. 199.

\bibitem[Majewski, Munn \& Hawley (1994)]{mmh94}
  Majewski, S. R., Munn, J. A. \& Hawley, S. L. 1994, \apjl, 427, L37
  
\bibitem[Majewski et al.\ (1996)Majewski, Munn \& Hawley]{mmh96}
  Majewski, S. R., Munn, J. A. \& Hawley, S. L. 1996, \apjl, 459, L73
 
\bibitem[Majewski et al.\ (2000)]{carinapaperII} Majewski, S. R.,
  Ostheimer, J. C., Patterson, R. J., Kunkel, W. E., Johnston, K. V. \&
  Geisler, D. 2000, \aj, 119, 760 

\bibitem[Majewski et al.(1999a)]{Maj1999a} Majewski, S.~R., 
Siegel, M.~H., Kunkel, W.~E., Reid, I.~N., Johnston, K.~V., Thompson, 
I.~B., Landolt, A.~U., \& Palma, C.\ 1999a, \aj, 118, 1709 

\bibitem[Majewski, Siegel, Patterson, \& Rood(1999b)]{1999ApJ...520L..33M} 
Majewski, S.~R., Siegel, M.~H., Patterson, R.~J., \& Rood, R.~T.\ 1999b, 
\apjl, 520, L33 

\bibitem[Marconi et al.(1998)]{1998A&A...330..453M} Marconi, G., Buonanno, 
R., Castellani, M., Iannicola, G., Molaro, P., Pasquini, L., \& Pulone, L.\ 
1998, \aap, 330, 453

\bibitem[Mart{\' i}nez-Delgado, Alonso-Garc{\' i}a, Aparicio, \& G{\' 
o}mez-Flechoso(2001)]{2001ApJ...549L..63M} Mart{\' i}nez-Delgado, D., 
Alonso-Garc{\' i}a, J., Aparicio, A., \& G{\' o}mez-Flechoso, M.~A.\ 2001a, 
\apjl, 549, L63 

\bibitem[Mart{\' i}nez-Delgado, Aparicio, G{\' o}mez-Flechoso, \& 
Carrera(2001b)]{MAGC01} Mart{\' i}nez-Delgado, D., Aparicio, 
A., G{\' o}mez-Flechoso, M.~{\' A}ngeles, \& Carrera, R.\ 2001b, \apjl, 549, 
L199 

\bibitem[]{} Mart\'{i}nez-Delgado, D., G\'{o}mez-Flechoso, M. A., \&
Aparicio, A. 2001b, in ASP Conf. Ser.,
Observed HR Diagrams and Stellar Evolution,
ed. T. Lejeune \& J. Fernandes (M01)

\bibitem[Mart{\' i}nez-Delgado, Zinn, Carrera, \& 
Gallart(2002)]{2002ApJ...573L..19M} Mart{\' i}nez-Delgado, D., Zinn, R., 
Carrera, R., \& Gallart, C.\ 2002, \apjl, 573, L19 

\bibitem[Mateo(1998)]{1998ARA&A..36..435M} Mateo, M.~L.\ 1998, \araa, 36, 435 

\bibitem[Mateo, Olszewski, \& Morrison(1998)]{1998ApJ...508L..55M} Mateo, 
M., Olszewski, E.~W., \& Morrison, H.~L.\ 1998, \apjl, 508, L55 

\bibitem[Mateo et al.(1995)]{ogle1995} Mateo, M., Udalski, A., 
Szymanski, M., Kaluzny, J., Kubiak, M., \& Krzeminski, W.\ 1995, \aj, 109, 
588 

\bibitem[Mayer et al.(2001)]{2001ApJ...559..754M} Mayer, L., Governato, F., 
Colpi, M., Moore, B., Quinn, T., Wadsley, J., Stadel, J., \& Lake, G.\ 
2001, \apj, 559, 754 

\bibitem[Mayer et al.(2002)]{2002MNRAS.336..119M} Mayer, L., Moore, B., 
Quinn, T., Governato, F., \& Stadel, J.\ 2002, \mnras, 336, 119 

\bibitem[Merrifield(2002)]{merrifield2002} Merrifield, M.~R.\ 2002, 
in ``The shapes of galaxies and their dark halos, Proceedings of the Yale 
Cosmology Workshop", ed. P. Natarajan (Singapore: World Scientific), p.170

\bibitem[Moore (1996)]{Moo96}  Moore, B. 1996, \apjl, 461, L13

\bibitem[Moore et al.(1999)]{1999ApJ...524L..19M} Moore, B., Ghigna, S., 
Governato, F., Lake, G., Quinn, T., Stadel, J., \& Tozzi, P.\ 1999, \apjl, 
524, L19 

\bibitem[Morrison(1993)]{1993AJ....106..578M} Morrison, H.~L.\ 1993, \aj, 
106, 578

\bibitem[Mould(1976)]{1976A&A....48..443M} Mould, J.~R.\ 1976, \aap, 48, 
443  

\bibitem[Murali \& Dubinski (1999)]{MD99}
Murali, C. \& Dubinski, J. 1999, \aj, 118, 9111

\bibitem[Navarro, Frenk \& White S. D. M. (1996)]{nfw96}
Navarro, J. F., Frenk, C. S. \& White S. D. M. 1996, \apj,
462, 563

\bibitem[Navarro, Frenk \& White S. D. M. (1997)]{nfw97}
 Navarro, J. F., Frenk, C. S. \& White S. D. M. 1997, \apj,
490, 493

\bibitem[Newberg et al.(2002)]{Newberg02} Newberg, H.~J.~et al.\ 
2002, \apj, 569, 245 

\bibitem[Nikolaev \& Weinberg(2000)]{2000ApJ...542..804N} Nikolaev, S.~\& 
Weinberg, M.~D.\ 2000, \apj, 542, 804

\bibitem[Odenkirchen et al.\,(2001a)]{ODEN01}
Odenkirchen, M. et al.\ 2001a, \apjl, 548, L165

\bibitem[Odenkirchen et al.(2001b)]{2001AJ....122.2538O} Odenkirchen, M.~et 
al.\ 2001b, \aj, 122, 2538 

\bibitem[Odenkirchen et al.\,(2003)]{ODEN03}
Odenkirchen, M. et al.\ 2003, \aj, in press (astro-ph/0307446) 

\bibitem[Olling(1997)]{1997dark.conf...44O} Olling, R.~P.\ 1997, Dark Matter in 
Astro- and Particle Physics : (DARK '96), eds. H.V.~Klapdor-Kleingrothaus, 
Y.~Ramachers, (Singapore: World Scientific), p. 44.

\bibitem[Ostheimer et al. (2003)]{Ost02}
Ostheimer, J.~C., Link, R., Majewski, S.~R., Patterson, R.~J. \& Crane, J.~D. 2003,
{\it in preparation}


\bibitem[Palma, Majewski, \& Johnston(2002)]{Palma02} Palma, 
C., Majewski, S.~R., \& Johnston, K.~V.\ 2002, \apj, 564, 736 

\bibitem[Palma, Majewski, \& Johnston(2003)]{Palma03} Palma, 
C., Majewski, S.~R., Siegel, M.S., Patterson, R.J., Ostheimer, J.C. \& 
Link, R.\ 2003, astro-ph/0205194 

\bibitem[Pe{\~ n}arrubia, Boily, Just, \& Kroupa(2000)]{2000AGM....17..P50P} Pe{\~ n}arrubia, J., Boily, C.~M., Just, 
A., \& Kroupa, P.\ 2000, Astronomische Gesellschaft Abstract Series, 
Vol.~17.~Abstracts of Contributed Talks and Posters presented at the Annual 
Scientific Meeting of the Astronomische Gesellschaft at Bremen, p. 50 

\bibitem[Pryor(1996)]{1996fogh.conf..424P} Pryor, C.\ 1996, ASP Conf.~Ser.~ 
92: Formation of the Galactic Halo...Inside and Out, 424 

\bibitem[Ratnatunga \& Freeman(1989)]{1989ApJ...339..126R} Ratnatunga, 
K.~U.~\& Freeman, K.~C.\ 1989, \apj, 339, 126 

\bibitem[Reid \& Majewski(1993)]{1993ApJ...409..635R} Reid, N.~\& Majewski, 
S.~R.\ 1993, \apj, 409, 635 

\bibitem[Richstone \& Tremaine(1986)]{1986AJ.....92...72R} Richstone, 
D.~O.~\& Tremaine, S.\ 1986, \aj, 92, 72 

\bibitem[Robin, Reyl{\' e}, \& Cr{\' e}z{\' e}(2000)]{2000A&A...359..103R} 
Robin, A.~C., Reyl{\' e}, C., \& Cr{\' e}z{\' e}, M.\ 2000, \aap, 359, 103 

\bibitem[Rocha-Pinto et al.(2003)]{helio} Rocha-Pinto, H. J., Majewski, S. R.,
Skrutskie, M. F., \& Crane, J. D. 2003, preprint (astro-ph/0307258) 

\bibitem[Rockosi et al.(2002)]{rockosi2002} Rockosi, C.~M.~et al.\ 
2002, \aj, 124, 349 

\bibitem[Sackett \& Pogge(1995)]{1995dama.conf..141S} Sackett, P.~D.~\& Pogge, R.~W.\ 1995, Dark Matter, 141 

\bibitem[Sackett, Rix, Jarvis, \& Freeman(1994)]{1994ApJ...436..629S} 
Sackett, P.~D., Rix, H., Jarvis, B.~J., \& Freeman, K.~C.\ 1994, \apj, 436, 
629

\bibitem[Samurovi{\' c}, {\' C}irkovi{\' c}, \& Milo{\v s}evi{\' 
c}-Zdjelar(1999)]{1999MNRAS.309...63S} Samurovi{\' c}, S., {\' C}irkovi{\' 
c}, M.~M., \& Milo{\v s}evi{\' c}-Zdjelar, V.\ 1999, \mnras, 309, 63 

\bibitem[Sarajedini \& Layden(1995)]{1995AJ....109.1086S} Sarajedini, A.~\& 
Layden, A.~C.\ 1995, \aj, 109, 1086 

\bibitem[Schlegel, Finkbeiner, \& Davis(1998)]{Schlegel1998} 
Schlegel, D.~J., Finkbeiner, D.~P., \& Davis, M.\ 1998, \apj, 500, 525 

\bibitem[Searle \& Zinn(1978)]{1978ApJ...225..357S} Searle, L.~\& Zinn, R.\ 
1978, \apj, 225, 357

\bibitem[Siegel et al.\,(2002)]
{siegel02} Siegel, M. H., Majewski, S. R., Reid, I. N. \& Thompson, I. 2002, ApJ, 578, 151 

\bibitem[Silvestri, Ventura, D'Antona, \& 
Mazzitelli(1998)]{1998ApJ...509..192S} Silvestri, F., Ventura, P., 
D'Antona, F., \& Mazzitelli, I.\ 1998, \apj, 509, 192 

\bibitem[Sohn et al.\(2003)]{sohn} Sohn, S., Majewski, S.R., Patterson, R.J. \& Siegel, M.H.
2003, {\it in prepartion}

\bibitem[Sparke(2002)]{2002sgdh.conf..178S} Sparke, L.~S.\ 2002, 
Proceedings of the Yale Cosmology 
Workshop "The Shapes of Galaxies and Their Dark Matter Halos", ~ ed. P. 
Natarajan (Singapore: World Scientific), p.178

\bibitem[Steinmetz \& Muller(1995)]{SM1995} Steinmetz, M.~\& 
Muller, E.\ 1995, \mnras, 276, 549

\bibitem[Sung et al.(1998)]{1998ApJ...505..199S} Sung, E., Han, C., Ryden, 
B.~S., Patterson, R.~J., Chun, M., Kim, H., Lee, W., \& Kim, D.\ 1998, 
\apj, 505, 199 

\bibitem[Totten \& Irwin(1998)]{1998MNRAS.294....1T} Totten, E.~J.~\& 
Irwin, M.~J.\ 1998, \mnras, 294, 1 

\bibitem[Totten, Irwin, \& Whitelock(2000)]{2000MNRAS.314..630T} Totten, 
E.~J., Irwin, M.~J., \& Whitelock, P.~A.\ 2000, \mnras, 314, 630 

\bibitem[Trager, Djorgovski, \& King(1995)]{TKD1995} Trager, 
S.~C., Djorgovski, S., \& King, I.~R.\ 1995, AJ, 109, 218

\bibitem[Tremaine (1993)]{tre93} Tremaine, S. 1993, in {\it Back to the
  Galaxy}, eds. S. S.  Holt \& F. Verter, (New York: AIP Conference
  Proceedings),  599


\bibitem[Unavane, Wyse, \& Gilmore(1996)]{1996MNRAS.278..727U} Unavane, M., 
Wyse, R.~F.~G., \& Gilmore, G.\ 1996, \mnras, 278, 727 

\bibitem[van der Marel(1991)]{1991MNRAS.248..515V} van der Marel, R.~P.\ 
1991, \mnras, 248, 515

\bibitem[van der Marel(2001)]{2001AJ....122.1827V} van der Marel, R.~P.\ 
2001, \aj, 122, 1827 

\bibitem[Velazquez \& White(1995)]{1995MNRAS.275L..23V} Vel\'azquez, H.~\& 
White, S.~D.~M.\ 1995, \mnras, 275, L23 

\bibitem[Vivas et al.\,(2001)]{vivas01}
Vivas, A. K. et al.\ 2001, \apjl 554, L33

\bibitem[von Hoerner(1957)]{1957ApJ...125..451V} von Hoerner, S.\ 1957, 
\apj, 125, 451 

\bibitem[Walcher et al.\(2002)]{walcher}
Walcher, J., Fried, J.W., Burkert, A. \& Klessen, R.S. 2002, astro-ph/0207467.

\bibitem[Warren, Quinn, Salmon, \& Zurek(1992)]{WQSZ1992} 
Warren, M.~S., Quinn, P.~J., Salmon, J.~K., \& Zurek, W.~H.\ 1992, \apj, 
399, 405

\bibitem[Weinberg(2000)]{weinberg2000} Weinberg, M.~D.\ 2000, \apj, 
532, 922

\bibitem[Weinberg \& Nikolaev (2001)]{WN01}
Weinberg, M. D., \& Nikolaev, S. 2001, \apj, 548, 712

\bibitem[Westfall et al.(2003)]{westfall2003} Westfall, K.~B., Majewski, S.~R.,
Ostheimer, J.~C., Frinchaboy, P.~M., Patterson, R.~J., \& 
Kunkel, W.~E.\ 2003, {\it in preparation}

\bibitem[Westfall et al.(2000)]{2000AAS...19713301W} Westfall, K.~B., 
Ostheimer, J.~C., Frinchaboy, P.~M., Patterson, R.~J., Majewski, S.~R., \& 
Kunkel, W.~E.\ 2000, American Astronomical Society Meeting, 197,  

\bibitem[Whitelock et al. (1999)]{WMIF99} 
Whitelock, P., Menzies, J., Irwin, M., \& Feast, M.\ 1999, IAU Symposium, 
192, 136 

\bibitem[Yanny et al.(2000)]{Yanny00} Yanny, B.~et al.\ 2000, 
\apj, 540, 825 

\bibitem[Yanny et al.(2003)]{yanny} Yanny, B.~et al. 2003, \apj, 588, 824 (Y03)


\bibitem[Yoss, Neese, \& Hartkopf(1987)]{1987AJ.....94.1600Y} Yoss, K.~M., 
Neese, C.~L., \& Hartkopf, W.~I.\ 1987, \aj, 94, 1600 

\bibitem[Zhao(1998)]{zhao98} Zhao, H.\ 1998, \apjl, 500, L149 

\bibitem[Zinnecker et al.(1988)]{1988IAUS..126..603Z} Zinnecker, H., 
Keable, C.~J., Dunlop, J.~S., Cannon, R.~D., \& Griffiths, W.~K.\ 1988, IAU 
Symp.~126: The Harlow-Shapley Symposium on Globular Cluster Systems in 
Galaxies, 126, 603 

\end{thebibliography}
\end{document}